\documentclass[opre,nonblindrev]{arxiv_version}

\DoubleSpacedXI 

\usepackage{endnotes}
\let\footnote=\endnote

\usepackage{natbib}
 \bibpunct[, ]{(}{)}{,}{a}{}{,}%
 \def\bibfont{\small}%
\TheoremsNumberedThrough     
\ECRepeatTheorems

\EquationsNumberedThrough    

\newcommand{\bbR}{\mathbb{R}}
\newcommand{\bzero}{\boldsymbol{0}}
\newcommand{\calH}{\mathcal{H}}
\newcommand{\blambda}{\boldsymbol{\lambda}}
\newcommand{\bmu}{\boldsymbol{\mu}}
\newcommand{\calK}{\mathcal{K}}
\newcommand{\calC}{\mathcal{C}}
\newcommand{\calA}{\mathcal{A}}
\newcommand{\bx}{\boldsymbol{x}}
\newcommand{\bxi}{\boldsymbol{\xi}}
\newcommand{\ba}{\boldsymbol{a}}
\newcommand{\bbZ}{\mathbb{Z}}
\newcommand\E[1]{\mathbb{E}\left[#1\right]}
\renewcommand\P[1]{\mathbb{P}\left[#1\right]}
\newcommand\inner[2]{\left \langle #1, #2 \right \rangle}
\newcommand{\bs}{\boldsymbol{s}}
\newcommand{\bu}{\boldsymbol{u}}
\newcommand{\bq}{\boldsymbol{q}}
\newcommand{\by}{\boldsymbol{y}}
\newcommand{\bnu}{\boldsymbol{\nu}}
\newcommand{\barq}{\bar{q}}
\newcommand{\bone}{\boldsymbol{1}}

\newcommand{\calV}{\mathcal{V}}

\usepackage{soul}
\newcommand{\calS}{\mathcal{S}}

\newcommand{\calE}{\mathcal{E}}

\newcommand{\calI}{\mathcal{I}}
\newcommand{\calJ}{\mathcal{J}}

\newcommand{\bbarq}{\boldsymbol{\bar{q}}}

\newcommand{\calT}{\mathcal{T}}
\newcommand{\calB}{\mathcal{B}}
\newcommand{\dfn}{\overset{\Delta}{=}}

\newcommand{\tk}{\tilde{k}}

\newcommand{\bomega}{\mathbf{\omega}}
\allowdisplaybreaks 
\usepackage{multirow}
\usepackage{bbm}
\usepackage{tikz}
\usepackage{amsmath}
\usepackage{amsfonts}
\usepackage{hyperref}
\usepackage{array}
\usepackage{stackengine}
\newcommand\xrowht[2][0]{\addstackgap[.5\dimexpr#2\relax]{\vphantom{#1}}}
\usepackage{soul}
\newcommand{\bz}{\boldsymbol{z}}
\usepackage{nicematrix}
\usepackage{colortbl}
\usepackage{algorithmic}
\usepackage[ruled,vlined]{algorithm2e}
\usepackage{etoolbox}
\usepackage{enumerate}
\usepackage{xcolor}

\definecolor{orange}{rgb}{0.90196078431, 0.62352941176, 0}
\definecolor{blue1}{rgb}{0.33725490196, 0.70588235294, 0.91372549019}
\definecolor{terquise}{rgb}{0, 0.61960784313, 0.45098039215}
\definecolor{red1}{rgb}{0.83529411764, 0.38431372549, 0}
\definecolor{pink}{rgb}{0.8, 0.47450980392, 0.65490196078}
\definecolor{yellow}{rgb}{0.8, 0.89411764705, 0.25882352941}
\definecolor{brown1}{rgb}{0, 0.21960784313, 0.31372549019}

\DeclareMathVersion{ttmath}
\SetSymbolFont{letters}{ttmath}{OT1}{\ttdefault}{m}{n}

\begin{document}


\RUNAUTHOR{Varma and Maguluri}

\RUNTITLE{Transportation Polytope, Parallel Server Systems}

\TITLE{Transportation Polytope and its Applications in Parallel Server Systems}

\ARTICLEAUTHORS{%
\AUTHOR{Sushil Mahavir Varma}
\AFF{Department of Industrial and Systems Engineering, Georgia Tech, Atlanta, GA, \EMAIL{sushil@gatech.edu}} 
\AUTHOR{Siva Theja Maguluri}
\AFF{Department of Industrial and Systems Engineering, Georgia Tech, Atlanta, GA, \EMAIL{siva.theja@gatech.edu}}
} 

\ABSTRACT{%
A parallel server system is a stochastic processing network with applications in manufacturing, supply chain, ride-hailing, call centers, etc. Heterogeneous customers arrive in the system, and only a subset of servers can serve any customer type given by the flexibility graph. The goal of the system operator is to minimize the delay that depends on the scheduling policy and the flexibility graph. A long line of literature focuses on designing near-optimal scheduling policies given a flexibility graph. On the contrary, we fix the scheduling policy to be the so-called MaxWeight scheduling given its superior delay performance and focus on designing near-optimal, sparse flexibility graphs. Our contributions are three fold.

First, we analyze the expected delay in the heavy-traffic asymptotic regime in terms of the properties of the flexibility graph, and use this result to translate the design question in terms of transportation polytope, the deterministic equivalent of parallel server queues. Second, we design the sparsest flexibility graph that achieves a given delay performance and shows the robustness of the design to demand uncertainty. Third, given the budget to add edges arrives sequentially in time, we present the optimal schedule for adding them to the flexibility graph. These results are obtained by proving new results for transportation polytopes and are of independent interest. In particular, translating the difficulties to a simpler model, i.e. transportation polytope, allows us to develop a unified framework to answer several design questions.
}%



\maketitle

%


\section{Introduction}
Many service and operational systems such as production systems \cite{zhong2019_process_flexibility}, call centers \cite{shumsky2004approximation}, computer systems \cite{Williams_CRP}, etc are modeled as parallel server system in the literature \cite{harrison1998heavy,kushner2000optimal, bell2001dynamic, skill_based_routing,Williams_CRP,zhong2019_process_flexibility,tezcan2010dynamic,amy_n_network}. In a parallel server system, multiple types of customers $\calI=\{1,2,\hdots,m\}$ arrive and depending on the capability of the servers $\calJ=\{1,2,\hdots,n\}$, they can only be served by a subset of the types of servers given by a flexibility graph $G(\calI \cup \calJ,\calE)$ with $\calE \subseteq \calI \times \calJ$. For example, in the case of production systems, servers are factories with limited manufacturing capabilities. As the set of compatible customers can be overlapping for different servers, scheduling decisions must be made. The question of interest is to design simple, yet approximately optimal scheduling policies such that the delay is minimized. Smaller delay leads to an operationally efficient system and improves the experience of the users.

As the exact analysis is challenging, the parallel server system has been studied in asymptotic regimes, like heavy traffic. Under the heavy traffic regime, the multidimensional state of the system collapses onto a lower-dimensional subspace which makes it more amenable to analysis. This phenomenon is known as state space collapse (SSC). Under certain conditions, known as complete resource pooling (CRP), the lower dimensional subspace is single dimensional. In other words, the system collapses onto a line, and thus behaves like a single server queue where 
all the resources are pooled together to meet the incoming demand. Under the CRP condition, it was shown in \cite{stolyar2004maxweight,zhong2019_process_flexibility} that, the so-called max-weight scheduling algorithm which gives priority to longer queues, minimizes the sum of queue length in heavy traffic. 
More recent work \cite{Hurtado_gen-switch_arxiv} characterized the expected queue length when CRP condition is not satisfied as well. As the dimension of SSC increases, servers idle while incompatible customers are waiting in the system due to poor resource pooling, leading to degraded delay performance. We formalize this in Section \ref{sec: HT_PSS}.

Thus, it is advantageous for the state space to collapse onto a low dimensional subspace. The dimension of the SSC primarily depends on the scheduling algorithm and the flexibility graph $G(\calI \cup \calJ,\calE)$ between customers and servers. For example, for a fully flexible system, i.e. $\calE=\calI \times \calJ$, the resources are trivially pooled together implying CRP. As max-weight has attractive delay performance \cite{stolyar2004maxweight,zhong2019_process_flexibility, Hurtado_gen-switch_arxiv}, we fix the scheduling algorithm and focus on designing the flexibility graph to minimize the dimension of the SSC, and we want to do this using as few edges as possible. More precisely, we consider the following three questions: 
\begin{itemize}
    \item [Q1.] What is the dimension of SSC under max-weight scheduling for a
    given flexibility graph? In particular, we aim to quantify the dimension in terms of the graph properties.
    \item [Q2.] How to design the sparsest flexibility graph that leads to SSC of a given 
    dimension? Is the design robust to demand uncertainty?    \item [Q3.] Suppose a flexibility graph is given, and the budget to add additional edges arrive sequentially in time, where should they be added, so that the SSC dimension of the resulting sequence of graphs is the lowest?
\end{itemize}
In the context of production systems, the results to these problems serve as a guidebook in designing the capabilities of the manufacturing plants and implementing upgrades for an established production system. Similarly, it is useful in guiding the training of the employees in call centers as the edges in parallel server system corresponds to the skill sets of the employees. 

In order to answer these questions, we establish concrete connections between parallel server queues and its fluid equivalent - transportation polytope. In particular, fluid supply $\bmu \in \bbZ_+^n$ from $n$ sources is to be transported to meet the demands $\bnu \in \bbZ_+^m$ at $m$. The set of all the feasible assignments $\bx \in \bbR_+^{m \times n}$ such that $x_{ij} = 0$ for all $(i, j) \notin \calE$ constitutes the transportation polytope with flexibility $G(\calI \cup \calJ, \calE)$, where $x_{ij}$ is the amount of type $i$ demand met using type $j$ supply. 
Support graph of a feasible assignment $\bx$ is a subgraph of $G(\calI \cup \calJ, \calE)$, with edges corresponding to the positive components of the assignment $\bx$. In the rest of the exposition, we use support graph and the corresponding feasible assignment interchangeably. 



\subsection{Main Contribution} \label{sec: main_contributions}
In this paper, we develop a unified framework to answer several interesting questions on designing process flexibility in parallel server queues. The novelty of our approach is in combining the asymptotic analysis of stochastic processing networks with the combinatorial theory of bipartite graphs and transportation polytope. We summarize the main results of the paper below.

\underline{\textbf{Heavy-Traffic Analysis:}} Define the so-called Effective Resource Pooled (ERP) number as the minimum number of connected components of support graphs of a transportation polytope. Then, we show that the dimension of SSC for parallel server queues is equal to the ERP number. In turn, we have
\begin{align*}
\E{\inner{\bone_m}{\bbarq}} \asymp \text{ERP number},
\end{align*}
where $\bbarq$ is the queue length in steady-state for the parallel server queues operating under MaxWeight scheduling. This result allows us to work with transportation polytope as opposed to parallel server queues, which is a much simpler system to analyze.

\underline{\textbf{Transportation Polytope:}} Next, we analyze connectivity of support graphs of transportation polytope by building on the long line of literature of matching in bipartite graphs and extreme points of transportation polytope. In particular, we prove the following two main results. First, for a fully flexibility transportation polytope (i.e. $\calE = \calI \times \calJ$), we present necessary and sufficient conditions for the existence of extreme point with a given degeneracy, or equivalently, the support graph has a given number of connected components.
We also present a polynomial time algorithm which outputs such an extreme point. 
Next, for a transportation polytope with a given flexibility (i.e. $G(\calI \cup \calJ, \calE)$), we characterize the support graph with minimum number of connected components and also present a polynomial time algorithm to construct such a feasible assignment.

\underline{\textbf{Putting the pieces together:}} Combining the two pieces mentioned above, we obtain novel results to analyze, design, and improve process flexibility for parallel server queues. In particular, we answer the three questions mentioned in the introduction.

For a transportation polytope, we say that an edge $(i, j) \in \calE$ is redundant if $x_{ij} = 0$ for all feasible assignments $\bx$. Denote the set of redundant edges by $\calE_r$. Then, any support graph is a sub-graph of $G(\calI \cup \calJ, \calE \backslash \calE_r)$. We denote $G(\calI \cup \calJ, \calE \backslash \calE_r)$ by \emph{CRP decomposition}, each connected component by \emph{CRP component}, and the number of connected components by the \emph{ERP number}. To answer \textbf{Q1}, we show that the CRP components forms an orthogonal basis of the subspace of SSC for the parallel server queues. Thus, the dimension of SSC is equal to the ERP number.

Next, we answer the question \textbf{(Q2)} of designing a new production system. Let $\bnu \in \bbR_+^m$ be the arrival rates of customers in heavy-traffic and $\bmu \in \bbR_+^n$ be the service rates. Then, we show that
$m+n-d+\mathbbm{1}\{d < d_\star\}$ edges in the flexibility graph are necessary and sufficient to ensure that the dimension of SSC is $d$, where
\begin{align*}
d_\star=\max\left\{1,m+n-\frac{\inner{\bone}{\bnu}}{\text{GCD}(\bnu,\bmu)}\right\}.
\end{align*}
Thus, for the special case of $d=1$, $m+n-1$ edges are sufficient if and only if $\inner{\bone}{\bnu} \geq (m+n-1)\text{GCD}(\bnu,\bmu)$, otherwise one extra edge is required. We also provide a simple algorithm to design the required flexibility graph. Lastly, we also show that the design is robust to demand uncertainties and quantify the robustness in terms of the so-called CRP-Gap which generalizes the GCG condition in \cite{zhong2019_process_flexibility}. An interesting observation is that the redundant edges do not contribute to the robustness.

Lastly, we answer the question \textbf{(Q3)} of improving an existing production system. We first address the simpler question, wherein, the budget to add a single edge is provided. We show that if adding an edge results in forming a cycle in the flexibility graph, then the dimension of SSC is reduced by the number of redundant edges in the cycle. Thus, the edge that results in a cycle with maximum number of redundant edges should be added. It is interesting to see that even though redundant edges are inconsequential in the present, they can lead to high gains in the future. Further, we extend this result for the planning problem, wherein, the budget to add edges arrive sequentially in time. We characterize the optimal solution for this problem and show that greedily minimizing the dimension of SSC may not be optimal. 
\subsection{Literature Review}
\subsubsection{Scheduling in Parallel Server Queues}
We first outline the line of work pertaining to the heavy traffic analysis in queueing theory. A well-studied framework is to use diffusion limits and study the resultant Brownian control problem. The seminal paper by Kingman \cite{kingman1962_brownian} analyzes the G/G/1 queue using diffusion limits. This method was generalized to analyze stochastic processing networks like heterogeneous customers \cite{harrison1988brownian}, generalized switch \cite{stolyar2004maxweight}, generalized Jackson networks \cite{gamarnik2006validity}, etc. This method is also employed in the literature to analyze the parallel server queues \cite{harrison1998heavy,kushner2000optimal,bell2001dynamic,mansto04a}. An alternate, more direct approach is the drift method. It was introduced in \cite{atilla} to analyze supermarket checkout model, and further generalized to analyze switch \cite{MagSri_SSY16_Switch}, and generalized switch \cite{Hurtado_gen-switch_arxiv}. The discrete time model of a parallel server queue is subsumed in the generalized switch model and is analyzed in \cite{Hurtado_gen-switch_arxiv}. Some other methods that can be used to analyze such queueing systems include transform method \cite{hurtado2020transform}, basic adjoint relationship (BAR) method \cite{braverman_BAR}, and Stein's method \cite{gurvich2014diffusion}.

There is a long line of literature on designing and analyzing scheduling policies in parallel server queues \cite{harrison1998heavy,kushner2000optimal, bell2001dynamic, skill_based_routing,Williams_CRP,zhong2019_process_flexibility,tezcan2010dynamic,amy_n_network, eyal_parallel_server_upper_bound, eyal_parallel_sever_lower_bound} and more general SPNs that subsumes parallel server queues \cite{ata2008heavy} using the methods outlined in the previous paragraph. N-Network is a special case of parallel server queue which has gotten significant interest in the literature \cite{harrison1998heavy,bell2001dynamic,osogami2005analysis,tezcan2010dynamic,down2010n,amy_n_network, jhunjhunwala2022heavy}. To quote from \cite{amy_n_network},
``N-system is one of the simplest parallel server system models that retains
much of the complexity inherent in more general models''.

Max-Weight scheduling algorithm is known to have a superior delay performance in the context of stochastic processing networks. Tassiulas and Ephremides  proposed the celebrated max-weight algorithm in their seminal paper \cite{TasEph_92}. This led to a huge surge of papers in the context of routing and scheduling in stochastic processing networks and the book \cite{srikantleibook}, presents an excellent exposition. Max-weight was shown to be nearly heavy traffic optimal in \cite{MagSri_SSY16_Switch} and the analysis was extended to the context of generalized switch in \cite{Hurtado_gen-switch_arxiv}. 
\subsubsection{Process Flexibility in Production Systems}
There is a long line of literature focused on understanding process flexibility for a single-period productin system model \cite{jordan1995principles, chou2010design, chou2011process,simchi2012understanding, wang2015process, chen2015optimal}. However, there has been a limited investigation on designing a flexibility graph for parallel server queues, which is a multi-period production system model \cite{varun_menu_design, zhong2019_process_flexibility}. The work closest to ours is \cite{zhong2019_process_flexibility}, which showed that $m+n$ edges in the flexibility graph are sufficient to ensure CRP, and constructed counter examples such that $m+n-1$ edges may not be sufficient. We, on the other hand, conduct a more fine tuned analysis by developing theory in the context of transportation polytope. We present necessary and sufficient conditions on the demand and supply rates such that $m+n-1$ edges are sufficient and also extend it to design non-CRP systems. We also present results pertaining to improving a given flexibility graph by adding edges, which is a crucial question for real life applications.
\subsubsection{Transportation Polytope}
Transportation problem was one of the first linear programming problems that were investigated in the literature \cite{kantorovich2006translocation, hitchcock1941distribution, koopmans1949optimum}. Koopmans received the Nobel Prize in Economics for his work in this area (see: \cite{hoffman2007transportation}). The books \cite{klee1968facets,emelichev1984polytopes,barg2014discrete} provides excellent survey of the results and open problems on transportation polytope. Some of the problems of interest are counting the number of vertices \cite{de2009graphs} and the number of faces of the transportation polytope \cite{pak2000number}. There is also a long line of work \cite{kim2010update,brightwell2006linear} analyzing the diameter of a transportation polytope which relates to the Hirsch conjecture for a general polytope \cite{balinski1984hirsch,santos2012counterexample}. 
\subsection{Outline of the Paper}
In Section \ref{sec: parallel_server}, we present the model and heavy-traffic delay analysis of Parallel Server Queues. Then, we answer Q1, i.e. establish connections with transportation polytope. We end the section by summarizing the goal of the paper. In Section \ref{sec: design_improve}, we answer Q2 and Q3. In particular, we first design the sparsest flexibility graph with a given ERP number and show robustness of the design to the demand uncertainty. Next, we improve an existing flexibility graph by sequentially adding edges in order to minimize a given function of the resultant sequence of ERP numbers. Next, in Section \ref{sec: algorithms}, we present two efficient algorithms pertaining to Q1 and Q2. First, we provide an algorithm that designs sparsest flexibility graph with a given ERP number. Next, we present a polynomial time algorithm to characterize the CRP decomposition of a given transportation polytope. We conclude this paper in Section \ref{sec: conclusion} and also mention possible future directions.
\subsection{Notation}
We denote the set of numbers $\{1,2,\hdots,n\}$ by $[n]$. All the vectors in the paper are boldfaced. Vector of ones and vector of zeros with dimension $n$ are denoted by $\bone_n$ and $\bzero_n$ respectively. For a vector $\by \in \bbZ_+^n$, we denote the greatest common divisor of its components by $\text{GCD}(\by)$. For two vectors $\by,\bz \in \bbR_+^n$, we denote the Euclidean inner product between them by $\inner{\by}{\bz}$. The vector formed by concatenating $\by$ and $\bz$ is denoted by $(\by, \bz)$. A matrix of ones and a matrix of zeros of dimension $m \times n$ are denoted by $\bone_{m \times n}$ and $\bzero_{m \times n}$ respectively. For two matrices $A,B \in \bbR_+^{m \times n}$, we denote the Hadamard product by $\inner{A}{B}$.
\section{Parallel Server Queues} \label{sec: parallel_server}
In this section, we present the parallel server queue model and establish concrete connections with the transportation polytope.
\subsection{Model}
Consider a discrete time queueing system with $\calI := \{1,2,\hdots, m\}$ customer types and $\calJ := \{1,2,\hdots,n\}$ server types. Type $i$ customer can be served by type $j$ server only if $(i, j) \in \calE$ for some $\calE \subseteq \calI \times \calJ$ as shown in Fig \ref{fig: parallel_server}. Now, we define the arrival and service process, and then present the queue as a discrete time Markov chain (DTMC).

The customer arrival process is i.i.d across time and is denoted by $\{\ba(k) \in \bbZ_+^m : k \in \bbZ_+\}$ where $a_i(k)$ is the number of type $i$ customer arrivals at time $k$ and $\E{a_i(1)}=\lambda_i$ for all $i \in \calI$. For simplicity, we consider arrivals to be independent across types and denote the variances by $\operatorname{Var}[a_i(1)]=\sigma_i^2$. All our results can be easily generalized to correlated arrivals. We assume that the arrival vector has a bounded support, i.e. there exists a constant $A_{\max}$, such that $|a_i(1)| \leq A_{\max}$ w.p. 1 for all $i \in \calI$.
\begin{figure}[h]
\FIGURE{
    \centering
    \begin{tabular}{c}
    \begin{tikzpicture}[scale=0.7]
\draw[black, very thick] (-3,0) -- (-1,0) -- (-1,1) -- (-3,1);
\draw[black, very thick] (-3,-1.5) -- (-1,-1.5) -- (-1,-0.5) -- (-3,-0.5);
\filldraw[black] (-2,-2) circle (2pt);
\filldraw[black] (-2,-2.3) circle (2pt);
\filldraw[black] (-2,-2.6) circle (2pt);
\draw[black, very thick] (-3,-4.1) -- (-1,-4.1) -- (-1,-3.1) -- (-3,-3.1);
\draw[black, thick, ->] (-4.5,0.5) -- (-3.4,0.5) node[anchor=south] {$\lambda_1$};
\draw[black, thick, ->] (-4.5,-1) -- (-3.4,-1) node[anchor=south] {$\lambda_2$};
\draw[black, thick, ->] (-4.5,-3.6) -- (-3.4,-3.6) node[anchor=south] {$\lambda_n$};
\draw[black, very thick] (2,0.5) circle (0.5cm);
\draw[black, very thick] (2,-1) circle (0.5cm);
\filldraw[black] (2,-2) circle (2pt);
\filldraw[black] (2,-2.3) circle (2pt);
\filldraw[black] (2,-2.6) circle (2pt);
\draw[black, very thick] (2,-3.6) circle (0.5cm);
\draw[black, thick, ->] (2.5,0.5) -- (3.6,0.5) node[anchor=south] {$\mu_1$};
\draw[black, thick, ->] (2.5,0.5) -- (3.6,0.5) node[anchor=south] {$\mu_1$};
\draw[black, thick, ->] (2.5,-1) -- (3.6,-1) node[anchor=south] {$\mu_2$};
\draw[black, thick, ->] (2.5,-3.6) -- (3.6,-3.6) node[anchor=south] {$\mu_m$};
\draw[black, thin, ->] (-0.9,0.5) -- (1.4,0.5);
\draw[black, thin, ->] (-0.9,0.5) -- (1.4,-0.85);
\draw[black, thin, ->] (-0.9,-1) -- (1.4,-1);
\draw[black, thin, dashed, ->] (-0.9,-1.1) -- (1.4,-3.45);
\draw[black, thin, ->] (-0.9,-3.6) -- (1.4,-3.6);
\end{tikzpicture}
\end{tabular}}{
\centering{A Parallel Server Queueing System}
\label{fig: parallel_server}}{}
\end{figure}
The potential service offered by type $j$ server is deterministic and is equal to $\mu_j$ for all $j \in \calJ$. It can be split to serve customers from any number of compatible queues. At time $k$, denote by $s_i(k)$, the effective potential service offered to queue $i$. For $\bs$ to be a feasible service vector, there must exist $\bx \in \bbZ_+^{m \times n}$ such that the following constraints are satisfied:
\begin{subequations}
\begin{align}
    s_i&=\sum_{j=1}^n x_{ij} \quad \forall i\in \calI \label{eq: consistency} \\
    \mu_j & = \sum_{i=1}^m x_{ij} \quad \forall j \in \calJ \label{eq: service_capacity}\\
    x_{ij}&=0 \quad \forall (i,j) \notin \calE. \label{eq: compatibility}
\end{align}
\label{eq: matching_constraints}
\end{subequations}
Let the set of feasible service vectors be denoted by $\calS$.
Let $x_{ij}$ be the potential number of type $i$ customers served by type $j$ server. The constraint \eqref{eq: consistency} ensures consistency of $\bs$, \eqref{eq: service_capacity} ensures that the total service offered by a server is equal to the potential service, and \eqref{eq: compatibility} ensures compatibility. Note that, if there are not enough customers waiting in the queue, a part of the service may be unused. Denote the unused service at queue $i$ by $u_i(k)$. Now, the queue evolution equation is as follows:
\begin{equation}
    \bq(k+1)=\bq(k)+\ba(k)-\bs(k)+\bu(k). \label{eq: queue_evolution}
\end{equation}
If $u_i(k)>0$, then the queue length at the start of the next time slot must be zero and vice versa. Thus, we have
\begin{equation}
    q_i(k+1)u_i(k)=0 \quad \forall i \in \calI. \label{eq: unused_service}
\end{equation}
The queue evolution \eqref{eq: queue_evolution} along with \eqref{eq: unused_service} implies that $\{\bq(k) : k \in \bbZ_+\}$ is a DTMC with $\bbZ_+^m$ as the state space. We assume that $\P{\ba=\bzero}>0$ which ensures irreducibility. Without loss of generality, we assume $\{\bq(k) : k \in \bbZ_+\}$ to be aperiodic, as otherwise, all the random variables can be scaled appropriately to ensure aperiodicity. Lastly, we define matching policy as the decision of the number of customers to be served, i.e. choosing $\bs(k) \in \calS$, given the state of the system $\bq(k)$. 

Note that, for a given $\bmu$, the arrival rates $\blambda$ must satisfy certain constraints to ensure the stability of the queueing system. The set of such `feasible' arrival rates is known as the capacity region $\calC$. In particular, for any $\blambda \in int(\calC)$, there exists a matching algorithm that stabilizes the queueing system. For the model under consideration, the capacity region \cite{Hurtado_gen-switch_arxiv} is as follows:
\begin{equation} \label{eq: capacity_region}
    \calC(\bmu)=\left\{\blambda \in \bbR_+^m: \sum_{i \in \tilde{\calI}}\lambda_i \leq \sum_{j: \exists i\in \tilde{\calI}, (i,j) \in \calE}\mu_j \ \forall \tilde{\calI} \subseteq \calI\right\}.
\end{equation}
In this paper, we consider a well know Max-Weight matching algorithm to make the service decisions, wherein, we pick the schedule that maximizes the total weighted queue lengths. In particular, we have
\begin{align*}
    \bs(k):= \arg \max_{\by \in \calS} \inner{\bq(k)}{\by}.
\end{align*}
The ties are broken at random. Max-Weight is known to be throughput optimal \cite{Hurtado_gen-switch_arxiv}, i.e. the underlying queueing system is stable for any $\blambda \in int(\calC)$. This implies the existence of a unique stationary distribution. The steady state quantities are denoted with a bar on the top. For example, $\bbarq$ denotes the limit in distribution of the random variables $\bq(k)$ as $k \uparrow \infty$.

As it is challenging to analyze this system in general, we focus on the heavy traffic asymptotic regime. Fix a vector $\bnu$ on the boundary of the capacity region $\calC$ and analyze a sequence of systems such that the mean arrival rate converges to $\bnu$.  We parametrize the queueing system by $\epsilon \in (0,1)$. The mean arrival rate is given by $\blambda^{(\epsilon)}=(1-\epsilon)\bnu$ and the heavy traffic is defined as the limit $\epsilon \downarrow 0$. Without loss of generality, consider $\inner{\bone_m}{\bnu}=\inner{\bone_n}{\bmu}$ as otherwise, some of the queues will be in light traffic and we can consider the part of the system which is in heavy traffic. All the quantities for the $\epsilon^{th}$ system is super-scripted by $\epsilon$. In particular, the queue length vector is denoted by $\bq^{(\epsilon)}(k)$, the arrivals by $\ba^{(\epsilon)}(k)$, and the effective potential service by $\bs^{(\epsilon)}(k)$, and the unused service by $\bu^{(\epsilon)}(k)$ respectively. 

The main advantage of considering the heavy traffic regime is state space collapse (SSC). In particular, in the heavy traffic limit, the queueing system behaves as if it lives in a lower dimensional subspace which makes the analysis tractable. It is known in the literature (e.g. see \cite{atilla,Hurtado_gen-switch_arxiv}) that if $\bnu$ belongs to the interior of a facet of the capacity region $\calC$, then the state space collapses onto a single dimension. This is known as complete resource pooling (CRP). More generally, given $\bnu \in Bo(\calC)$, let $\{\calI_l\}_{l=1}^{d'}$ be the sets corresponding to which, the capacity constraints are binding, i.e. $\sum_{i \in \calI_l} \nu_i = \sum_{j: \exists i \in \calI_l, (i,j) \in \calE} \mu_j$. Then, the state space collapses to the normal cone \cite{Hurtado_gen-switch_arxiv} of the binding constraints given by
\begin{align}
    \calK=\left\{\bx \in \bbR_+^m: \bx=\sum_{l=1}^{d'} \xi_l \mathbbm{1}\{\calI_l\}, \ \bxi \in \bbR_+^k\right\}. \label{eq: ssc_cone}
\end{align}
Specifically, as $\epsilon \downarrow 0$ the queue length vector $\bbarq^{(\epsilon)} \in \calK$ with high probability. In addition, we denote the affine hull of the cone $\calK$ by $\calH$. 
\subsection{Heavy-Traffic Analysis with a Given Flexibility Graph} \label{sec: HT_PSS}
In this section, we characterize the mean steady-state queue length in terms of the flexibility graph $G(\calI \cup \calJ, \calE)$. We start by characterizing the subspace of SSC. Note that, as $\calH$ is the subspace spanning the cone $\calK$, we can consider a subset of $\{\calI_l\}_{l=1}^{d^\prime}$ to form a basis of $\calH$. By carefully analyzing properties of $\calK$, we obtain the following result.
\begin{lemma} \label{lemma: orthogonal_basis}
There exists a disjoint cover $\{\calI_l\}_{l=1}^d$ of $\calI$ for some $d \in \bbZ_+$ such that
\begin{align}
     \calH=\left\{\bq \in \bbR_+^m : \bq=\sum_{l=1}^{d} \xi_l \mathbbm{1}\{\calI_l\}, \bxi \in \bbR^{d} \right\}. \label{eq: ssc_subspace}
\end{align}
\end{lemma}
The above result provides structural insights, i.e., it ensures the existence of a basis that corresponds to a disjoint cover of $\calI$. We use this lemma along with the heavy traffic theory of generalized switch \cite{Hurtado_gen-switch_arxiv,hurtadolange2020logarithmic} to characterize the mean queue length. We formally present it in the following proposition.
\begin{proposition} \label{prop: HT}
Given $(\bnu,\bmu,\calE)$, let $\{\calI_l\}_{l=1}^d$ be a disjoint cover of $\calI$ that characterizes $\calH$ as in \eqref{eq: ssc_subspace}. Then, we have
\begin{align}
    \bigg|\sum_{l=1}^{d} \frac{1}{|\calI_l|}\left( \left(\sum_{i \in \calI_l} \nu_i\right)\E{\sum_{i \in \calI_l} \barq_i^{(\epsilon)}}-\sum_{i \in \calI_l} \frac{(\sigma_i^2)^{(\epsilon)}}{2\epsilon}\right)\bigg|
    \leq \beta\log\left(\frac{1}{\epsilon}\right). \label{eq: theorem_expected_sum}
\end{align}
Assuming that $(\sigma_i^2)^{(\epsilon)} \rightarrow \sigma_i^2$ for all $i \in \calI$ as $\epsilon \downarrow 0$, we have
\begin{align}
   \lim_{\epsilon \downarrow 0}\epsilon\sum_{l=1}^{d} \frac{1}{|\calI_l|} \left(\sum_{i \in \calI_l} \nu_i\right)\E{\sum_{i \in \calI_l} \barq_i^{(\epsilon)}}=\sum_{l=1}^{d} \frac{1}{|\calI_l|}\sum_{i \in \calI_l} \frac{\sigma_i^2}{2} \label{eq: limiting_queue_length}
\end{align}
\end{proposition}
Note that, when $d=1$, i.e. when the so-called CRP condition is satisfied, we recover \cite[Proposition 1]{zhong2019_process_flexibility} in the limit as $\epsilon \downarrow 0$. In fact, for the pre-limit system, i.e. for a finite $\epsilon$, Proposition \ref{prop: HT} provides a logarithmic error as compared to polynomial error in \cite[Proposition 1]{zhong2019_process_flexibility}. To gain further insights into Proposition \ref{prop: HT} for $d \geq 1$, consider the projection of $\bbarq$ on $\calH$ given by
\begin{align}
    \bbarq_{||\calH}= \sum_{l=1}^d\left(\frac{1}{|\calI_l|}\sum_{i \in \calI_l} \barq_i\right)\mathbbm{1}\{\calI_l\}. \label{eq: projection_subspace}
\end{align}
By state space collapse, the queue lengths reside in $\calH$ with high probability, i.e. $\bbarq \approx \bbarq_{||\calH}$. Thus, for $d=1$, all the queues are approximately equal in heavy traffic. In particular, $\barq_{|| \calH}$ is a constant vector as $d=1$ and $\calI_1=\calI$. Now, for the general case when $d>1$, observe that, each set $\{\calI_l\}_{l=1}^d$ exhibits a similar behavior. Thus, $\{\calI_l\}_{l=1}^d$ essentially characterizes the heavy-traffic queue lengths.

In the next section, we establish connections between parallel server queues and the transportation polytope. This allow us to characterize $\{\calI_l\}_{l=1}^d$ and design efficient algorithms.
\subsection{Connections to Transportation Polytope}
In this section, we establish concrete connections between parallel server queues and transportation polytope. In particular, we characterize $\calH$ in terms of the properties of transportation polytope. We start by defining transportation polytope as follows:
\begin{definition}
    Given $(\bnu, \bmu, \calE)$ such that $\inner{\bone}{\bnu} = \inner{\bone}{\bmu}$, we define the corresponding transportation polytope $\calT_\calE^{(\bnu, \bmu)}$ as follows:
    \begin{align} \label{eq: chi_capacity_region}
 \calT_\calE^{(\bnu, \bmu)} = \left\{\bx \in \bbR_+^{m \times n} : \sum_{j=1}^n x_{ij}=\nu_i \ \forall i \in \calI, \sum_{i=1}^m x_{ij}=\mu_j \ \forall j \in \calJ, x_{ij}=0 \ \forall (i,j) \notin \calE \right\}.
\end{align} 
\end{definition}
Transportation polytope can be interpreted as the set of feasible matrices with given row and column sums as shown in Figure~\ref{fig: trans_matrix}. It can also be interpreted as satisfying a static demand $\bnu \in  \bbZ_+^m$ using the available supply $\bmu \in \bbZ_+^n$ as shown in Figure~\ref{fig: trans_bipartite}. In particular, $x_{ij}$ is the amount of type $i$ demand satisfied by type $j$ supply. To ensure compatibility of supply-demand pairs, $x_{ij} = 0$ for all $(i, j) \notin \calE$ is enforced as constraints. The feasibility of transportation polytope is closely related to Hall's matching condition for bipartite graphs. We present it in the following lemma. 
\begin{lemma} \label{lemma: non_empty}
    The transportation polytope $\calT_\calE^{(\bnu, \bmu)}$ is non-empty if and only if $\bnu \in \calC(\bmu)$ and $\inner{\bone_m}{\bnu} = \inner{\bone_n}{\bmu}$, where $\calC(\bmu)$ is the capacity region of parallel server queues as defined in \eqref{eq: capacity_region}.
\end{lemma}
The proof of the above lemma follows immediately from \cite[Lemma 2.5]{gregoryhall}. Nonetheless, it establishes a connection between feasibility of transportation polytope and stability of parallel server queues. 
\begin{figure}[!hbt]
    \begin{minipage}[b]{0.56\textwidth}
    \FIGURE{
    \centering
    \begin{tabular}{c} \begin{tikzpicture}[scale=0.7]
\draw[black, very thick] (0,0) circle (0.75);
\node[black,very thick] at (0,0) {$\nu_2$};
\draw[black, very thick] (0,2) circle (0.75);
\node[black,very thick] at (0,2) {$\nu_1$};
\fill[black] (0,-1.3) circle (0.05);
\fill[black] (0,-1.5) circle (0.05);
\fill[black] (0,-1.7) circle (0.05);
\draw[black, very thick] (0,-3) circle (0.75);
\node[black,very thick] at (0,-3) {$\nu_m$};
\draw[black, very thick] (4.5,0) circle (0.75);
\node[black,very thick] at (4.5,0) {$\mu_2$};
\draw[black, very thick] (4.5,2) circle (0.75);
\node[black,very thick] at (4.5,2) {$\mu_1$};
\fill[black] (4.5,-1.3) circle (0.05);
\fill[black] (4.5,-1.5) circle (0.05);
\fill[black] (4.5,-1.7) circle (0.05);
\draw[black, very thick] (4.5,-3) circle (0.75);
\node[black,very thick] at (4.5,-3) {$\mu_n$};
\draw[black,thin]  (0.75, 2) edge[<->]  (3.75,2);
\draw[black,thin]  (0.75, 0) edge[<->]  (3.75,2);
\draw[black,thin]  (0.75, -3) edge[<->]  (3.75,2);
\draw[black,thin]  (0.75, 2) edge[<->]  (3.75,0);
\draw[black,thin]  (0.75, 0) edge[<->]  (3.75,0);
\draw[black,thin]  (0.75, -3) edge[<->]  (3.75,0);
\draw[black,thin]  (0.75, 2) edge[<->]  (3.75,-3);
\draw[black,thin]  (0.75, 0) edge[<->]  (3.75,-3);
\draw[black,thin]  (0.75, -3) edge[<->]  (3.75,-3);
\node[black, align=center] at (0,3.5) {\footnotesize Demand};
\node[black, align=center] at (4.5,3.5) {\footnotesize Supply};
\end{tikzpicture}
    \end{tabular}}{
 \centering{Bipartite Graph Representation of the Transportation Problem}
 \label{fig: trans_bipartite}}{}
\end{minipage}
\begin{minipage}[b]{0.42\textwidth}
\FIGURE{
\centering
\begin{tabular}{|p{20pt}|p{20pt}|p{20pt}|p{20pt}|c}
\cline{1-4} \xrowht{20pt}
    \hfil $x_{11}$ & \hfil $x_{12}$ & \hfil $\hdots$ & \hfil $x_{1n}$ & $\nu_1$  \\
    \cline{1-4} \xrowht{20pt}
    \hfil $x_{21}$ & \hfil $x_{22}$ & \hfil $\hdots$ & \hfil $x_{2n}$ & $\nu_2$ \\
    \cline{1-4} \xrowht{20pt}
    \hfil $\vdots$ &  \hfil $\vdots$ &  \hfil$\vdots$ &  \hfil $\vdots$ & \\
    \cline{1-4}
     \hfil $x_{m1}$ & \hfil $x_{m2}$ & \hfil$\hdots$ & \hfil $x_{mn}$ & $\nu_m$ \xrowht{20pt}\\
     \cline{1-4} 
     \multicolumn{1}{c}{$\mu_1$} & \multicolumn{1}{c}{$\mu_2$} & \multicolumn{1}{c}{$\hdots$} & \multicolumn{1}{c}{$\mu_n$} &
\end{tabular}}{\centering{Matrix Representation of Transportation Problem}
 \label{fig: trans_matrix}}{}
\end{minipage}
\end{figure}
Now, we define various properties of a transportation polytope and relate them to state space collapse. We start by defining CRP condition as follows:
\begin{definition}[CRP Condition] \label{def: CRP_condition}
For all $\tilde{\calI} \subsetneq \calI$, we have
\begin{align*}
    \sum_{i \in \tilde{\calI}} \nu_i < \sum_{j: (i,j) \in \calE} \mu_j, \quad \textit{and} \quad \sum_{i \in \calI} \nu_i = \sum_{j \in \calJ} \mu_j.
\end{align*}
\end{definition}
It is known \cite{Hurtado_gen-switch_arxiv} that CRP condition implies SSC onto a one-dimensional sub-space, i.e. $d = 1$ in \eqref{eq: ssc_subspace}. We show that CRP Condition is equivalent to the connectivity of so-called support graph in the transportation polytope. In particular, for a given $\bx \in \calT_\calE^{(\bnu, \bmu)}$, we define the support graph corresponding to $\bx$ as $\calB(\bx)\dfn G(\calI \cup \calJ, \{(i,j) \in \calE : x_{ij} > 0\})$. With a little abuse of notation, we also denote $\{(i,j) \in \calE : x_{ij} > 0\}$ by $\calB(\bx)$. Now, we have the following result.
\begin{lemma} \label{lemma: CRP_connectedness}
Given $(\bnu,\bmu,\calE)$, the following are true:
\begin{enumerate}
    \item There exists an $\bx \in \calT_{\calE}^{(\bnu, \bmu)}$ such that $G(\calI \cup \calJ,\calB(\bx))$ is connected if and only if the CRP condition is satisfied. 
    \item If the CRP condition is satisfied, then, there exists $\bx \in \calT_{\calE}^{(\bnu, \bmu)}$ such that $x_{ij}>0$ for all $(i,j) \in \calE$.
\end{enumerate}
\end{lemma}
Thus, if the CRP condition is not satisfied, then, there might exist $(i, j) \in \calE$ such that $x_{ij} = 0$ for all $\bx \in \calT_\calE^{(\bnu, \bmu)}$. We denote such edges as redundant edges and define it formally as follows.
\begin{definition}[Redundant Edge] \label{def: redundant} Given $(\bnu,\bmu,\calE)$, an edge $(i,j) \in \calE$ is redundant if $x_{ij}=0$, or equivalently $(i, j) \notin \calB(\bx)$ for all $\bx \in \calT_{\calE}^{(\bnu, \bmu)}$. The set of redundant edges is denoted by $\calE_r \subseteq \calE$.
\end{definition}
If the CRP condition is satisfied, then we have $\calE_r = \emptyset$ by Lemma \ref{lemma: CRP_connectedness} (2). Now, we show the converse in the following lemma.
\begin{lemma}[Existence of redundant edges] \label{lemma: existence_redundant}
Given $(\bnu,\bmu,\calE)$, if the CRP condition is not satisfied, then either $G(\calI \cup \calJ,\calE)$ is not connected or there exists a redundant edge.
\end{lemma}
Observe that, to characterize the connectivity of the support graph $\calB(\bx)$ for any $\bx \in \calT_\calE^{(\bnu, \bmu)}$, one can consider the connected components in $G(\calI \cup \calJ,\calE \backslash \calE_r)$ which motivates the following definition.
\begin{definition}[CRP Decomposition and ERP Number] \label{def:  CRP_decomposition}
Given $(\bnu,\bmu,\calE)$, let $\calE_r$ be the set of redundant edges of $\calT_\calE^{(\bnu, \bmu)}$. The set of connected components of $G(\calI \cup \calJ, \calE \backslash \calE_r)$ is defined as the CRP decomposition of $\calT_\calE^{(\bnu, \bmu)}$. We call each connected component a CRP component. The number of CRP components is denoted by the ERP number.
\end{definition}
Now, we show that for any support graph, the number of connected components is at least the ERP number. In addition, this lower bound is achieved.
\begin{proposition} \label{theo: ERP_CRP}
Given $(\bnu,\bmu,\calE)$, for any $\bx \in \calT_{\calE}$, the support graph has at least ERP number of connected components. In addition, there exists $\bx \in \calT_{\calE}$ such that the corresponding support graph has ERP number of connected components
\end{proposition}
Using the above defined properties of transportation polytope, we now characterize the subspace $\calH$ that corresponds to SSC of parallel server queues. In particular, we show that CRP decomposition constitutes an orthogonal basis of $\calH$. This is formally presented in the following theorem.
\begin{theorem} \label{theo: sub_space_SSC}
Given $(\bnu,\bmu,\calE)$, let $\{G_l(\calI_l \cup \calJ_l, \calE_l)\}_{l=1}^d$ be the CRP decomposition of $\calT_\calE^{(\bnu, \bmu)}$. Then, we have
\begin{align*}
    \calH=\left\{\bq \in \bbR_+^m : \bq=\sum_{l=1}^{d} \xi_l \mathbbm{1}\{\calI_l\}, \bxi \in \bbR^{d} \right\}.
\end{align*}
The above equation implies that the dimension of the sub-space $\calH$ is equal to the ERP number.
\end{theorem}
This establishes a concrete connection between transportation polytope and $\calH$, via the CRP decomposition and ERP number. We believe such a connection is fundamental between a stochastic processing network and its corresponding static problem. We present a remark which outlines the occurrence of redundant edges in the context of ride-hailing.
\begin{remark}
The definition of the non-basic activities given in \cite[Section 4]{basicedges_ridehailing} is equivalent to the definition of redundant edges. In particular, each connected component by considering only basic edges in \cite{basicedges_ridehailing} corresponds to the CRP components in this paper. In addition, the dimension of the underlying Brownian control problem in \cite{basicedges_ridehailing} is essentially equal to the ERP number which is equal to the dimension of the SSC presented in this paper. 
\end{remark}
We conclude this section by showing that Proposition \ref{prop: HT} can also be interpreted as removing the set of redundant edges from the graph $G(\calI \cup \calJ,\calE)$ such that the linear combination of expected sum of queue length remains the same.
\begin{corollary} \label{corollary: removing_edges}
Under the notations given in Proposition \ref{prop: HT}, \eqref{eq: theorem_expected_sum} is satisfied for the bipartite graph $G(\calI \cup \calJ, \bigcup_{l=1}^{d}\calE_l)$, arrival rates $\bnu \in \bbR_{+}^m$ and service rates $\bmu \in \bbR_{+}^n$.
\end{corollary}
\subsection{Back to Parallel Server Queues: Goal of the Paper} \label{sec: goal}
Now, using the heavy-traffic analysis (Proposition \ref{prop: HT}) along with the connection to transportation polytope (Theorem \ref{theo: sub_space_SSC}), we immediately obtain the following corollary.
\begin{corollary} \label{cor: HT}
Given $(\bnu,\bmu,\calE)$, let $\{G_l(\calI_l \cup \calJ_l, \calE_l)\}_{l=1}^d$ be the CRP decomposition of $\calT_\calE^{(\bnu, \bmu)}$. Then, assuming that $(\sigma_i^2)^{(\epsilon)} \rightarrow \sigma_i^2$ for all $i \in \calI$ as $\epsilon \downarrow 0$, we have
\begin{align} \nonumber
    \lim_{\epsilon \downarrow 0}\epsilon\sum_{l=1}^{d} \frac{1}{|\calI_l|} \left(\sum_{i \in \calI_l} \nu_i\right)\E{\sum_{i \in \calI_l} \barq_i^{(\epsilon)}}=\sum_{l=1}^{d} \frac{1}{|\calI_l|}\sum_{i \in \calI_l} \frac{\sigma_i^2}{2}.
\end{align}
\end{corollary}
Now, using the above corollary, we immediately obtain the following upper and lower bound on expected sum of queue length in heavy-traffic:
\begin{align}
   \left(\frac{\min_{i \in \calI} \sigma_i^2}{2 \max_{i \in \calI} \nu_i}\right) \operatorname{ERP} \leq \lim_{\epsilon \downarrow 0} \epsilon\E{\sum_{i \in \calI} \barq_i^{(\epsilon)}} \leq \left(\frac{\max_{i \in \calI} \sigma_i^2}{2 \min_{i \in \calI} \nu_i}\right) \operatorname{ERP}. \label{eq: lower_upper_bound_erp}
\end{align}
Thus, the expected sum of queue length in heavy traffic scales linearly with the ERP number. This motivates the design of flexibility graphs of parallel server systems that minimizes the ERP number. Our focus in the next section is to develop machinery for transportation polytope to minimize the ERP number.
\section{Sparse Near-Optimal Production Systems} \label{sec: design_improve}
Given the MaxWeight scheduling policy and $(\bnu, \bmu)$, we focus on designing the flexibility graph $G(\calI \cup \calJ, \calE)$ to minimize the ERP number. In particular, we focus on two key objectives. First one is to \emph{design a robust, sparse production system with a given ERP number}. In particular, given the ERP number we minimize $|\calE|$ in Section~\ref{sec: design}, and show robustness to demand uncertainty, i.e. $\bnu$ in Section~\ref{sec: robustness}. The second objective is to \emph{improve an existing production system, where the budget to add edges to $\calE$ arrives sequentially in time.} In particular, we address the question of adding a single edge in Section \ref{sec: add_one_edge}, and adding multiple edges in Section \ref{sec: add_multiple_edges}.
\subsection{Clean Slate Design} \label{sec: design}
Motivated by \eqref{eq: lower_upper_bound_erp}, we consider the problem of designing the sparsest flexibility graph with a given ERP number. In this section, we present a tight characterization of the minimum number of edges required to attain a given ERP number. This is presented in the following theorem.
\begin{theorem} \label{theo: min_edges}
Given $\bnu \in \bbZ_+^m, \bmu \in \bbZ_+^n$, there exists $d_\star \leq d^\star \leq m$ such that the following is true.
\begin{enumerate}[(a)]
    \item No $\calE$ exists such that ERP number of $\calT_\calE^{(\bnu, \bmu)}$ is greater than $d^\star$. \label{theo: min_edges_1}
    \item For $d \in [d^\star]$, ERP number of $\calT_\calE^{(\bnu, \bmu)}$ equal to $d$ implies $|\calE| \geq m+n-d+\mathbbm{1}\{d<d_\star\}$ \label{theo: min_edges_2}
    \item For $d \in [d^\star]$, $\exists \ \calE$ s.t. the ERP number of $\calT_\calE^{(\bnu, \bmu)}$ is $d$ and $|\calE| = m+n-d+\mathbbm{1}\{d<d_\star\}$ where, \label{theo: min_edges_3}
\end{enumerate}
\begin{align*}
    d_\star=\max\left\{1,m+n-\frac{\inner{\bone}{\bnu}}{\text{GCD}(\bnu,\bmu)}\right\}.
\end{align*} 
\end{theorem}
Observe that at least $m+n-d$ edges are required to obtain a graph with $m+n$ vertices and $d$ connected components. As the ERP number is at least the number of connected components, thus, at least $m+n-d$ edges are required to ensure the ERP number is $d$. The above theorem says that either $m+n-d$ or $m+n-d+1$ edges are sufficient. Moreover, the theorem also provides the conditions under which one needs an extra edge. This is pictorially represented in Fig.~\ref{fig: min_edges_representation}.
\begin{figure}[bth!]
    \FIGURE{
    \begin{tikzpicture}
        \draw[very thick] (-3.5, 0) -- (10, 0);
        \node at (-3.5, -0.3) {$1$};
        \node at (10, -0.3) {$m$};
        \node at (1.5, -0.3) {$d_\star$};
        \node at (6.5, -0.3) {$d^\star$};
        \draw[thick] (1.7, -0.75) -- (4.8, -0.75) -- (4.8, -1.25) -- (1.7, -1.25) -- (1.7, -0.75) node at (3.25, -1) {ERP Number $(d)$};
        \draw[thin] (-3.5, 0) -- (-3.5, 3);
        \draw[thin] (1.5, 0) -- (1.5, 3);
        \draw[thin] (6.5, 0) -- (6.5, 3);
        \draw[thin] (10, 0) -- (10, 3);
        \node[align=center] at (8.25, 1.5) {No flexibility \\[-0.35cm] graph exists};
        \node[align=center] at (4, 1.5) {\hl{$m+n-d$} edges \\ [-0.35cm]are necessary and sufficient};
        \node[align=center] at (-1, 1.5) {\hl{$m+n-d+1$} edges \\ [-0.35cm]are necessary and sufficient };
    \end{tikzpicture}
    }{
    Given $(\bnu, \bmu, d)$, we illustrate the number of edges required to ensure that the ERP number is $d$.
    \label{fig: min_edges_representation}}{}
\end{figure}
As a side remark, note that, Theorem \ref{theo: min_edges} can be extended to the case when $\bnu, \bmu$ are real vectors by considering generalized GCD (For $a_1,a_2, \hdots, a_k \in \bbR_{+}$, GCD is defined as the maximum $c \in \bbR_{+}$ such that there exists $b_1, b_2, \hdots, b_k \in \bbZ_+$ with $a_l = cb_l$ for all $l \in [k]$.). We present the simplest case in the theorem for the ease of exposition.

Now, note that, the following result immediately follows for the special case of $d=1$ in Theorem~\ref{theo: min_edges}.
\begin{corollary} \label{cor: min_edges}
Given $(\bnu, \bmu)$, there exists $G(\calI \cup \calJ, \calE)$ with $|\calE| = m+n-1$ such that $\calT_\calE^{(\bnu, \bmu)}$ satisfies the CRP condition if and only if 
\begin{align*}
    \inner{\bone}{\bnu} \geq (m+n-1)\operatorname{GCD}(\bnu, \bmu).
\end{align*} 
\end{corollary}
This is an improvement over the state of the art \cite{zhong2019_process_flexibility}. The authors in \cite{zhong2019_process_flexibility} show that $m+n$ edges are sufficient to ensure CRP and there exists certain $(\bnu,\bmu)$ such that $m+n-1$ are not sufficient. On the other hand, we explicitly characterize $(\bnu,\bmu)$ such that $m+n-1$ are sufficient to ensure CRP. Furthermore, as Max-Weight is delay optimal \cite{zhong2019_process_flexibility} under the CRP condition, we have the following corollary:
\begin{corollary} \label{corollary: optimality_CRP_max_weight}
Given $(\bnu,\bmu)$ such that $\sum_{i=1}^m \nu_i \geq (m+n-1)\text{GCD}(\bnu,\bmu)$, then there exists a feasibility graph $G(\calI \cup \calJ,\calE)$ with $|\calE|=m+n-1$ such that max-weight minimizes the limiting expected sum of queue lengths as $\epsilon \rightarrow 0$, i.e.
\begin{align*}
    \lim_{\epsilon \downarrow 0} \mathbb{E}^{\operatorname{MaxWeight}}\left[\sum_{i=1}^m \barq^{(\epsilon)}_i\right] \leq \lim_{\epsilon \downarrow 0} \mathbb{E}^{\operatorname{ALG}}\left[\sum_{i=1}^m \barq^{(\epsilon)}_i\right] \quad \forall \operatorname{ALG} \in \textit{scheduling policies}.
\end{align*}
\end{corollary}
The above result is promising as it guarantees the existence of a graph with $m+n-1$ edges with heavy-traffic optimal delay. In other words, a carefully designed sparse graph provides all the benefits of a complete graph but requires fewer edges by order of magnitude.

Recall that, Theorem \ref{theo: min_edges} states that there exists $d^\star \leq m$ such that maximum possible ERP number is $d^\star$. In the next lemma, we characterize $d^\star$ in terms of $G(\calI \cup \calJ, \calE)$ and $(\bnu, \bmu)$.
\begin{lemma} \label{lemma: e_star}
Given $(\bnu, \bmu)$, there exists $\calT_\calE^{(\bnu, \bmu)} \neq \emptyset$ such that the ERP number is $d$ if and only if there exists a disjoint cover $\{\calI_l\}_{l=1}^{d}$ of $\calI$ and $\{\calJ_l\}_{l=1}^{d}$ of $\calJ$ such that $\sum_{i \in \calI_l} \nu_i=\sum_{j \in \calJ_l} \mu_j$ for all $l \in [d]$.
\end{lemma}
In other words, $d^\star$ is the maximum cardinality of a disjoint cover of $\calI$ and $\calJ$ such that the total demand is equal to the total supply for each set in the cover. Another way to interpret $d^\star$ is as follows. Consider the fixed cost transportation problem - a unit cost is incurred if a non-zero type $i$ demand is met using type $j$ supply. In particular, we have
\begin{align*}
    x^\star = \arg\min_{\bx \in \calT} \sum_{i=1}^m \sum_{j=1}^n \mathbbm{1}\{x_{ij}>0\}.
\end{align*}
Now, consider the support graph corresponding to $\bx^\star$, i.e. $\calB(\bx^\star)$. As $\bx^\star$ is the optimal solution the fixed cost transportation problem, $\calB(\bx^\star)$ is the sparsest flexibility graph such that $\calT_{\calB(\bx^\star)} \neq \emptyset$. One can show that $\calT_{\calB(\bx^\star)}$ has ERP number equal to $d^\star$ but we omit the details here for brevity.
\subsubsection{Proof of Theorem \ref{theo: min_edges}}
In this section, we prove Theorem \ref{theo: min_edges} by exploiting the properties of extreme points of transportation polytope. Proof of \eqref{theo: min_edges_1} follows by Lemma \ref{lemma: e_star} by setting $d^\star$ equal to the maximum cardinality of a disjoint cover of $\calI$ and $\calJ$ such that the total demand is equal to the total supply for each set in the cover. Now, to prove \eqref{theo: min_edges_2} and \eqref{theo: min_edges_3}, we establish the following result for extreme points of a transportation polytope. 
\begin{proposition} \label{prop: degeneracy}
Consider a transportation polytope $\calT_{\calI \times \calJ}^{(\bnu, \bmu)}$ defined by $(\bnu,\bmu, \calI \times \calJ)$. There exists an extreme point such that the support graph has $d$ connected components if and only if $d \in \{d_\star,\hdots,d^\star\}$, where 
\begin{align}
    d_\star=\max\left\{1,m+n-\frac{\sum_{i=1}^m \nu_i}{\text{GCD}(\bnu,\bmu)}\right\}. \label{eq: d_star}
\end{align} 
\end{proposition}
The above proposition is a new result in the context of transportation polytope, and thus, it is of independent interest. Furthermore, we use the following result that establishes connection between support graph and extreme points.
\begin{lemma}[Theorem 4, \cite{klee1968facets}] \label{lemma: forest}
The point $\bx \in \calT_{\calI \times \calJ}$ is an extreme point if and only if $\calB(\bx)$ is a spanning forest. Moreover, $\bx$ is a non-degenerate extreme point if and only if $\calB(\bx)$ is a spanning tree.
\end{lemma}
If $\calB(\bx)$ contains a cycle, then $\bx$ cannot be an extreme point as, we can perturb the components of $\bx$ corresponding to the cycle by $\pm \epsilon$ while ensuring feasibility. Now, we consider two cases.

\textbf{Case I $(d^\star \geq d \geq d_\star)$:} As ERP number is at least the number of connected components of $G(\calI \cup \calJ, \calE)$, we should have $|\calE| \geq m+n-d$. Now, we construct a flexibility graph such that $|\calE| = m+n-d$ and the ERP number is $d$. By Proposition \ref{prop: degeneracy}, there exists an extreme point $\bx \in \calT_{\calI \times  \calJ}^{(\bnu, \bmu)}$ such that $\calB(\bx)$ has $d$ connected components. Now, as $\bx$ is an extreme point, by Lemma \ref{lemma: forest}, $\calB(\bx)$ is a forest. Thus, we have $|\calB(\bx)| = m+n-d$. Now, consider the transportation polytope $\calT_{\calB(\bx)}^{(\bnu, \bmu)}$, i.e. set $\calE = \calB$. Note that, it is non-empty, the flexibility graph has $m+n-d$ edges, and the ERP number is $d$ as all the edges are non-redundant. This completes the construction.

\textbf{Case II $(d < d_\star)$:} By Proposition \ref{prop: degeneracy}, there does not exist an extreme point with $d$ connected components. Thus, by Lemma \ref{lemma: forest}, $\bx \in \calT_{\calI \times \calJ}^{(\bnu, \bmu)}$ with $d$ connected components must have at least one cycle - an additional edge is required. Thus, we should have $|\calE| \geq m+n-d+1$ for all $\calT_\calE^{(\bnu, \bmu)}$ with ERP number equal to $d$. The existence of $\calE$ such that the equality holds is proved by construction and the algorithm is outlined in Section \ref{sec: alg_design}. We present it in the following claim.
\begin{claim}
    There exists $\calE$ such that $|\calE|=m+n-d+1$ and ERP number of $\calT_\calE^{(\bnu, \bmu)}$ is $d$.
\end{claim}
This completes the proof of Theorem \ref{theo: min_edges}.
\subsection{Robustness to the Demand Uncertainty} \label{sec: robustness}
In this section, we show that small perturbations of the demand cannot increase the ERP number of a transportation polytope. Given $(\bnu, \bmu, \calE)$, let $\calT_{\calE}^{(\bnu, \bmu)}$ be the corresponding transportation polytope and $\calE_r$ be the set of redundant edges. Now, define the CRP decomposition gap (CRP-Gap) of $\calT_{\calE}^{(\bnu, \bmu)}$ as follows:
\begin{align}
    \delta_{\calT_{\calE}^{(\bnu, \bmu)}} = \min \left\{ \sum_{j : \exists i \in \calC, (i, j)\in \calE} \mu_j - \sum_{i \in \calC} \nu_i : \ \sum_{j : \exists i \in \calC, (i, j)\in \calE \backslash \calE_r} \mu_j - \sum_{i \in \calC} \nu_i > 0\right\}, \label{eq: crp_gap}
\end{align}
We leave out the trivial case of $\sum_{j : \exists i \in \calC, (i, j)\in \calE \backslash \calE_r} \mu_j = \sum_{i \in \calC} \nu_i$ for all $\calC \subseteq \calI$ to ensure that $\delta_{\calT_{\calE}^{(\bnu, \bmu)}}$ is well defined. Note that the CRP-Gap is always positive by definition. For the special case when $\calT_{\calE}^{(\bnu, \bmu)}$ satisfy the CRP condition, the CRP-Gap is equivalent to the generalized chaining gap (GCG) \cite{zhong2019_process_flexibility}. In other words, CRP-Gap generalizes GCG. We also remark that CRP-Gap is closely related to the maximum deficiency of a graph \cite{ore1955graphs}. Now, we formalize the robustness of non-redundancy of edges by showing that CRP-Gap quantifies the maximum allowable perturbation of the demand.
\begin{theorem} \label{theo: strong_connectivity}
For $\hat{\bnu} \in \bbR_+^m$, let $\calT_{\calE}^{(\hat{\bnu}, \bmu)}$ be a transportation polytope. Then, ERP number of $\calT_{\calE}^{(\hat{\bnu}, \bmu)}$ is at most the ERP number of $\calT_{\calE}^{(\bnu, \bmu)}$ for all $\hat{\bnu} \in \Lambda_{\bnu}$, where 
\begin{equation*}
    N_{\bnu} = \left\{\hat{\bnu} \in \bbR_+^m: \exists \bomega \in \bbR^m, \ \hat{\bnu} = \bnu + \bomega, \|\bomega\|_1 < 2\delta_{\calT_{\calE}^{(\bnu, \bmu)}}, \calT_{\calE}^{(\hat{\bnu},\bmu)} \neq \emptyset\right\}.
\end{equation*}
\end{theorem}
Thus, small perturbations of the demand can only decrease the ERP number. This shows robustness of delay performance to the demand uncertainty. 
\subsubsection{Impact of Redundant Edges}
A more natural definition of CRP-gap would not involve redundant edges at all. In particular, consider the following alternative definition of CRP-gap:
\begin{align}
\tilde{\delta}_{\calT_{\calE}^{(\bnu, \bmu)}} = \min \left\{ \sum_{j : \exists i \in \calC, (i, j)\in \calE} \mu_j - \sum_{i \in \calC} \nu_i : \ \sum_{j : \exists i \in \calC, \colorbox{green}{$(i, j)\in \calE$}} \mu_j - \sum_{i \in \calC} \nu_i > 0\right\}, \label{eq: alternate_crp_gap}
\end{align}
where the difference is highlighted in green. However, $\tilde{\delta}_{\calT_{\calE}^{(\bnu, \bmu)}}$ exhibits Braess' paradox which is undesirable for the definition of robustness. In particular, adding an edge may reduce the CRP-gap. We illustrate this in Fig. \ref{fig: braess_paradox}.
\begin{figure}[hbt!]
    \FIGURE{
    \begin{minipage}[b]{0.45\textwidth}
   \centering
\begin{tikzpicture}[scale=0.6]
\draw[very thick] (3.5,2.5) circle (0.75);
\draw[very thick] (0,2.5) circle (0.75);
\draw[very thick] (3.5,0) circle (0.75);
\draw[very thick] (0,0) circle (0.75);
\draw[very thick] (3.5,-2.5) circle (0.75);
\draw[very thick] (0,-2.5) circle (0.75);
\draw[very thick]  (0.75, 2.5) --  (2.75,2.5);
\draw[very thick]  (0.75, 0) --  (2.75,0);
\draw[very thick]  (0.75, -2.5) --  (2.75,-2.5);
\draw[very thick]  (0.75, -2.5) --  (2.75, 0);
\draw[thick, ->] (-3, 2.5) -- (-0.75, 2.5) node at (-2.5, 2.9) {$\nu_1=\xi$};
\draw[thick, ->] (-3, 0) -- (-0.75, 0) node at (-2.5, 0.4) {$\nu_2=1$};
\draw[thick, ->] (-3, -2.5) -- (-0.75, -2.5) node at (-2.5, -2.1) {$\nu_3=1$};
\draw[thick, ->] (4.25, 2.5) -- (6.5, 2.5) node at (6, 2.9) {$\mu_1=\xi$};
\draw[thick, ->] (4.25, 0) -- (6.5, 0) node at (6, 0.4) {$\mu_2=1.1$};
\draw[thick, ->] (4.25, -2.5) -- (6.5, -2.5) node at (6, -2.1) {$\mu_2=0.9$};
\draw[rounded corners, red] (-1, -1) rectangle (1, 1);
\draw[rounded corners, red] (2.5, -1) rectangle (4.5, 1);
\end{tikzpicture}
\end{minipage}
\begin{minipage}[b]{0.45\textwidth}
   \centering
\begin{tikzpicture}[scale=0.6]
\draw[very thick] (3.5,2.5) circle (0.75);
\draw[very thick] (0,2.5) circle (0.75);
\draw[very thick] (3.5,0) circle (0.75);
\draw[very thick] (0,0) circle (0.75);
\draw[very thick] (3.5,-2.5) circle (0.75);
\draw[very thick] (0,-2.5) circle (0.75);
\draw[very thick]  (0.75, 2.5) --  (2.75,2.5);
\draw[very thick]  (0.75, 0) --  (2.75,0);
\draw[very thick]  (0.75, -2.5) --  (2.75,-2.5);
\draw[very thick]  (0.75, -2.5) --  (2.75, 0);
\draw[very thick, terquise] (0.75, 0) -- (2.75, 2.5);
\draw[thick, ->] (-3, 2.5) -- (-0.75, 2.5) node at (-2.5, 2.9) {$\nu_1=\xi$};
\draw[thick, ->] (-3, 0) -- (-0.75, 0) node at (-2.5, 0.4) {$\nu_2=1$};
\draw[thick, ->] (-3, -2.5) -- (-0.75, -2.5) node at (-2.5, -2.1) {$\nu_3=1$};
\draw[thick, ->] (4.25, 2.5) -- (6.5, 2.5) node at (6, 2.9) {$\mu_1=\xi$};
\draw[thick, ->] (4.25, 0) -- (6.5, 0) node at (6, 0.4) {$\mu_2=1.1$};
\draw[thick, ->] (4.25, -2.5) -- (6.5, -2.5) node at (6, -2.1) {$\mu_2=0.9$};
\draw[rounded corners, red] (-1, -3.5) rectangle (1, 1);
\draw[rounded corners, red] (2.5, -3.5) rectangle (4.5, 3.5);
\end{tikzpicture}
\end{minipage}
    }{The alternate expression of CRP-gap as defined in \eqref{eq: alternate_crp_gap} reduces from 0.1 (left) to $\xi$ (right) for an arbitrary $\xi > 0$. The box represents $\calC$ and $j : \exists i \in \calC, (i, j) \in \calE$ for which the minimum in \eqref{eq: alternate_crp_gap} is attained.
    \label{fig: braess_paradox}}{}
\end{figure}
Thus, we work with the definition of CRP-gap given in \eqref{eq: crp_gap}. 

An interesting observation is that redundant edges do not help in making the system robust. This is presented in the following result.
\begin{proposition} \label{proposition: redundant_edges_not_help}
    Let $\calE_r$ be the set of redundant edges of $\calT_\calE^{(\bnu, \bmu)}$. Then, we have
    \begin{align*}
    \delta_{\calT_\calE^{(\bnu, \bmu)}} = \delta_{\calT_{\calE \backslash \calE_r}^{(\bnu, \bmu)}}.
    \end{align*}
    Thus, redundant edges do not contribute towards robustness.
\end{proposition}
Even though the robustness is not improved by redundant edges, we show that it can help in other ways. In the next section, we consider the problem of improving the ERP number by adding new edges, and show that having redundant edges can lead to large gains by adding just one extra edge.
\subsection{Improve a Production System by Adding One Edge} \label{sec: add_one_edge}
In this section, we consider the following question. Given $(\bnu, \bmu, \calE)$, and the budget to add one edge to $\calE$, which edge should be added to minimize the ERP number of the resultant transportation polytope? For the ease of notation, we denote the ERP number of $\calT_\calE^{(\bnu, \bmu)}$ by $n_{\calE}^{\operatorname{ERP}}$. 

We start by constructing CRP-graph $D(V_{crp},\calE_{crp})$ by contracting all the non-redundant edges of $G(\calI \cup \calJ, \calE)$, i.e. $\calE \backslash \calE_r$. Thus, each vertex in $D$ corresponds to a CRP component and each edge corresponds to a redundant edge. More formally, let the ERP number be $d$, and $\{G_l(\calI_l \cup \calJ_l, \calE_l)\}_{l=1}^d$ be the CRP decomposition of $\calT_\calE^{(\bnu, \bmu)}$. Then, we have
\begin{align}
    V_{crp} &= \{1,2,\hdots,d\}, \quad \calE_{crp} = \left\{(l_1,l_2) \in [d] \times [d]:\exists i \in \calI_{l_1}, j \in \calJ_{l_2}, \textit{s.t. } (i,j) \in \calE\right\}. \label{eq: crp_graph}
\end{align}
\begin{figure}[!hbt]
\FIGURE{
 \begin{minipage}[b]{0.28\textwidth}
   \centering
\begin{tikzpicture}[scale=0.6]
\draw[very thick] (3.5,2.5) circle (0.75) node at (3.5, 2.5) {$2$};
\draw[very thick] (0,2.5) circle (0.75) node at (0, 2.5) {$1$};
\draw[very thick] (3.5,0) circle (0.75) node at (3.5, 0) {$4$};
\draw[very thick] (0,0) circle (0.75) node at (0, 0) {$3$};
\draw[very thick] (3.5,-2.5) circle (0.75) node at (3.5, -2.5) {$6$};
\draw[very thick] (0,-2.5) circle (0.75) node at (0, -2.5) {$5$};
\draw[very thick] (0,-5) circle (0.75) node at (0, -5) {$7$};
\draw[very thick] (3.5,-5) circle (0.75) node at (3.5, -5) {$8$};
\draw[very thick]  (0.75, 2.5) --  (2.75, 0);
\draw[very thick]  (0.75, 2.5) --  (2.75,2.5);
\draw[very thick]  (0.75, 0) --  (2.75,0);
\draw[very thick]  (0.75, -2.5) --  (2.75,-2.5);
\draw[very thick]  (0.75, -5) --  (2.75,-5);
\draw[very thick]  (0.75, -2.5) --  (2.75,-5);
\draw[very thick]  (0.75, 2.5) --  (2.75,-5);
\end{tikzpicture}
\end{minipage}
 \begin{minipage}[b]{0.28\textwidth}
   \centering
\begin{tikzpicture}[scale=0.6]
\draw[very thick, orange] (3.5,2.5) circle (0.75) node at (3.5, 2.5) {$2$};
\draw[very thick, orange] (0,2.5) circle (0.75) node at (0, 2.5) {$1$};
\draw[very thick, blue1] (3.5,0) circle (0.75) node at (3.5, 0) {$4$};
\draw[very thick, blue1] (0,0) circle (0.75) node at (0, 0) {$3$};
\draw[very thick, terquise] (3.5,-2.5) circle (0.75) node at (3.5, -2.5) {$6$};
\draw[very thick, terquise] (0,-2.5) circle (0.75) node at (0, -2.5) {$5$};
\draw[very thick, red] (0,-5) circle (0.75) node at (0, -5) {$7$};
\draw[very thick, red] (3.5,-5) circle (0.75) node at (3.5, -5) {$8$};
\draw[very thick]  (0.75, 2.5) --  (2.75, 0);
\draw[very thick, orange]  (0.75, 2.5) --  (2.75,2.5);
\draw[very thick, blue1]  (0.75, 0) --  (2.75,0);
\draw[very thick, terquise]  (0.75, -2.5) --  (2.75,-2.5);
\draw[very thick, red]  (0.75, -5) --  (2.75,-5);
\draw[very thick]  (0.75, -2.5) --  (2.75,-5);
\draw[very thick]  (0.75, 2.5) --  (2.75,-5);
\end{tikzpicture}
\end{minipage}
\begin{minipage}[b]{0.4\textwidth}
   \centering
\begin{tikzpicture}[scale=0.7]
\draw[very thick, orange] (5, 0) circle (0.75) node at (5, 0) {$\{1, 2\}$};
\draw[very thick, blue1] (7.5, -2.5) circle (0.75) node at (7.5, -2.5) {$\{3, 4\}$};
\draw[very thick, terquise] (0, 0) circle (0.75) node at (0, 0) {$\{5, 6\}$};
\draw[very thick, red] (2.5, -2.5) circle (0.75) node at (2.5, -2.5) {$\{7, 8\}$};
\draw[very thick, ->] (0.75/1.414, -0.75/1.414) -- (2.5 - 0.75/1.414, -2.5 + 0.75/1.414);
\draw[very thick, ->] (5 - 0.75/1.414, -0.75/1.414) -- (2.5 + 0.75/1.414, -2.5 + 0.75/1.414);
\draw[very thick, ->] (5 + 0.75/1.414, -0.75/1.414) -- (7.5 - 0.75/1.414, -2.5 + 0.75/1.414);
\draw[white] (0, -5) -- (1, -5);
\end{tikzpicture}
\end{minipage}
}{
Illustration of CRP-graph with $\bnu = \bmu = \mathbf{1}_4$ and $G$ is given by the left figure. The middle figure highlights the CRP components in Orange, Blue, Green, and Red. The right figure is the CRP-graph. \label{fig: illustration_crp_graph}
}{}
\end{figure}
Note that, $|V_{crp}|$ is equal to the ERP number and $|\calE_{crp}|$ is equal to the number of redundant edges. We show that the constructed graph $D(V_{crp},\calE_{crp})$ is a directed acyclic graph (DAG) in the following lemma.
\begin{lemma} \label{lemma: DAG}
Let $\{G_l(\calI_l \cup \calJ_l, \calE_l)\}_{l=1}^d$ be the CRP decomposition of $\calT_{\calE}^{(\bnu ,\bmu)}$. Then, $D(V_{crp}, \calE_{crp})$ defined as in \eqref{eq: crp_graph} is a directed acyclic graph (DAG).
\end{lemma}
Now, observe that adding an edge $(i, j) \notin \calE$ to the flexibility graph $G(\calI \cup \calJ$ such that $i \in \calI_{l_1}$ and $j \in \calJ_{l_2}$ corresponds to adding the edge $(l_1, l_2)$ in $D(V_{crp}, \calE_{crp})$. In the next theorem, we show that by adding such an edge, the ERP number is reduced by one less than the length of the cycle formed in $D(V_{crp}, \calE_{crp} \cup (l_1, l_2))$.
\begin{theorem} \label{theo: add_one_edge}
Let $(i, j) \notin \calE$ be the edge added to $G(\calI \cup \calJ, \calE)$. Also, let $\{G_l(\calI_l \cup \calJ_l, \calE_l)\}_{l=1}^d$ be the CRP decomposition of $\calT_{\calE}^{(\bnu ,\bmu)}$, and $D(V_{crp}, \calE_{crp})$ be the corresponding CRP-graph defined as in \eqref{eq: crp_graph}. Also, let $l_1, l_2 \in [d]$ be such that $i \in \calI_{l_1}$ and $j \in \calJ_{l_2}$, and $\{\calA_l\}_{l=1}^k$ be the set of cycles in $D(V_{crp}, \calE_{crp} \cup (l_1, l_2))$. Then, we have
\begin{align*}
n^{\operatorname{ERP}}_{\calE \cup (i, j)} = n^{\operatorname{ERP}}_{\calE} - \max\left\{\bigg| \bigcup_{l=1}^k \calA_l \bigg| - 1, 0\right\}
\end{align*}
In words, the ERP number is decreased by one less than the cumulative length of the cycles (0 if none exists) in $D(V_{crp}, \calE_{crp} \cup (l_1, l_2))$.
\end{theorem}
To prove the theorem, we substantially use the fact that a topological sorting exists for a DAG. The details are deferred to Appendix \ref{app: adding_one_edge}. As a side note, the above theorem can also be interpreted as follows. The ERP is reduced by the number of redundant edges in the cycle formed in $G(\calI \cup \calJ, \calE)$ by adding $(i, j)$. 

It is evident that the ERP number is minimized when $\bigg| \bigcup_{l=1}^k \calA_l \bigg|$ is maximized. Intuitively, only the edges that connects a sink vertex to a source vertex in the CRP-graph can maximize $\bigg| \bigcup_{l=1}^k \calA_l \bigg|$. We show this in the following corollary.
\begin{corollary} \label{corollary: min_erp_number}
    Under the notations defined as in Theorem \ref{theo: add_one_edge}, let $\calA_\star, \calA^\star$ be the set of sink and source vertices respectively in $D(V_{crp}, \calE_{crp})$. Then, there exists $l_\star \in \calA_\star$ and $l^\star \in \calA^\star$ such that for all $(i^\star, j^\star) \in \calI_{l_\star} \times \calJ_{l^\star}$, we have
    \begin{align*}
        n_{\calE \cup (i^\star, j^\star)}^{\operatorname{ERP}} \leq n_{\calE \cup (i, j)}^{\operatorname{ERP}} \quad \forall (u, v) \in \calI \times \calJ.
    \end{align*}
\end{corollary}
The above corollary reduces the search of the edge that minimizes the ERP number. In particular, one can implement Depth-First Search (DFS) to find the topological sorting of the CRP-graph which gives us $\calA_\star, \calA^\star$. As the number of sources and sinks are generally sufficiently small, one can enumerate all possible combinations to find the optimal $l_\star \in \calA_\star$ and $l^\star \in \calA^\star$.
\subsection{Improve a Production System by Adding Multiple Edges} \label{sec: add_multiple_edges}
In this section, we investigate the following problem. Given $(\bnu, \bmu, \calE)$, and the budget to add edges arrive sequentially in time, what is the sequence of edges that minimizes a given function of resultant ERP numbers? An interesting observation is that a greedy algorithm, motivated by Theorem \ref{theo: add_one_edge} may not be optimal in an online setting. Figure \ref{fig: greedy_not_optimal} illustrates that adding edges sequentially to greedily minimize the ERP number may not be optimal in the long run.
\begin{figure}[!hbt]
\FIGURE{
   \begin{minipage}[b]{0.28\textwidth}
   \centering
   \begin{tabular}{c} \begin{tikzpicture}[scale=0.6]
\draw[very thick, orange] (3.5,2.5) circle (0.75) node at (3.5, 2.5) {$2$};
\draw[very thick, orange] (0,2.5) circle (0.75) node at (0, 2.5) {$1$};
\draw[very thick, blue1] (3.5,0) circle (0.75) node at (3.5, 0) {$4$};
\draw[very thick, blue1] (0,0) circle (0.75) node at (0, 0) {$3$};
\draw[very thick, terquise] (3.5,-2.5) circle (0.75) node at (3.5, -2.5) {$6$};
\draw[very thick, terquise] (0,-2.5) circle (0.75) node at (0, -2.5) {$5$};
\draw[very thick, red] (0,-5) circle (0.75) node at (0, -5) {$7$};
\draw[very thick, red] (3.5,-5) circle (0.75) node at (3.5, -5) {$8$};
\draw[very thick]  (0.75, 2.5) --  (2.75, 0);
\draw[very thick, orange]  (0.75, 2.5) --  (2.75,2.5);
\draw[very thick, blue1]  (0.75, 0) --  (2.75,0);
\draw[very thick, terquise]  (0.75, -2.5) --  (2.75,-2.5);
\draw[very thick, red]  (0.75, -5) --  (2.75,-5);
\draw[very thick]  (0.75, -2.5) --  (2.75,-5);
\draw[very thick]  (0.75, 2.5) --  (2.75,-5);
\end{tikzpicture}
    \end{tabular}
   \end{minipage}
   \begin{minipage}[b]{0.68\textwidth}
   \centering
   \begin{tabular}{|c|c|c|c|c|} \hline
    & \multicolumn{2}{c|}{Greedy based on Theorem \ref{theo: add_one_edge}} & \multicolumn{2}{c|}{Optimal} \\ 
    \hline
    Time & Edge added & ERP Number & Edge Added  & ERP Number  \\ \hline
    0 & - & 4 & - & 4 \\ \hline
    1 & (7, 6) & 3 & (3, 6) & 4 \\ \hline
    2 & (3, 2) & 2 & (2, 7) & 1 \\ \hline
    \end{tabular}
   \end{minipage}}{
   \centering{Sequentially adding edges to greedily minimize the ERP number versus minimizing the ERP number in the long run.}
   \label{fig: greedy_not_optimal}}{}
\end{figure}

Motivated by this observation, we consider the following optimization problem. Let $f_i : \bbZ_+ \rightarrow \bbR$ be arbitrary non-decreasing functions for all $i \in [K]$. Given $(\bnu, \bmu)$, we consider the following objective:
\begin{align}
    \min \sum_{i=1}^K f_i(n^{\text{ERP}}_{\calE_i}) \quad
    \text{subject to} \quad \calE_{i-1} \subseteq \calE_{i}, \ |\calE_i| \leq |\calE_{i-1}| + 1 \quad \forall i \in [K]. \tag{OBJ} \label{eq: min_erp_schedule}
\end{align}
The optimization problem \eqref{eq: min_erp_schedule} is intractable in its current form as we are optimizing over a sequence of graphs. We achieve tractability by characterizing the structural properties of the optimal solution of \eqref{eq: min_erp_schedule}, which in turn, leverages the result of optimally adding a single edge (see Theorem \ref{theo: add_one_edge}).
The result is presented below.
\begin{theorem} \label{theo: opt_sol_multiple_long_chains}
Consider a non-empty transportation polytope $\calT_{\calE_0}^{(\bnu, \bmu)}$ with no redundant edges. Then, there exists an optimal solution of \eqref{eq: min_erp_schedule} of the following form. Let
\begin{align}
    \mathcal{K} \overset{\Delta}{=} \left\{\mathbf{k} \in [K]^{p+2}: k_0 = 0, \ k_i \leq \eta+i-1 \ \forall i \in [p], \ k_{p+1}=K, \  k_{i} < k_{i+1} \ \forall i \in [p-1]\right\} \label{eq: long_chain_set}
\end{align}
for some $p \in \bbZ_+$. Then, we have
\begin{align}
    \calE_k = \begin{cases}
    \calE_{k-1} \cup (i_{k-|\{k_l \leq k\}|}, j_{k-|\{k_l \leq k\}| + 1}) &\textit{if } k \notin \mathcal{K} \\
    \calE_{k-1} \cup (i_{k_l-l+1}, j_{k_{l-1}-l+2}) &\textit{if } k = k_l \quad \forall l \in [p].
    \end{cases} \label{eq: opt_soln}
\end{align}
where $\{G_0^l\}_{l=1}^{\eta}$ are connected components of $G_0 \dfn G(\calI \cup \calJ, \calE_0)$, and $i_l \in \calI_l$, $j_l \in \calJ_l$ for all $l \in [\eta]$.
\end{theorem}
Note that, \eqref{eq: opt_soln} corresponds to sequentially adding edges between the connected components of $G_0$. All the newly added edges in $G_0$ are redundant unless they form a cycle, as a consequence of Theorem \ref{theo: add_one_edge}. Thus, to consistently reduce the ERP number, cycle forming edges are added. In particular, $\calK$ corresponds to the time epochs when cycles are formed and $p$ corresponds to the number of cycles formed. Even though larger $p$ consistently reduces the ERP number, it also results in requiring more edges to attain CRP. We now dwell more on this trade-off.

Consider the example illustrated in Fig~\ref{fig: example_sequential_edges}, i.e. for some $\eta \in \bbZ_+$, let $G_0 = \left([\eta] \times [\eta], \{(i, i): \ i \in [\eta]\}\right)$ and $\bnu = \bmu = \mathbf{1}_{\eta}$. Note that at least $\eta$ more edges are required to ensure CRP by Theorem \ref{theo: min_edges}. In particular, CRP is attained by sequentially adding $\{(i, i+1)\}_{i=1}^{\eta-1}$ to $G_0$ and at the last step, $(\eta, 1)$ is added to form a connected cycle. 
\begin{figure}[!hbt]
\FIGURE{
   \begin{tabular}{c} \begin{tikzpicture}[scale=0.7]
\draw[black, very thick] (0, 0) circle (0.75);
\draw[black, very thick] (2, 0) circle (0.75);
\draw[black, very thick] (4, 0) circle (0.75);
\draw[black, very thick] (6, 0) circle (0.75);
\draw[black, very thick] (8, 0) circle (0.75);
\draw[black, very thick] (10, 0) circle (0.75);
\draw[black, very thick] (12, 0) circle (0.75);
\draw[black, very thick] (14, 0) circle (0.75);
\draw[black, very thick] (16, 0) circle (0.75);
\draw[black, very thick] (0, 4) circle (0.75);
\draw[black, very thick] (2, 4) circle (0.75);
\draw[black, very thick] (4, 4) circle (0.75);
\draw[black, very thick] (6, 4) circle (0.75);
\draw[black, very thick] (8, 4) circle (0.75);
\draw[black, very thick] (10, 4) circle (0.75);
\draw[black, very thick] (12, 4) circle (0.75);
\draw[black, very thick] (14, 4) circle (0.75);
\draw[black, very thick] (16, 4) circle (0.75);
\draw[black,thick]  (0, 0.75) edge[-]  (0, 3.25);
\draw[black,thick]  (2, 0.75) edge[-]  (2, 3.25);
\draw[black,thick]  (4, 0.75) edge[-]  (4, 3.25);
\draw[black,thick]  (6, 0.75) edge[-]  (6, 3.25);
\draw[black,thick]  (8, 0.75) edge[-]  (8, 3.25);
\draw[black,thick]  (10, 0.75) edge[-]  (10, 3.25);
\draw[black,thick]  (12, 0.75) edge[-]  (12, 3.25);
\draw[black,thick]  (14, 0.75) edge[-]  (14, 3.25);
\draw[black,thick]  (16, 0.75) edge[-]  (16, 3.25);
\draw[blue,thick]  (0, 0.75) edge[-]  (2, 3.25);
\draw[blue,thick]  (2, 0.75) edge[-]  (4, 3.25);
\draw[blue,thick]  (4, 0.75) edge[-]  (6, 3.25);
\draw[blue,thick]  (6, 0.75) edge[-]  (8, 3.25);
\draw[blue,thick]  (8, 0.75) edge[-]  (10, 3.25);
\draw[blue,thick]  (10, 0.75) edge[-]  (12, 3.25);
\draw[blue,thick]  (12, 0.75) edge[-]  (14, 3.25);
\draw[blue,thick]  (14, 0.75) edge[-]  (16, 3.25);
\draw[blue,thick]  (6, 0.75) edge[-]  (0, 3.25);
\draw[blue,thick]  (12, 0.75) edge[-]  (6, 3.25);
\draw[blue,thick]  (16, 0.75) edge[-]  (12, 3.25);
\end{tikzpicture}
\end{tabular}}{
   \centering{Graph $G_K$ corresponding to \eqref{eq: opt_soln} with $K=11$, $p=3$, $k_1=4$, $k_2=8$, $k_3=11$, and $G_0$ correspond to the black edges, i.e. $G_0 = ([9] \times [9], \{(i, i) : i \in [9]\})$.}
   \label{fig: example_sequential_edges}}{}
\end{figure}
This approach is optimal if the objective only depends on the ERP number at the end of the time horizon. We state this in the following result.
\begin{corollary} \label{corollary: last_erp_number_min}
An optimal solution of \eqref{eq: min_erp_schedule} for $f_i(\cdot) = 0 \ \forall i \in [K-1]$ is given by Theorem \ref{theo: opt_sol_multiple_long_chains} with $p=1$ and $k_1 = \min\left\{\eta, K\right\}$.
\end{corollary}
Even though setting $p=1$ minimizes the ERP number using the least number of edges, the ERP number for all intermediate graphs is equal to $\eta$ as a consequence of Theorem \ref{theo: add_one_edge}. To ensure consistent progress, edges that form a cycle are added once in a while. Fig. \ref{fig: example_sequential_edges} illustrates this using an example. If the objective is to minimize the sum of the ERP number, then there is a trade-off between the ERP number of earlier graphs versus later graphs. The parameter $p$ in \eqref{eq: opt_soln} captures this trade-off. We resolve this trade-off by calculating the optimal value of $p$ and $\mathbf{k}$ in the following result.
\begin{corollary} \label{corollary: total_erp_number_min}
An optimal solution of \eqref{eq: min_erp_schedule} for $f_i(x) = x \ \forall i \in [K]$ is given by Theorem \ref{theo: opt_sol_multiple_long_chains}, with
\begin{align}
    p =  \arg\min_{\bar{p} \in [\eta]} \left(\bar{p}+\mathbbm{1}\left\{\frac{K}{\bar{p}+1} < \frac{\eta-1}{\bar{p}} + \frac{1}{2} \right\}\right) \left(g(\bar{p})^2 + \operatorname{frac}\left(g(\bar{p})\right) - \operatorname{frac}\left(g(\bar{p})\right)^2\right) - \frac{1}{6}\bar{p}(\bar{p}+1)(2\bar{p}+1) \span \nonumber \\
   k_i = \left\lfloor i g(p) - \frac{(i-1)i}{2} \right\rfloor \quad \forall i \in [p] \span \label{eq: k_i_optimal}
\end{align}
where $g(p) = \min\left\{\frac{\eta-1}{p}+\frac{1}{2}, \frac{K}{p+1}\right\}+ \frac{p}{2}$. 
Solving the above, we get
\begin{align*}
    p &= \left(1 \pm \frac{1}{\sqrt{2}}\right)\left(1+o(1)\right)\sqrt{\min\left\{\eta, K\right\}} \\
    k_i &= i\left(\frac{5}{2} \mp \sqrt{2} \pm \frac{1}{2\sqrt{2}}\right)\left(1+o(1)\right)\sqrt{\min\left\{\eta, K\right\}} - \frac{1}{2}i(i-1) \quad \forall i \in [p].
\end{align*}
\end{corollary}
Note that the optimization problem to find the optimal value of $p$ can be solved by enumerating all values of $p$ in $O(\eta)$ time. For $K$ large enough, the solution of Corollary \ref{corollary: total_erp_number_min} adds a total of $p = \Theta(\sqrt{\eta})$ cycle forming edges to ensure a consistent reduction in the ERP number over time. Thus, additional $\Theta(\sqrt{\eta})$ edges are required to attain CRP compared to the result in Corollary \ref{corollary: last_erp_number_min}. The benefit of the solution of Corollary \ref{corollary: total_erp_number_min} is that it reduces the sum of the ERP number to $\Theta\left(\eta^{3/2}\right)$ as opposed to $\Theta\left(\eta^2\right)$ for Corollary \ref{corollary: last_erp_number_min}. We summarize this trade-off in Table~\ref{tab: trade_off_erp}.
\begin{table}[bth!]
\TABLE{
    \centering{Illustrating the trade-off between achieving CRP using least possible number of edges vs minimizing the sum of the ERP number} \label{tab: trade_off_erp}}{
    \centering
    \begin{tabular}{|c|c|c|c|} \hline
         & min $k$ s.t. $n_{\calE_k}^{\text{ERP}} = 1$ & $\sum_{i=1}^K G_i^{\text{ERP}}$ & Number of cycle-forming edges \\ \hline
       Solution of Corollary \ref{corollary: last_erp_number_min}  & $\eta$ & $\Theta\left(\eta^2\right)$ & 1 \\ \hline
       Solution of Corollary \ref{corollary: total_erp_number_min} & $\eta + \Theta\left(\sqrt{\eta}\right)$ & $\Theta\left(\eta^{3/2}\right)$ & $\Theta\left(\sqrt{\eta}\right)$ \\ \hline
    \end{tabular}}{}
\end{table}
\section{Efficient Algorithms for Production System Design} \label{sec: algorithms}
In this section, we provide polynomial time algorithms to design a new production system, characterize CRP decomposition of a given production system, and improve the ERP number by adding an edge to a given production system.
\subsection{Clean Slate Design} \label{sec: alg_design}
Theorem \ref{theo: min_edges} guarantees the existence of sparse graphs with near-optimal delay performance. In this section, we present an algorithm that constructs these sparse graphs.
\begin{algorithm}
\caption{Constructing Sparsest Flexibility with a given ERP Number}\label{alg: feasible_tree}
{\fontsize{10}{13}\selectfont
    \SetAlgoLined
\SetAlgoLined
\SetKwInOut{Input}{Input}
\SetKwInOut{Initialize}{Initialization}
\Input{$\bnu$, $\bmu$}
$\bx \leftarrow \text{Algorithm \ref{alg: extreme_points}}(\bnu,\bmu)$\;
\While{$|\calB(\bx)|<m+n-d_\star$}{
Choose $i_1,i_2 \in \calI$, $j_1,j_2 \in \calJ$ such that $(i_1,j_1),(i_2,j_2) \in \calB(\bx)$, $x_{i_1j_1}\neq x_{i_2j_2}$, and $i_1$ and $i_2$ are not connected\;
\If{No such $i_1,i_2,j_1,j_2$ exists}{\textbf{Break}\;}
$x_{i_kj_k} \leftarrow x_{i_kj_k}-\min\{x_{i_1j_1},x_{i_2j_2}\} \ \forall k \in\{1,2\}$\; $x_{i_{3-k}j_k} \leftarrow \min\{x_{i_1j_1},x_{i_2j_2}\} \ \forall k \in \{1,2\}$ \;}
\If{$d<d_\star$}{
Let $(i_l, j_l)_{l=1}^{d_\star} \in \calB(\bx)$ be such that $i_{l_1}$ and $i_{l_2}$ are not connected for all $l_1 \neq l_2 \in [d_\star]$\;
Define $e=d_\star-d+1$\;
$$
    x_{ij}\leftarrow \begin{cases}
    x_{ij}-\frac{1}{2}\min_{l \in [e]} \{x_{i_lj_l}\} &\textit{if } (i,j) \in \{(i_l,j_l): l \in [e]\} \\
    \frac{1}{2}\min_{l \in [e]}\{x_{i_lj_l}\} &\textit{if } (i,j) \in \{(i_l,j_{l+1}): l \in [e-1]\} \cup \{(i_{e},j_1)\} \\
    x_{ij} &\textit{otherwise}.
    \end{cases}
$$}
\textbf{Return:}  $\calB(\bx)$
}
\end{algorithm}
Given $(\bnu, \bmu)$, let $\calT \dfn \calT_{\calI \times \calJ}$ be the corresponding transportation polytope. We initialize the algorithm with an arbitrary extreme point $\bx^0 \in \calT$ by using Algorithm \ref{alg: extreme_points} (see Appendix \ref{app: prelim}). The corresponding support graph $\calB(\bx)$ is a forest by \cite[Theorem 4]{klee1968facets}. The algorithm has two phases. 

In the first phase at iteration $k \in \bbZ_+$, we jump to a neighboring extreme point such that $|\calB(\bx^k)| = |\calB(\bx^{k-1})| + 1$. The transformation is illustrated in Fig.~\ref{fig: transformation}. This process is repeated until $\calB(\bx^k)$ is a forest with $\max\{d, d_{\star}\}$ connected components, where $d_{\star}$ is the minimum number of connected components of an extreme point of $\calT$. If $d \geq d_{\star}$, the algorithm terminates. Otherwise, let $\bx_{\star}$ be the output of phase 1 and now, we proceed to phase 2.
\begin{figure}[!hbt]
\FIGURE{
   \begin{minipage}[b]{0.32\textwidth}
   \begin{tabular}{c} \begin{tikzpicture}[scale=0.6]
\draw[black, very thick] (0,0) circle (1);
\node[black,very thick] at (0,0) {$\nu_1=2$};
\draw[black, very thick] (0,-2.7) circle (1);
\node[black,very thick] at (0,-2.7) {$\nu_2=1$};
\draw[black, very thick] (4,0) circle (1);
\node[black,very thick] at (4,0) {$\mu_1=2$};
\draw[black, very thick] (4,-2.7) circle (1);
\node[black,very thick] at (4,-2.7) {$\mu_2=1$};
\draw[black,very thick]  (1, 0) edge[-]  (3,0);
\draw[black,very thick]  (1, -2.7) edge[-]  (3,-2.7);
\end{tikzpicture}
    \end{tabular}
   \end{minipage}
   \begin{minipage}[b]{0.32\textwidth}
   \begin{tabular}{c} \begin{tikzpicture}[scale=0.6]
\draw[black, very thick] (0,0) circle (1);
\node[black,very thick] at (0,0) {$\nu_1=2$};
\draw[black, very thick] (0,-2.7) circle (1);
\node[black,very thick] at (0,-2.7) {$\nu_2=1$};
\draw[black, very thick] (4,0) circle (1);
\node[black,very thick] at (4,0) {$\mu_1=2$};
\draw[black, very thick] (4,-2.7) circle (1);
\node[black,very thick] at (4,-2.7) {$\mu_2=1$};
\draw[black, very thick]  (1, 0) edge[-]  (3,0);
\draw[red,very thick]  (1, -2.7) edge[-]  (3,-2.7);
\draw[terquise,very thick] (1,-2.7) edge[-] (3,0);
\draw[terquise,very thick] (1,0) edge[-] (3,-2.7);
\end{tikzpicture}
\end{tabular}
 \end{minipage}
   \begin{minipage}[b]{0.32\textwidth}
   \begin{tabular}{c} \begin{tikzpicture}[scale=0.6]
\draw[black, very thick] (0,0) circle (1);
\node[black,very thick] at (0,0) {$\nu_1=2$};
\draw[black, very thick] (0,-2.7) circle (1);
\node[black,very thick] at (0,-2.7) {$\nu_2=1$};
\draw[black, very thick] (4,0) circle (1);
\node[black,very thick] at (4,0) {$\mu_1=2$};
\draw[black, very thick] (4,-2.7) circle (1);
\node[black,very thick] at (4,-2.7) {$\mu_2=1$};
\draw[black,very thick]  (1, 0) edge[-]  (3,0);
\draw[black,very thick] (1,-2.7) edge[-] (3,0);
\draw[black,very thick] (1,0) edge[-] (3,-2.7);
\end{tikzpicture}
    \end{tabular}
   \end{minipage}}{
   \centering{Illustrating One Step of Algorithm \ref{alg: feasible_tree} - $(i_1,j_1)=(1,1)$, $(i_2,j_2)=(2,2)$, $x_{11}=2$, $x_{22}=1$. As $x_{11} > x_{22}$, $(1,2)$ and $(2,1)$ (Blue) are added, and $(2,2)$ (Red) is removed.}
   \label{fig: transformation}}{}
\end{figure}

In the second phase, $d_{\star} - d$ connected components of $\bx_{\star}$ are converted to a single connected component by connecting them with a cycle of edges. A cycle is formed to ensure that the edges that are added are not redundant. This results in a feasible $\bx \in \calT$ such that $\calB(\bx)$ has $d$ connected components with $m+n-1+\mathbbm{1}\{d < d_\star\}$ edges. One extra edge is required whenever phase 2 is executed as a cycle is formed. This completes the construction. The correctness of the algorithm is proved in the following proposition.
\begin{proposition} \label{prop: non_degenerate}
Given $(\bnu, \bmu, d)$, Algorithm \ref{alg: feasible_tree} outputs a feasible $\bx \in \calT_{\calI \times \calJ}$ such that $|\calB(\bx)|=m+n-d + \mathbbm{1}\{d < d_{\star}\}$ where $d_{\star}$ is given by \eqref{eq: d_star}.
\end{proposition}
\subsection{Finding the CRP Decomposition}
\subsubsection{Structural Properties of CRP Decomposition}
We start by illustrating the concept of redundant edges and CRP decomposition in Figure \ref{fig: CRP_decomp}. The redundant edges are highlighted in blue and the connected components that constitutes the CRP decomposition are highlighted in magenta, green, and orange. For the sub-graph highlighted in orange, $\{\mu_4,\mu_5\}$ is the only feasible supply to meet the demand $\{\nu_4,\nu_5\}$. As $\nu_4+\nu_5 = \mu_4+\mu_5$, we must have $x_{14}=x_{25}=0$ for any feasible $\bx \in \calT_\calE$, as otherwise, there won't be enough remaining supply to meet the demand $\{\nu_4,\nu_5\}$. Thus, $\{(1,4),(2,5)\}$ are redundant edges. We can then consider the graph $\calE \backslash \{(1,4),(2,5)\}$ and repeat the same procedure by considering the sub-graph highlighted in green. This observation is formalized in the following proposition.
\begin{figure}[!hbt]
\FIGURE{
   \begin{minipage}[b]{0.48\textwidth}
   \centering
   \begin{tabular}{c} \begin{tikzpicture}[scale=0.7]
\draw[black, very thick] (0,0) circle (0.75);
\node[black,very thick] at (0,0) {$\nu_1=1$};
\draw[black, very thick] (0,-2) circle (0.75);
\node[black,very thick] at (0,-2) {$\nu_2=1$};
\draw[black, very thick] (0,-4) circle (0.75);
\node[black,very thick] at (0,-4) {$\nu_3=2$};
\draw[black, very thick] (0,-6) circle (0.75);
\node[black,very thick] at (0,-6) {$\nu_4=2$};
\draw[black, very thick] (0,-8) circle (0.75);
\node[black,very thick] at (0,-8) {$\nu_5=1$};
\draw[black, very thick] (3.5,0) circle (0.75);
\node[black,very thick] at (3.5,0) {$\mu_1=2$};
\draw[black, very thick] (3.5,-2) circle (0.75);
\node[black,very thick] at (3.5,-2) {$\mu_2=1$};
\draw[black, very thick] (3.5,-4) circle (0.75);
\node[black,very thick] at (3.5,-4) {$\mu_3=1$};
\draw[black, very thick] (3.5,-6) circle (0.75);
\node[black,very thick] at (3.5,-6) {$\mu_4=1$};
\draw[black, very thick] (3.5,-8) circle (0.75);
\node[black,very thick] at (3.5,-8) {$\mu_5=2$};
\node[blue,very thick] at (1.75,0.5) {$\calE$}; 
\draw[blue,thick]  (0.8, 0) edge[-]  (2.7,-2);
\draw[blue,thick]  (0.8, 0) edge[-]  (2.7,-4);
\draw[blue,thick]  (0.8, -2) edge[-]  (2.7,0);
\draw[blue,thick]  (0.8, -2) edge[-]  (2.7,-6);
\draw[blue,thick]  (0.8, -4) edge[-]  (2.7,0);
\draw[blue,thick]  (0.8, -4) edge[-]  (2.7,-4);
\draw[blue,thick]  (0.8, -6) edge[-]  (2.7,-6);
\draw[blue,thick]  (0.8, -6) edge[-]  (2.7,-8);
\draw[blue,thick]  (0.8, -8) edge[-]  (2.7,-8);
\draw[blue,thick]  (0.8, 0) edge[-]  (2.7,-8);
\end{tikzpicture}
    \end{tabular}
   \end{minipage}
   \begin{minipage}[b]{0.48\textwidth}
   \centering
   \begin{tabular}{c} \begin{tikzpicture}[scale=0.7]
\draw[magenta, very thick] (0,0) circle (0.75);
\node[black,very thick] at (0,0) {$\nu_1=1$};
\draw[black!50!green, very thick] (0,-2) circle (0.75);
\node[black,very thick] at (0,-2) {$\nu_2=1$};
\draw[black!50!green, very thick] (0,-4) circle (0.75);
\node[black,very thick] at (0,-4) {$\nu_3=2$};
\draw[orange, very thick] (0,-6) circle (0.75);
\node[black,very thick] at (0,-6) {$\nu_4=2$};
\draw[orange, very thick] (0,-8) circle (0.75);
\node[black,very thick] at (0,-8) {$\nu_5=1$};
\draw[black!50!green, very thick] (3.5,0) circle (0.75);
\node[black,very thick] at (3.5,0) {$\mu_1=2$};
\draw[magenta, very thick] (3.5,-2) circle (0.75);
\node[black,very thick] at (3.5,-2) {$\mu_2=1$};
\draw[black!50!green, very thick] (3.5,-4) circle (0.75);
\node[black,very thick] at (3.5,-4) {$\mu_3=1$};
\draw[orange, very thick] (3.5,-6) circle (0.75);
\node[black,very thick] at (3.5,-6) {$\mu_4=1$};
\draw[orange, very thick] (3.5,-8) circle (0.75);
\node[black,very thick] at (3.5,-8) {$\mu_5=2$};
\node[blue,very thick] at (1.75,0.5) {$\calE$}; 
\draw[magenta,thick]  (0.8, 0) edge[-]  (2.7,-2);
\draw[blue,thick]  (0.8, 0) edge[-]  (2.7,-4);
\draw[black!50!green,thick]  (0.8, -2) edge[-]  (2.7,0);
\draw[blue,thick]  (0.8, -2) edge[-]  (2.7,-6);
\draw[black!50!green,thick]  (0.8, -4) edge[-]  (2.7,0);
\draw[black!50!green,thick]  (0.8, -4) edge[-]  (2.7,-4);
\draw[orange,thick]  (0.8, -6) edge[-]  (2.7,-6);
\draw[orange,thick]  (0.8, -6) edge[-]  (2.7,-8);
\draw[orange,thick]  (0.8, -8) edge[-]  (2.7,-8);
\draw[blue,thick]  (0.8, 0) edge[-]  (2.7,-8);
\end{tikzpicture}
    \end{tabular}
   \end{minipage}}{
   \centering{Illustration of CRP decomposition with $G_1, G_2, G_3$ highlighted by orange, green, and magenta respectively. Blue colored edges belongs to $G$ but not in $G_1, G_2, G_3$.}
   \label{fig: CRP_decomp}}{}
\end{figure}
\begin{proposition} \label{prop: characterization_CRP_decomp}
Let $\{\calI_l\}_{l=1}^d$ be a disjoint cover of $\calI$. Then, the set of sub-graphs $\{G_l\}_{l=1}^d$ is a CRP decomposition if $G_l$ satisfy the CRP condition with $G_l=G(\calI_l \cup \calJ_l,\calE_l)$ and
\begin{align*}
    \calJ_0\dfn\emptyset, \quad \calJ_l\dfn\left\{j \in \calJ : \exists i \in \calI_l, (i,j) \in \calE\right\} \backslash \bigcup_{l'=0}^{l-1} \calJ_{l'}, \quad
    \calE_l\dfn\calE \cap \calI_l \times \calJ_l \quad \forall l \in [d].
\end{align*}
\end{proposition}
Intuitively, we sequentially decompose the graph $G(\calI \cup \calJ,\calE)$ to obtain $G_1,G_2,\hdots$ such that each sub-graph satisfy the CRP condition as illustrated in Figure \ref{fig: CRP_decomp}. Note that, Proposition \ref{prop: characterization_CRP_decomp} is the converse of Lemma \ref{lemma: DAG} as it says that any decomposition such that each component satisfy CRP condition, and the corresponding CRP-graph as defined in \eqref{eq: crp_graph} is a DAG, ensures that the decomposition is a CRP decomposition. This allow us to easily verify if any given decomposition is CRP or not.
To prove the proposition, we first show that the edges connecting the sub-graphs $\{G_l\}_{l=1}^d$ are redundant. This is presented in the following lemma:
\begin{lemma} \label{lemma: chi_zero}
Given $(\bnu,\bmu)$, let $\{G_l\}_{l=1}^d$ be sub-graphs of $G(\calI \cup \calJ,\calE)$ satisfying the conditions given in Proposition \ref{prop: characterization_CRP_decomp}. Then, for any $\bx \in \calT_\calE$, we have $x_{ij}=0$ for all $(i,j) \notin \bigcup_{l=1}^d \calE_l$.
\end{lemma}
To conclude the proof, we construct a solution $\bx \in \calT_\calE$ such that $x_{ij}>0$ for all $(i,j) \in \calE_l$, $l \in [d]$.
\subsubsection{Algorithm to Characterize Redundant Edges}
We now present Algorithm \ref{alg: CRP_decomp} to characterize the set of redundant edges. We start by introducing the notion of alternating paths. For a feasible $\bx \in \calT_\calE$, a path is a $\calB(\bx)$ alternating path if odd edges of the path belongs to $\calB(\bx)$. In particular, $\{l_1, l_2, \hdots, l_k\}$ is a $\calB(\bx)$ alternating path if and only if $\{l_1, l_2, \hdots, l_k\}$ is a path in the graph $G(\calI \cup \calJ, \calE)$ and $(l_{2i-1}, l_{2i}) \in \calB(\bx)$ for all $i \in \left[\lfloor k/2 \rfloor\right]$. In addition, the set of $\calB(\bx)$ reachable demand edges from $j \in [n]$ is the set $\calC \subseteq \calI$ such that there exists a $\calB(\bx)$ alternating path from $j$ to $i$ for all $i \in \calC$. Now, we present our algorithm to find the CRP decomposition below.

\begin{algorithm}
\caption{Finding a CRP Decomposition} \label{alg: CRP_decomp}
{\fontsize{10}{13}\selectfont
\SetAlgoLined
\SetKwInOut{Input}{Input}
\Input{$\bnu, \bmu, \calE$}
Set an arbitrary $\bx \in \calT_\calE$ and let $\calE_r = \emptyset$\;
\For{$\lambda \in \calJ$}{
$\calC \leftarrow \calB(\bx)\textrm{ reachable demand vertices from } \lambda$\;
$N(\calC) \leftarrow \left\{j : \exists i \in \calC, (i, j) \in \calE\right\}$\;
$\calE_r \leftarrow \calE_r \cup \{(i, j): i \notin \calC, j \in N(\calC)\}$\;
}}
\textbf{Return:} $\calE_r$
\end{algorithm}
To gain intuition, let $\bx \in \calT_\calE$ be such that $x_{ij} > 0$ for all $(i, j) \in \calE \backslash \calE_r$. Then, for $\lambda \in \calJ_l$, the set of $\calB(\bx)$ reachable demand vertices is equal to $\bigcup_{i=1}^l \calI_i$. For the example shown in Figure \ref{fig: CRP_decomp}, one possible instance of Algorithm \ref{alg: CRP_decomp} is given in Table.~\ref{tab: crp_decomp} which corroborates with the intuition.
\begin{table}[bth!]
\TABLE{An instance of Algorithm \ref{alg: CRP_decomp} for the example illustrated in Fig. \ref{fig: CRP_decomp}.
    \label{tab: crp_decomp}}{\centering
    \begin{tabular}{|c|c|c|c|} \hline
       $j$  & $\calC^j$  & $N(\calC^j)$ & $\calE_r^j$ \\ \hline
        1 & $\{2, 3, 4, 5\}$ & $\{1, 3, 4, 5\}$ & $\{(1, 3), (1, 5)\}$ \\ \hline
        2 & $\{1, 2, 3, 4, 5\}$ & $\{1, 2, 3, 4, 5\}$ & $\emptyset$ \\ \hline
        3 & $\{2, 3, 4, 5\}$ & $\{1, 3, 4, 5\}$ & $\{(1, 3), (1, 5)\}$ \\ \hline
        4 & $\{4, 5\}$ & $\{4, 5\}$ & $\{(1, 4), (1, 5)\}$ \\ \hline
        5 & $\{4, 5\}$ & $\{4, 5\}$ & $\{(1, 4), (1, 5)\}$ \\ \hline
    \end{tabular}}{}
\end{table}
Thus, Algorithm \ref{alg: CRP_decomp} is essentially constituting the disjoint cover $\{\calI_l\}_{l=1}^d$ as defined in Proposition \ref{prop: characterization_CRP_decomp}. Now, we formally show the correctness of Algorithm \ref{alg: CRP_decomp} in the following proposition.
\begin{proposition} \label{prop: alg_crp_decomp}
For a given $(\bnu, \bmu, \calE)$, Algorithm \ref{alg: CRP_decomp} outputs the set of redundant edges $\calE_r$ of $\calT_\calE$. The run-time of the algorithm is at most $O((m+n)(m+n+|\calE|))$.
\end{proposition}
\section{Conclusion and Future Work} \label{sec: conclusion}
We present a framework that serves as a guidebook in designing flexibility graphs for parallel server systems. In particular, we show that state space collapse for parallel server queues is equivalent to the CRP decomposition for transportation polytope. This result allows us to analyze transportation polytope instead of parallel server queues, which is a simpler model. 

The paper presents two classes of results. First, we design the sparsest flexibility graph for a given ERP number, develop a polynomial time algorithm for it, and show the robustness of the design to demand uncertainty. Second, given the budget to add edges to the flexibility graph arrives sequentially in time, we present an optimal schedule that minimizes a given objective depending on the resultant sequence of ERP numbers. 

The presented framework leaves the possibility of several relevant extensions. For instance, a customer-dependent service rate for parallel server queues is a natural first extension. Other applications include designing flexibility in a multi-hop stochastic processing network. 
\section{Acknowledgement}
This work was partially supported by NSF grants EPCN-2144316 and CMMI-2140534.

\bibliographystyle{informs2014}
\bibliography{references}

\ECSwitch
\ECHead{E-Companion}
\renewcommand{\theHsection}{A\arabic{section}}
\begin{APPENDICES}{}

\section{Preliminaries for Transportation Polytope} \label{app: prelim}
We start by presenting Algorithm \ref{alg: extreme_points} that can be used to enumerates all possible extreme points of the transportation polytope. 
\begin{algorithm}
\caption{Extreme Points of Transportation Polytope}\label{alg: extreme_points}
{\fontsize{10}{13}\selectfont
    \SetAlgoLined
\SetKwInOut{Input}{Input}
\SetKwInOut{Initialize}{Initialization}
\Input{$\bnu$, $\bmu$}
\Initialize{$\tilde{\calI}=\calI$, $\tilde{\calJ}=\calJ$, $\tilde{\bx}=\bzero_{m \times n}$}
\While{$\tilde{\calI} \neq \emptyset$ and $\tilde{\calJ} \neq \emptyset$}{
Choose $i \in \tilde{\calI}$, $j \in \tilde{\calJ}$ and set $\tilde{x}_{ij}=\min\{\nu_i,\mu_j\}$\;
\If{$\nu_i \leq \mu_j$}{$\tilde{\calI} \leftarrow \tilde{\calI} \backslash \{i\}$\;}
\If{$\nu_i\geq \mu_j$}{$\tilde{\calJ} \leftarrow \tilde{\calJ} \backslash \{j\}$\;}
$\nu_i \leftarrow \nu_i-\min\{\nu_i,\mu_j\}$, $\mu_j \leftarrow \mu_j-\min\{\nu_i,\mu_j\}$
}
\textbf{Return:}  $\tilde{\bx}$\;
}
\end{algorithm}
In each iteration, the algorithm arbitrarily chooses one supply and one demand node, and greedily meets the maximum possible demand using the available supply. This is repeated until all the demand is met and all the supply has been exhausted. One can repeat this in all possible ways to list down all possible extreme points and we state this as a lemma.
\begin{lemma}[Corollary 8.1.4, \cite{brualdi2006combinatorial}] \label{lemma: alg_extreme_points}
The extreme points of $\calT$ are exactly those matrices that results by carrying out Algorithm \ref{alg: extreme_points} in all possible ways.
\end{lemma}
By using the above lemma, the following immediately follows.
\begin{lemma}[Corollary 2.11, \cite{barg2014discrete}] \label{lemma: integer}
If $\bnu$ and $\bmu$ are integer vectors, then all the extreme points of the transportation polytope $\calT$ are integer matrices. 
\end{lemma}
\section{Heavy-Traffic Analysis of Parallel Server Queues}
\subsection{Orthogonal Basis of State Space Collapse}
We require the following three lemmas to prove Lemma \ref{lemma: orthogonal_basis}.
\begin{lemma} \label{lemma: base_case_chi_zero}
Given $\bnu \in \calC(\bmu)$, if $\tilde{\calI} \subsetneq \calI$ be such that
\begin{align*}
    \sum_{i \in \tilde{\calI}} \nu_i=\sum_{j : \exists i \in \tilde{\calI}, (i,j) \in \calE} \mu_j
\end{align*}
then for any $\bx \in \calT$, $x_{ij}=0$ for all $i \in \calI \backslash \tilde{\calI}$ and $j \in \{j: \exists i \in \tilde{\calI},(i,j) \in \calE\}$.
\end{lemma}

\begin{lemma} \label{lemma: intersection}
Given $\bnu \in \calC(\bmu)$, if $\mathbbm{1}\{\calI_1\}, \mathbbm{1}\{\calI_2\} \in \calK$, then $\mathbbm{1}\{\calI_1 \cap \calI_2\} \in \calK$.
\end{lemma}
 
\begin{lemma} \label{lemma: refinement}
Given $\bnu \in \calC(\bmu)$, if for some $\calI_1, \calI_2 \subseteq \calI$,  $\mathbbm{1}\{\calI_1\}, \mathbbm{1}\{\calI_2\} \in \calK$, then $\mathbbm{1}\{\calI_1 \backslash \calI_2\} \in \calH$, $\mathbbm{1}\{\calI_2 \backslash \calI_1\} \in \calH$, and $\mathbbm{1}\{\calI_1 \cap \calI_2\} \in \calH$.
 \end{lemma}
We defer the proof of these three lemmas to the end of this section. Now, we present the proof of Lemma \ref{lemma: orthogonal_basis}.
\proof{Proof of Lemma \ref{lemma: orthogonal_basis}}
Observe that $\calH$ is the affine hull of $\calK$ which is the conic combination of $\{\mathbbm{1}\{\calI_l\}\}_{l=1}^{d^\prime}$. Thus, $\calH$ is a linear combination of $\{\mathbbm{1}\{\calI_l\}\}_{l=1}^{d^\prime}$. Now, using the inclusion-exclusion principle for sets, we construct disjoint sets as follows:
\begin{align*}
    \tilde{\calI}_{\calA} = \bigcap_{l \in \calA} \calI_l \backslash \bigcap_{l \notin \calA} \calI_l \quad \forall \calA \subseteq \calI.
\end{align*}
Note that, span of $\{\tilde{\calI}_\calA\}_{\calA \subseteq \calI}$ is equal to the span of $\{\calI_l\}_{l=1}^{d^\prime}$. To complete the proof, consider a linearly independent subset of $\{\tilde{\calI}_\calA\}_{\calA \subseteq \calI}$ and note that it forms the required orthogonal basis of $\calH$. This completes the construction. \hfill $\square$
\endproof
\proof{Proof of Lemma \ref{lemma: base_case_chi_zero}} Consider an arbitrary $\bx \in \calT_\calE$. We have
\begin{align*}
    \sum_{j: \exists i \in \tilde{\calI}, (i,j) \in \calE} \mu_j &= \sum_{j: \exists i \in \tilde{\calI}, (i,j) \in \calE} \sum_{i \in \calI} x_{ij} \\
    &= \sum_{j: \exists i \in \tilde{\calI}, (i,j) \in \calE} \sum_{i \in \tilde{\calI}} x_{ij}+\sum_{j: \exists i \in \tilde{\calI}, (i,j) \in \calE} \sum_{i \in \calI \backslash \tilde{\calI}} x_{ij} \\
    &= \sum_{i \in \tilde{\calI}} \sum_{j: \exists i \in \tilde{\calI}, (i,j) \in \calE} x_{ij}+\sum_{j: \exists i \in \tilde{\calI}, (i,j) \in \calE} \sum_{i \in \calI \backslash \tilde{\calI}} x_{ij} \\
    &\overset{*}{=} \sum_{i \in \tilde{\calI}} \sum_{j \in \calJ} x_{ij}+\sum_{j: \exists i \in \tilde{\calI}, (i,j) \in \calE} \sum_{i \in \calI \backslash \tilde{\calI}} x_{ij} \\
    &=\sum_{i \in \tilde{\calI}} \nu_i+\sum_{j: \exists i \in \tilde{\calI}, (i,j) \in \calE} \sum_{i \in \calI \backslash \tilde{\calI}} x_{ij} \\
    &\overset{**}{=} \sum_{j: \exists i \in \tilde{\calI}, (i,j) \in \calE} \mu_j+\sum_{j: \exists i \in \tilde{\calI}, (i,j) \in \calE} \sum_{i \in \calI \backslash \tilde{\calI}} x_{ij},
\end{align*}
where $(*)$ follows as $(i, j)\notin \calE$ for all $i \in \tilde{\calI}$ and $j \notin \{j:\exists i \in \tilde{\calI}, (i, j) \in \calE\}$. In particular, $\{j:\exists i \in \tilde{\calI}, (i, j) \in \calE\}$ is the neighborhood of $\tilde{\calI}$. Lastly, $(**)$ follows by the statement of the lemma. Now, we have
\begin{align*}
   \sum_{j: \exists i \in \tilde{\calI}, (i,j) \in \calE} \sum_{i \in \calI \backslash \tilde{\calI}} x_{ij} = 0 \Rightarrow x_{ij}=0 \ \forall i \notin \tilde{\calI}, j \in \calJ :  \exists i \in \tilde{\calI}, (i,j) \in \calE.
\end{align*}
This proves the lemma.
\hfill $\square$
\endproof
\proof{Proof of Lemma \ref{lemma: intersection}}
 \begin{align*}
   \sum_{j: \exists \tilde{i} \in \calI_1 \cap \calI_2, (\tilde{i}, j) \in \calE} \mu_j &\overset{*}{\geq} \sum_{i \in \calI_1 \cap \calI_2} \nu_i \\ 
   &= \sum_{i \in \calI_1 \cap \calI_2} \sum_{j: \exists \tilde{i} \in \calI_1 \cap \calI_2, (\tilde{i},j) \in E} x_{ij}  \\
    &\overset{*}{=} \sum_{j: \exists \tilde{i} \in \calI_1 \cap \calI_2, (\tilde{i}, j) \in \calE} \sum_{i \in \calI_1 \cap \calI_2} x_{ij} \\
    &\geq \sum_{j: \exists \tilde{i} \in \calI_1 \cap \calI_2, (\tilde{i}, j) \in \calE} \sum_{i \in \calI} x_{ij} - \sum_{j: \exists \tilde{i} \in \calI_1 \cap \calI_2, (\tilde{i}, j) \in \calE} \sum_{i \notin \calI_1} x_{ij}-\sum_{j: \exists \tilde{i} \in \calI_1 \cap \calI_2, (\tilde{i}, j) \in \calE} \sum_{i \notin \calI_2} x_{ij} \\
    & \overset{**}{=} \sum_{j: \exists \tilde{i} \in \calI_1 \cap \calI_2, (\tilde{i}, j) \in \calE} \mu_j
 \end{align*}
where $(*)$ follows by Lemma \ref{lemma: non_empty} as $\bnu \in \calC$. Further, $(**)$ follows by Lemma \ref{lemma: base_case_chi_zero} as $x_{ij} = 0$ for $i \in \calI \backslash \calI_k$ and $j: \exists \tilde{i} \in \calI_k, (\tilde{i}, j) \in \calE$ for $k \in \{1, 2\}$. This completes the proof.
\hfill $\square$
\endproof
\proof{Proof of Lemma \ref{lemma: refinement}}
 By Lemma \ref{lemma: intersection} and the fact that $\calK \subset \calH$, we have $\mathbbm{1}\{\calI_1 \cap \calI_2\} \in \calH$. Now, as $\calH$ is a subspace, we have $\mathbbm{1}\{\calI_1\}-\mathbbm{1}\{\calI_1 \cap \calI_2\}=\mathbbm{1}\{\calI_1 \backslash \calI_2\} \in \calH$. Similarly, we also have $\mathbbm{1}\{\calI_2 \backslash \calI_1\} \in \calH$. This completes the proof.
\hfill $\square$
\endproof
\subsection{Expected Queue Length in Heavy-Traffic}
\proof{Proof of Proposition \ref{prop: HT}}
By applying \cite[Theorem 1]{hurtadolange2020logarithmic} for our context, we get
\begin{align*}
    \bigg|\E{\inner{\bq_{||\calH}}{\bnu}} - \frac{1}{2\epsilon}\sum_{i=1}^n H_{ii}(\sigma_i^2)^{(\epsilon)}\bigg| \leq \beta \log\left(\frac{1}{\epsilon}\right),
\end{align*}
where $\mathbf{H} \in \bbR^{m \times m}$ is the projection matrix corresponding to the subspace $\calH$. To complete the proof, we simplify the left hand side (LHS) of the above equation as follows.

By Lemma \ref{lemma: orthogonal_basis}, $\{\calI_l\}_{l=1}^d$ is a disjoint cover of $\calI$ such that it forms a basis for $\calH$. Thus, we can directly write the projection $\bq_{||\calH}$ of $\bq \in \bbR^m$ as follows:
\begin{align}
    \bq_{||\calH}&=\sum_{l=1}^{d} \frac{\inner{\bq}{\xi_l\mathbbm{1}\{\calI_l\}}}{\xi_l^2|\calI_l|}\xi_l\mathbbm{1}\{\calI_l\}=\sum_{l=1}^{d} \left(\frac{1}{|\calI_l|}\sum_{i \in \calI_l} q_i\right) \mathbbm{1}\{\calI_l\}. \nonumber
\end{align}
Now, by multiplying both sides by $\bnu$, we have
\begin{align}
    \inner{\bq_{||\calH}}{\bnu}= \sum_{l=1}^{d}\frac{1}{|\calI_l|}\left(\sum_{i \in \calI_l} \nu_i\right) \sum_{i \in \calI_l} q_i. \label{eq: q_parallel}
\end{align}
Next, the projection matrix $H$ corresponding to the subspace $\calH$ is such that
\begin{align*}
    H_{ii}=\begin{cases}\frac{1}{|\calI_l|} &\textit{ if } \exists l \in [d] \textit{ such that } i \in \calI_l. \\
    0 &\textit{otherwise}.
    \end{cases}
\end{align*}
Thus, 
\begin{align}
    \sum_{i=1}^n H_{ii}(\sigma_i^2)^{(\epsilon)}=\sum_{l=1}^{d} \sum_{i \in \calI_l} \frac{(\sigma_i^2)^{(\epsilon)}}{|\calI_l|}. \label{eq: variances}
\end{align}
This completes the proof.
\hfill $\square$
\endproof
\section{CRP Decomposition}
\subsection{Characterizing CRP Decomposition}
\proof{Proof of Lemma \ref{lemma: chi_zero}}
Let $\{\calI_l\}_{l=1}^d$ be a disjoint cover of $\calI$ and $G_l=G(\calI_l \cup \calJ_l,\calE_l)$ be the corresponding sub-graphs satisfying the hypothesis of Lemma \ref{lemma: chi_zero}. Now, we will prove the lemma using induction as follows:

\underline{\textbf{Induction hypothesis:}} For any $\bx \in \calT$ and $l\leq d$, we have
\begin{align*}
    x_{ij}=0 \quad \forall i \notin \calI_{l'}, \ j \in  \calJ_{l'} \quad \forall l' \leq l.
\end{align*}

\underline{\textbf{Base Case:}} This follows directly by Lemma \ref{lemma: base_case_chi_zero} as $\{j : \exists i \in \calI_1, (i,j) \in \calE\}=\calJ_1$ and $\sum_{i \in \calI_1}\nu_i=\sum_{j \in \calJ_1} \mu_j$ as $G_1$ satisfy the CRP condition.

\underline{\textbf{Induction Step:}} Note that by Induction hypothesis, $\bx\in \calT_\calE$ if and only if $\bx \in \calT_{\calE'}$ where
\begin{align*}
    \calE'=\calE \big\backslash \left\{(i,j) \in \calE: i \in \calI_{l+1}, j \in \bigcup_{l' \leq l} \calJ_{l'}\right\}.
\end{align*} 
Now, consider the graph $G(\calI \cup \calJ,\calE')$. The neighbors of $\calI_{l+1}$ with respect to $G(\calI \cup \calJ, \calE')$ is given by
\begin{align*}
    \{j : \exists i \in \calI_{l+1}, (i,j) \in \calE'\}&=\left\{j : \exists i \in \calI_{l+1}, (i,j) \in \calE \backslash \left\{(i,j) \in \calE: i \in \calI_{l+1},j \in \bigcup_{l' \leq l} \calJ_{l'}\right\}\right\} \\
    &=\{j : \exists i \in \calI_{l+1}, (i,j) \in \calE\} \backslash \bigcup_{l' \leq l} \calJ_{l'} \\
    &=\calJ_{l+1}.
\end{align*} 
As $G_{l+1}$ satisfy the CRP condition by the hypothesis of the lemma, we have $\sum_{i \in \calI_{l+1}} \nu_i=\sum_{j \in \calJ_{l+1}} \mu_j$. Thus, by considering the graph $G(\calI \cup \calJ, \calE')$ and applying Lemma \ref{lemma: base_case_chi_zero}, for all $\bx \in \calT_{\calE'}=\calT_{\calE}$ we have
\begin{align*}
    x_{ij}=0 \quad \forall i \notin \calI_{l+1}, \ j \in \calJ_{l+1}.
\end{align*}
This completes the induction step. Thus, induction implies that for any $\bx \in \calT_\calE$, we have
\begin{align*}
    x_{ij}=0 \quad \forall i \notin \calI_l \ j \in \calJ_l \quad \forall l \in [d].
\end{align*}
The proof is completed by noting that $\{(i,j) \in \calE : i \notin \calI_l \ j \in \calJ_l \quad \forall l \in [d]\}=\calE \backslash \bigcup_{l=1}^d \calE_l$.
\hfill $\square$
\endproof
\proof{Proof of Proposition \ref{prop: characterization_CRP_decomp}}
We first show that $\{G_l\}_{l=1}^d$ as defined in the statement of the proposition is a CRP decomposition. By Lemma \ref{lemma: chi_zero}, we know that $\calE \backslash \bigcup_{l=1}^d \calE_l \subseteq \calE_r$. Now, we show the converse which completes the proof. Consider the transportation polytope $\calT_l$ defined by the graph $G_l(\calI_l \cup \calJ_l, \calE_l)$, demand $\{\nu_i\}_{i\in \calI_l}$, and supply $\{\mu_j\}_{j \in \calJ_l}$.
As $G_l$ satisfy the CRP condition, there exists $\bx^l \in \calT_l$ such that $x^l_{ij}>0$ for all $(i,j) \in \calE_l$ by Lemma \ref{lemma: CRP_connectedness}. Now, define $\bx$ as follows:
\begin{align*}
    x_{ij}=\begin{cases}
    x^l_{ij} &\textit{if } (i,j) \in \calE_l \ \forall l \in [d] \\
    0 &\textit{otherwise}.
    \end{cases}
\end{align*}
Note that, $\bx \in \calT$ and $x_{ij}>0$ for all $(i,j) \in \bigcup_{l=1}^d \calE_l$. This implies that any edge $(i,j) \in \bigcup_{l=1}^d \calE_l$ is not redundant which implies $\calE_r \subseteq \calE \backslash \bigcup_{l=1}^d \calE_l$. Thus, we have $\calE_r = \calE \backslash \bigcup_{l=1}^d \calE_l$. Now, by Definition \ref{def:  CRP_decomposition}, $\{G_l\}_{l=1}^d$ as defined in the statement of the proposition is the CRP decomposition. This completes the proof.
\hfill $\square$
\endproof
\subsection{Connection to State Space Collapse}
\proof{Proof of Theorem \ref{theo: sub_space_SSC}}  We first show that for a CRP decomposition $\{G_l=G(\calI_l \cup \calJ_l,\calE_l)\}_{l=1}^{d}$ as defined in Proposition \ref{prop: characterization_CRP_decomp}, $\mathbbm{1}\{\calI_l\} \in \calH$ for all $l \in [d]$. As $\{G_l\}_{l \in [d]}$ satisfies the CRP condition, we have
\begin{align*}
    \sum_{i \in \bigcup_{l'=1}^l \calI_{l'}} \nu_i =\sum_{j \in \bigcup_{l'=1}^l \calJ_{l'}} \mu_j = \sum_{j: \exists i \in \bigcup_{l'=1}^l \calI_{l'}, (i,j) \in \calE} \mu_j
\end{align*}
where the last equality follows as 
\begin{align*}
   \bigcup_{l'=1}^l \calJ_l &\subseteq \bigcup_{l'=1}^l\left\{j: \exists i \in  \calI_{l'}, (i,j) \in \calE\right\}\subseteq\bigcup_{l'=1}^l\left(\bigcup_{l''\leq l'} \calJ_{l''}\right)=\bigcup_{l'=1}^l \calJ_{l'}.
\end{align*}
Thus, we have $\bigcup_{l'=1}^{l} \calI_{l'} \in \calK$ for all $l \in [d]$. By Lemma \ref{lemma: refinement}, we immediately conclude that $\mathbbm{1}\{\calI_l\} \in \calH$ for all $l \in [d]$. This implies that
\begin{align*}
     \left\{\sum_{l=1}^d \xi_l \mathbbm{1}\{C_l\}\right\} \subseteq \calH.
\end{align*}
Now, we show the converse to complete the proof. 
Consider the union of the disjoint graphs $\{G_l\}_{l=1}^d$ given by $G(\calI \cup \calJ, \bigcup_{l=1}^d\calE_l)$ and denote its heavy traffic cone by $\calK'$ and its affine hull by $\calH'$. Note that if $\tilde{\calI} \notin \calK'$, then as $\bigcup_{l=1}^d\calE_l\subseteq \calE$, we must have $\tilde{\calI} \notin \calK$. Thus, we conclude that $\calK \subseteq \calK'$ as well as $\calH \subseteq \calH'$. In particular, $\left\{\sum_{l=1}^d \xi_l \mathbbm{1}\{\calI_l\}\right\} \subseteq \calH'$. We claim that
\begin{align*}
    \calH' = \left\{\sum_{l=1}^d \xi_l \mathbbm{1}\{\calI_l\}\right\}.
\end{align*}
As $G_l$ satisfies the CRP condition, we have $\tilde{\calI} \notin \calK'$ for all $\tilde{\calI} \subset \calI_l$. Now, consider any $\tilde{\calI} \subset \calI$ such that $\tilde{\calI} \cap \calI_l \notin \{\emptyset,\calI_l\}$. Then $\tilde{\calI} \notin \calK'$, as otherwise, $\tilde{\calI} \cap \calI_l \subset \calI_l \in \calK'$ by Lemma \ref{lemma: intersection}. This is a contradiction as $\tilde{\calI} \notin \calK$ for all $\tilde{\calI} \subset \calI_l$. Thus,
\begin{align*}
    \tilde{\calI} \notin \calK' \quad \textit{if } \exists \ l \textit{ such that } \tilde{\calI} \cap \calI_l \notin \{\emptyset,\calI_l\}.
\end{align*}
Thus any $\tilde{\calI} \in \calK'$ must be union of a subset of sets of $\{\calI_l\}_{l=1}^d$. This implies that $\{\calI_l\}_{l=1}^d$ forms a basis of $\calH'$. Thus, we have
\begin{align}
    \left\{\sum_{l=1}^d \xi_l \mathbbm{1}\{\calI_l\}\right\} \subseteq \calH \subseteq \calH' = \left\{\sum_{l=1}^d \xi_l \mathbbm{1}\{\calI_l\}\right\}. \label{eq: equal_sub_space}
\end{align}
This completes our proof.
\hfill $\square$
\endproof
\subsection{Further Results on CRP Decomposition}
\proof{Proof of Lemma \ref{lemma: CRP_connectedness}}
The first part of the Lemma follows similar to the arguments presented in \cite[Assumption 2.4]{CRP_queueing} and we omit the details for brevity. Now, we prove the second part of the lemma. Define the transportation polytope $\calT_\calE^\epsilon$ with type $i$ demand equal to $\nu_i - \epsilon|\{j: (i,j) \in \calE\}|$ and type $j$ supply equal to $\mu_j-\epsilon|\{i: (i,j) \in \calE\}|$.
As CRP condition is satisfied for $\calT_\calE^{(\bnu, \bmu)}$, there exists $\delta>0$ such that
\begin{align*}
    \sum_{i \in \tilde{\calI}} \nu_i < \sum_{j: \exists i \in \tilde{\calI}, (i,j) \in \calE} \mu_j+\delta \quad \forall \tilde{\calI} \subsetneq \calI.
\end{align*}
For $\epsilon<\delta/|\calE|$, $\calT_\calE^\epsilon$ is feasible by Lemma \ref{lemma: non_empty}. Let $\bx^\epsilon \in \calT^{\epsilon}_{\calE}$ and define $\bx$ as follows:
\begin{align*}
    x_{ij}=\begin{cases}
    x^\epsilon_{ij}+\epsilon &\textit{if } (i,j) \in \calE \\
    0 &\textit{otherwise}.
    \end{cases}
\end{align*}
Observe that $\bx \in \calT_{\calE}^{(\bnu, \bmu)}$ by construction and $x_{ij}>0$ for all $(i,j) \in \calE$. This completes the proof of the second part of the lemma.
\hfill $\square$
\endproof
\proof{Proof of Lemma \ref{lemma: existence_redundant}}
We prove the contrapositive. Assume that $G(\calI \cup \calJ,\calE)$ is connected and there does not exist a redundant edge. Then, there exists $\bx^{ij} \in \calT_{\calE}^{(\bnu, \bmu)}$ for all $(i,j) \in \calE$ such that $x_{ij}^{ij}>0$. Define $\bx=\frac{1}{|\calE|}\sum_{(i,j) \in \calE}\bx^{ij}$ and note that as $\calT_{\calE}^{(\bnu, \bmu)}$ is a polytope, $\bx \in \calT_{\calE}^{(\bnu, \bmu)}$ and $x_{ij}>0$ for all $(i,j) \in \calE$. Thus, $\calB(\bx)=\calE$. As $G(\calI \cup \calJ,\calE)$ is connected, $G(\calI \cup \calJ,\calB(\bx))$ is connected which implies that CRP condition is satisfied using Lemma \ref{lemma: CRP_connectedness}. This completes the proof.
\hfill $\square$
\endproof
\proof{Proof of Proposition \ref{theo: ERP_CRP}}
Note that for any $\bx \in \calT_{\calE}^{(\bnu, \bmu)}$, $\calB(\bx) \subseteq \calE \backslash \calE_r$ as $x_{ij}=0$ for all $(i,j) \in \calE_r$. This implies that any support graph has at least ERP number of connected components. Now, as $(i,j) \in \calE \backslash \calE_r$ is not redundant by definition, there exists $\bx^{ij} \in \calT_{\calE}^{(\bnu, \bmu)}$ such that $x_{ij}^{ij}>0$. Define $\bx=\frac{1}{|\calE\backslash \calE_r|}\sum_{(i,j) \in \calE\backslash \calE_r}\bx^{ij}$ and note that as $\calT_{\calE}^{(\bnu, \bmu)}$ is a polytope, $\bx \in \calT_{\calE}^{(\bnu, \bmu)}$ and $x_{ij}>0$ for all $(i,j) \in \calE \backslash \calE_r$. Thus, $\calB(\bx)=\calE \backslash \calE_r$. Thus, there exists $\bx \in \calT_{\calE}^{(\bnu, \bmu)}$ such that the support graph has ERP number of connected components. This completes the proof. 
\hfill $\square$
\endproof
\proof{Proof of Corollary \ref{corollary: removing_edges}}
By \eqref{eq: equal_sub_space}, the subspace $\calH$ of $G(\calI \cup \calJ, \calE)$ is same as the subspace $\calH'$ of $G(\calI \cup \calJ,\bigcup_{l=1}^d \calE)$. Thus, the result follows as the parallel component of the queue length will be same in both the cases, i.e. $\bbarq_{||\calH}=\bbarq_{||\calH'}$.
\hfill $\square$
\endproof
\subsection{Correctness of Algorithm \ref{alg: CRP_decomp}}
\proof{Proof of Proposition \ref{prop: alg_crp_decomp}}
Let $\calE_r^l$ be the set of edges after $l \in [n]$ iterations of Algorithm \ref{alg: CRP_decomp}, $\calE_r$ be the set of redundant edges of $\calT_\calE^{(\bnu, \bmu)}$, and $\calC^j$ be the set of $\calB(\bx)$ reachable demand vertices from $j \in [n]$. In particular, we have $N(\calC^{j^\prime}) = \{j : \exists i \in N(\calC^{j^\prime}), (i, j) \in \calE\}$, and $\calE_r^l = \{(i, j) \in \calE : i \notin \calC^l, j \in N(\calC^l)\}$. We first show that the set of edges $\bigcup_{l=1}^n \calE_r^l$ are redundant.
\begin{claim} \label{claim: redundant_edges_are_removed}
For all $j^\prime \in [n]$, we have
\begin{align*}
    \sum_{i \in \calC^{j^\prime}} \nu_i = \sum_{j \in N(\calC^{j^\prime})} \mu_{j}.
\end{align*}
\end{claim}
By the above claim, definition of $\calE^l_r$, and Lemma \ref{lemma: base_case_chi_zero}, we have $\calE_r^l \subseteq \calE_r$ for all $l \in [n]$. Thus we have $\bigcup_{l=1}^n \calE_r^l \subseteq \calE_r$.
Now, we show that no more redundant edges are left after the termination of Algorithm \ref{alg: CRP_decomp}. After $n$ iterations of Algorithm \ref{alg: CRP_decomp}, let $i \in \calI$ be such that $(i, j) \in \calE_r \backslash \bigcup_{l=1}^n \calE_r^l$ for some $j \in \calJ$. Note that, there exists a $\calB(\bx)$ alternating path from $j$ to $i$. As otherwise, $i \notin \calC^j$ which implies that $(i, j) \in \calE_r^{j}$ as $j \in N(\calC^j)$ by definition. This is a contradiction. Now, denote the $\calB(\bx)$ alternating path from $j$ to $i$ by $P$. Let $2\epsilon = \min_{(i, j) \in \calB(\bx)} x_{ij}$, and define
\begin{align*}
    x^\prime_{i^\prime j^\prime} = \begin{cases}
    \epsilon &\textit{if } (i^\prime, j^\prime) = (i, j) \\
    x^\prime_{i^\prime j^\prime} - \epsilon &\textit{if } (j^\prime, i^\prime) \in P  \\
    x^\prime_{i^\prime j^\prime} + \epsilon &\textit{if } (i^\prime, j^\prime) \in P \\
    x^\prime_{i^\prime j^\prime} &\textit{ otherwise}.
    \end{cases}
\end{align*}
By the definition of $\calB(\bx)$ alternating path, $(j^\prime, i^\prime) \in P$ implies $(j^\prime, i^\prime) \in \calB(\bx)$. Thus, by the definition of $\epsilon$, we have $\bx^\prime \geq \bzero_{m \times n}$. In addition, as $\epsilon$ is added and subtracted from alternating edges corresponding to a cycle in $\calE$, we have $\bx^\prime \in \calT_\calE^{(\bnu, \bmu)}$ and $x^\prime_{ij} > 0$. This is a contradiction as $(i, j)$ is assumed to be a redundant edge. This implies that $\calE_r \backslash \bigcup_{l=1}^n \tilde{\calE}_r^l = \emptyset$ which is same as $\calE_r \subseteq \bigcup_{l=1}^n \tilde{\calE}_r^l$. This completes the proof.
\hfill $\square$
\endproof
\proof{Proof of Claim \ref{claim: redundant_edges_are_removed}}
Assume that, there exist $j \in [n]$ such that 
\begin{align*}
    \sum_{i \in \calC^j} \nu_i < \sum_{N(\calC^j)} \mu_{j^\prime}.
\end{align*}
Recall that $N(\calC^j) = \left\{j^\prime : \exists i \in \calC^j, (i, j^\prime) \in \calE\right\}$ and note that, there exist $j^\prime \in N(\calC^j)$ such that $x_{i^\prime j^\prime} > 0$ for some $i^\prime \notin \calC^j$. Also, by definition, there exists $i^{\prime \prime} \in \calC^j$ such that $(i^{\prime \prime}, j^\prime) \in \calE$. Now, denote by $P$, the $\calB(\bx)$ alternating path from $j$ to $i^{\prime \prime}$. If $j^\prime \notin P$, then consider the $\calB(\bx)$ alternating path $P \cup \{i^{\prime \prime}, j^\prime, i^\prime\}$ which implies that $i^\prime \in \calC^j$. This is a contradiction. On the other hand, if $j^\prime \in P$, then consider the path $P[j, j^{\prime}]$ that terminates $P$ at $j^\prime$. Now, $P[j, j^{\prime}] \cup \{i^\prime\}$ is a $\calB(\bx)$ alternating path from $j$ to $i^\prime$ which is a contradiction. This completes the proof of the claim. \hfill $\square$
\endproof

\section{Clean Slate Design}
\subsection{Correctness of Algorithm \ref{alg: feasible_tree}}
To prove Proposition \ref{prop: non_degenerate}, we require the following lemma which we state and prove before proving Proposition \ref{prop: non_degenerate}.
\begin{lemma} \label{lemma: iterates_alg_extrem_point}
Let $\bx^0$ be an extreme point of $\calT_{\calI \times \calJ}^{(\bnu, \bmu)}$ with $|\calB(\bx^0)|=\tilde{l}_0$ for some $\tilde{l}_0 \in \bbZ_+$. Then the iterates $\{\bx^k\}_{k=1}^{m+n-d_\star-\tilde{l}_0}$ of Algorithm \ref{alg: feasible_tree} are extreme points of $\calT_{\calI \times \calJ}^{(\bnu, \bmu)}$ and $|\calB(\bx^k)|=\tilde{l}_0+k$.
\end{lemma}
\proof{Proof of Lemma \ref{lemma: iterates_alg_extrem_point}}
For the ease of notation, we denote $\calT_{\calI \times \calJ}^{(\bnu, \bmu)}$ by simply $\calT$ for the rest of the proof. We prove this lemma by induction. Denote the assignment obtained after $k$ iteration of Algorithm \ref{alg: feasible_tree} by $\bx^k$. 

\underline{\textbf{Induction Hypothesis:}} After $0 \leq k< m+n-d_\star-\tilde{l}_0$ iterations of the while loop in Algorithm \ref{alg: feasible_tree}, we have $\bx^k \in \calT$ and 
the graph $\calB(\bx^k)$ is a forest with $\tilde{l}_k=|\calB(\bx^k)|=\tilde{l}_0+k$.

\underline{\textbf{Base Case:}} We initialize $\bx$ by the output of Algorithm \ref{alg: extreme_points}. Thus, by Lemma \ref{lemma: alg_extreme_points}, $\bx^0$ is an extreme point of $\calT$ and by Lemma \ref{lemma: forest}, $\calB(\bx^0)$ is a forest. Lastly, $|\calB(\bx^0)|=\tilde{l}_0$ by definition. Thus, the base case is satisfied.

\underline{\textbf{Induction Step:}} This is proved in three steps.

\textbf{Step 1 (Existence):} By contradiction, we show that if $|\calB(\bx^k)|<m+n-d_\star$, then there exists $i_1,i_2 \in \calI$, $j_1,j_2 \in \calJ$ such that $(i_1,j_1),(i_2,j_2) \in \calB(\bx^k)$, $x_{i_1j_1} \neq x_{i_2j_2}$, and $i_1$ and $i_2$ are not connected in $\calB(\bx^k)$. Assume no such $i_1,i_2,j_1,j_2$ exists. Then, there must exist $c \in \bbZ_+$ such that $\bx = c \mathbbm{1}_{m \times n}$. Thus, we have
\begin{align}
    \inner{\bone_m}{\bnu}=\inner{\bone_{m \times n}}{\bx}=|\calB(\bx^k)|c \overset{(a)}{\leq} |\calB(\bx^k)|\text{GCD}(\bnu,\bmu) \overset{(b)}{<}(m+n-d_\star) \text{GCD}(\bnu,\bmu), \label{eq: contradiction}
\end{align}
where $(a)$ follows as $c \geq GCD(\bnu, \bmu)$ as $\bnu, \bmu$ are linear combination of components of $\bx$. Lastly, $(b)$ follows as $|\calB(\bx^k)| < m+n-d_\star$ as otherwise, the algorithm would have terminated at step $k$. Now, note that \eqref{eq: contradiction} implies that $d_\star<m+n-\inner{\bone_m}{\bnu}/\text{GCD}(\bnu,\bmu)$ which contradicts the definition of $d_\star$. This completes Step 1.

\textbf{Step 2 (Feasibility):} Observe that the transformation on $\bx^k$ to obtain $\bx^{k+1}$ ensures non-negativity and also that the row sums and the column sums remains unaltered. This implies that $\bx^{k+1} \in \calT$.

\textbf{Step 3 (Extreme Point):} Lastly, we show that $\calB(\bx^{k+1})$ is a forest, where 
\begin{align*}
    \calB(\bx^{k+1})=\calB(\bx^{k}) \cup \{(i_1,j_2),(i_2,j_1)\} \backslash \{(i_1,j_1)\}.
\end{align*}
Note that, by construction, we have $|\calB(\bx^{k+1})|=|\calB(\bx^{k})|+1$. Let $\{G_l^k\}_{l=1}^{m+n-\tilde{l}_k}$ be the connected components of $\calB(\bx^k)$, where $G_1^k\dfn G(\calI_1^k \cup \calJ_1^k,\calE_1^k), G_2^k \dfn G(\calI_2^k \cup \calJ_2^k,\calE_2^k)$ be such that $(i_1,j_1) \in \calE_1^k$, $(i_2,j_2) \in \calE_2^k$, and $x_{i_1j_1}^k \neq x_{i_2j_2}^k$. By one iteration of Algorithm \ref{alg: feasible_tree}, $G_1^k \cup G_2^k$ is transformed to obtain $G_{1 \cup 2}^{k+1}$ with $\calI_1^k \cup \calI_2^k$ demand nodes, $\calJ_1^k \cup \calJ_2^k$ supply nodes, and edge set $\calE_1^k \cup \calE_2^k \cup \{(i_1,j_2),(i_2,j_1)\} \backslash \{(i_1,j_1)\}$. Thus, we have $\calB(\bx^{k+1}) = \bigcup_{l=3}^{m+n-\tilde{l}_k} G_l \cup G_{1 \cup 2}^{k+1}$.

Note that, $G_{1 \cup 2}^{k+1} \cup \{(i_1, j_1)\}$ is a connected graph with exactly one cycle as $G_1$ and $G_2$ are trees. In particular, the cycle is $\{(i_1, j_1), (i_1, j_2), (i_2, j_2), (i_1, j_2)\}$. Now, by removing one edge from the cycle, i.e. removing $(i_1, j_1)$ preserves connectivity. Thus, $G_{1 \cup 2}^{k+1}$ is a connected graph. Now, note that,
\begin{align*}
    |\calE_1^k \cup \calE_2^k \cup \{(i_1,j_2),(i_2,j_1)\} \backslash \{(i_1,j_1)\}|&=|\calE_1^k|+|\calE_2^k|+2-1 \\
    &=|\calI_1^k|+|\calJ_1^k|-1+|\calI_2^k|+|\calJ_2^k|-1+1 \\
    &=|\calI_1^k|+|\calJ_1^k|+|\calI_2^k|+|\calJ_2^k|-1,
\end{align*}
where the second equality follows as $G_1^k,G_2^k$ are trees. This shows that $G_{1 \cup 2}^{k+1}$ is a tree. This implies that, $\calB(\bx^{k+1})$ is a forest. Now, by Lemma \ref{lemma: forest}, we conclude that $\bx^{k+1}$ is an extreme point which completes this step of the proof.





By Step 1-3, the induction step is complete. 
\hfill $\square$
\endproof
\proof{Proof of Proposition \ref{prop: non_degenerate}} We prove this proposition by considering two cases depending on the value of $d$.

\textbf{Case I $(d^\star \geq d \geq d_\star)$:} Initialize Algorithm \ref{alg: feasible_tree} with an extreme point $\bx^0 \in \calT_\calE^{(\bnu, \bmu)}$ such that $\calB(\bx^0)$ is a forest with $d_0 \geq d$ connected components. Then, by Lemma \ref{lemma: alg_extreme_points}, after $d - d_0$ iterations of the while loop in Algorithm \ref{alg: feasible_tree}, we obtain an extreme point $\bx^{d-d_0} \in \calT_\calE^{(\bnu, \bmu)}$ such that $\calB(\bx^{d-d_0})$ is a forest with $d$ connected components. This implies $|\calB(\bx^{d-d_0})| = m+n-d$ which terminates Algorithm \ref{alg: feasible_tree} and returns $\calB(\bx^{d-d_0})$. This completes Case I.

\textbf{Case II $(d<d_\star)$:} Initialize Algorithm \ref{alg: feasible_tree} with arbitrary extreme point $\bx^0 \in \calT_\calE^{(\bnu, \bmu)}$ such that $\calB(\bx^0)$ is a forest with $d_0$ connected components. Then, by Lemma \ref{lemma: alg_extreme_points}, after $d_\star - d_0$ iterations, we obtain an extreme point $\bx_\star \in \calT_\calE^{(\bnu, \bmu)}$ such that $\calB(\bx_\star)$ is a forest with $d_\star$ connected components. Now, as $d < d_\star$, Algorithm \ref{alg: feasible_tree} transforms $\bx_\star$ as follows:
\begin{align*}
    x_{ij}=\begin{cases}
    x^\star_{ij}-\frac{1}{2}\min_{l \in [e]} \{x_{i_lj_l}^\star\} &\textit{if } (i,j) \in \{(i_l,j_l): l \in [e]\} \\
    \frac{1}{2}\min_{l \in [e]}\{x_{i_lj_l}^\star\} &\textit{if } (i,j) \in \{(i_l,j_{l+1}): l \in [e-1]\} \cup \{(i_{e},j_1)\} \\
    x_{ij}^\star &\textit{otherwise},
    \end{cases}
\end{align*}
where $e=d_\star-d+1$. Note that,
\begin{align*}
    \calB(\bx) &= \calB(\bx_\star) \cup \{(i_l,j_{l+1}): l \in [e-1]\} \cup \{(i_{e},j_1)\} \\
     &= \bigcup_{l=e+1}^{d_\star} G_l \cup \tilde{G},
\end{align*}
where $\tilde{G} = \bigcup_{l=1}^e G_l \cup \{(i_l,j_{l+1}): l \in [e-1]\} \cup \{(i_{e},j_1)\}$. Note that, $\tilde{G}$ is a connected graph as it contains the trees $\{G_l\}_{l=1}^e$ connected to each other by $\{(i_l,j_{l+1}): l \in [e-1]\} \cup \{(i_{e},j_1)\}$. Thus, $\calB(\bx)$ has $d_\star-e+1=d$ connected components. Note that $|\calB(\bx)|=m+n-d_\star+d_\star-d+1-e=m+n-d+1$. This completes Case II as the Algorithm \ref{alg: feasible_tree} terminates and returns $\calB(\bx)$.

Combining both the cases, the proof is complete.
\hfill $\square$
\endproof
\subsection{Existence of Extreme Points with Given Connectivity}
In this section, we present the proof of Proposition \ref{prop: degeneracy}. First, we present a basic result along with its proof about extreme points of $\calT$.
\begin{lemma} \label{lemma: integer_x}
For a given $\bnu \in \bbZ_+^n,\bmu \in \bbZ_+^m$, any extreme point of $\calT$ is component wise divisible by $\text{GCD}(\bnu,\bmu)$.
\end{lemma}
\proof{Proof of Lemma \ref{lemma: integer_x}} Consider the following transportation polytope with scaled demand and supply:
\begin{align*}
    \calT'=\left\{\by \in \bbR_+^{n \times m}:\frac{\nu_i}{\text{GCD}(\bnu,\bmu)}=\sum_{j=1}^n y_{ij} \forall i \in \calI, \ \frac{\mu_j}{\text{GCD}(\bnu,\bmu)}=\sum_{i=1}^m y_{ij} \forall j \in \calJ\right\}.
\end{align*}
Note that, $\by^\star$ is a basic feasible solution of $\calT'$ if and only if $\bx^\star=\by^\star \text{GCD}(\bnu,\bmu)$ is a basic feasible solution of $\calT$. This can be verified by using the definition of a basic feasible solution of a polytope. By Lemma \ref{lemma: integer}, we know that $\by^\star$ is an integral matrix. This completes the proof.
\hfill $\square$
\endproof
Now, we use the above result to prove Proposition \ref{prop: degeneracy} below.
\proof{Proof of Proposition \ref{prop: degeneracy}}
First, we show that there exists an extreme point $\bx \in \calT$ with $|\calB(\bx)|=m+n-d$ for all $d \in \{d_\star,\hdots,d^\star\}$. By definition of $d^\star$, there exists an extreme point $\bx^{\star} \in \calT$ such that $\calB(\bx^{\star})$ is a forest with $d^\star$ connected components. Initialize $\bx=\bx^{\star}$ in Algorithm \ref{alg: feasible_tree}. Now, by Lemma \ref{lemma: iterates_alg_extrem_point}, Algorithm \ref{alg: feasible_tree} generates extreme points $\{\bx^k\}_{k=1}^{d^\star-d_\star}$ such that $\calB(\bx^k)$ is a forest with $m+n-d^\star+k$. This completes one part of the proof.

Now, we show that there does not exist extreme point $\bx$ with $|\calB(\bx)| > m+n-d_\star$ if $d \notin \{d_\star,\hdots,d^\star\}$. By the definition of $d^\star$, no such $\bx$ exists if $d>d^\star$. Now, assume that there exists an extreme point $\bx \in \calT$ such that $|\calB(\bx)|>m+n-d_\star$. Then, we have
\begin{align}
    \inner{\bone_m}{\bnu}=\inner{\bone_{m \times n}}{\bx} \overset{(*)}{\geq} |\calB(\bx^k)|\text{GCD}(\bnu,\bmu) >(m+n-d_\star) \text{GCD}(\bnu,\bmu). \label{eq: contradiction_prop}
\end{align}
By Lemma \ref{lemma: integer_x}, the extreme points of $\calT$ are component wise divisible by $\text{GCD}(\bnu,\bmu)$. Thus, if $x_{ij}>0$, then we must have $x_{ij} \geq \text{GCD}(\bnu,\bmu)$ which implies $(*)$. By \eqref{eq: contradiction_prop}, we get $d_\star>m+n-\inner{\bone_m}{\bnu}/\text{GCD}(\bnu,\bmu)$. Thus, by the definition of $d_\star$ given by \eqref{eq: d_star}, we obtain $d_\star=1$ implying that $|\calB(\bx)|>m+n-1$. This is a contradiction as $\calB(\bx)$ is a forest by Lemma \ref{lemma: forest}.
\hfill $\square$
\endproof
\subsection{Proof of Lemmas and Corollaries}
\proof{Proof of Corollary \ref{cor: min_edges}}
If $\inner{\bone_n}{\bnu} \geq (m+n-1)\text{GCD}(\bnu,\bmu)$, then $d_\star=1$. By Theorem \ref{theo: min_edges} \ref{theo: min_edges_3}, there exists $\calE$ such that $|\calE|=m+n-1$ and ERP number of $\calT_\calE^{(\bnu, \bmu)}$ is 1, i.e. CRP condition is satisfied. This completes one part of the proof.

Conversely, assume that there exists $\calE$ such that $|\calE| = m+n -1$ and $\calT_\calE^{(\bnu, \bmu)}$ satisfies the CRP condition, i.e. ERP number is 1. Then, by Theorem \ref{theo: min_edges} \ref{theo: min_edges_2}, we must have $d_\star = 1$. By the definition of $d_\star$, we have $\inner{\bone_n}{\bnu} \geq (m+n-1)\text{GCD}(\bnu,\bmu)$. This completes the proof.
\hfill $\square$
\endproof
\proof{Proof of Corollary \ref{corollary: optimality_CRP_max_weight}}
If $\sum_{i=1}^m \nu_i \geq (m+n-1)\text{GCD}(\bnu,\bmu)$, then by Corollary \ref{cor: min_edges}, there exists $\bx \in \calT$ such that $|\calB(\bx)|=m+n-1$ and $G(\calI \cup \calJ,\calB(\bx))$ is a tree. Let $\calE=\calB(\bx)$ and note that $\calT_\calE^{(\bnu, \bmu)}$ satisfy the CRP condition. Then, we upper bound the sum of queue lengths under max-weight by Proposition \ref{prop: HT} and we lower bound the sum of queue lengths under any policy by \cite[Corollary 1]{zhong2019_process_flexibility} to complete the proof.
\hfill $\square$
\endproof
\proof{Proof of Lemma \ref{lemma: e_star}}
Let $\calE$ be such that the ERP number of $\calT_\calE^{(\bnu, \bmu)}$ is $d$. Then, by Proposition \ref{theo: ERP_CRP}, there exists $\bx \in \calT_\calE^{(\bnu, \bmu)}$ such that $\calB(\bx)$ has $d$ connected components. By feasibility of $\bx$, each connected component $\{G(\calI_l\cup\calJ_l,\calE_l)\}_{l=1}^{d}$ of $G(\calI \cup \calJ,\calB(\bx))$ must be such that $\sum_{i \in \calI_l} \nu_i=\sum_{j \in \calJ_l}\mu_j$. Thus, $\{\calI_l\}_{l=1}^{d}$ is a disjoint cover of $\calI$ and $\{\calJ_l\}_{l=1}^{d}$ is a disjoint cover of $\calJ$ that satisfies the conditions of Lemma \ref{lemma: e_star}.

To prove the converse, let $\{\calI_l\}_{l=1}^{d}$ and $\{\calJ_l\}_{l=1}^{d}$ by the disjoint covers of $\calI$ and $\calJ$ respectively, such that the conditions of Lemma \ref{lemma: e_star} are satisfied. Now, define $\calE = \bigcup_{l=1}^d \calI_l \times \calJ_l$ and note that the ERP number of $\calT_\calE^{(\bnu, \bmu)}$ is $d$. In particular, it is easy to check that $\{G_l(\calI_l \cup \calJ_l, \calI_l \times \calJ_l)\}_{l=1}^d$ is the CRP decomposition of $\calT_\calE^{(\bnu, \bmu)}$. This completes the proof.
\hfill $\square$
\endproof
\subsection{Robustness to Demand Uncertainty}
\proof{Proof of Theorem \ref{theo: strong_connectivity}}
Let $\hat{\bnu} \in N_{\bnu}$, and $\calC \subseteq \calI$ be such that $\sum_{j : \exists i \in \calC, (i, j)\in \calE \backslash \calE_r} \mu_j - \sum_{i \in \calC} \nu_i > 0$. Then, we have
\begin{align}
    \sum_{i \in \calC} \hat{\nu}_i = \sum_{i \in \calC} \nu_i + \sum_{i \in \calC} \omega_i \overset{*}{\leq} \sum_{j : \exists i \in \calC, (i, j) \in \calE} \mu_j - \delta_{\calT_{\calE}^{(\bnu, \bmu)}} + \sum_{i \in \calC} \omega_i \overset{**}{<} \sum_{j : \exists i \in \calC, (i, j) \in \calE} \mu_j. \label{eq: robust_strict_ineq}
\end{align}
where $(*)$ follows by the definition of $\delta_{\calT_{\calE}^{(\bnu, \bmu)}}$. Next, $(**)$ holds due to the following: as $\calT_{\calE}^{(\hat{\bnu}, \bmu)} \neq \emptyset$, we have
\begin{align*}
\sum_{j \in \calJ} \mu_j = \sum_{i \in \calI} \hat{\nu}_i = \sum_{i \in \calI} \nu_i + \sum_{i \in \calI} \omega_i = \sum_{j \in \calJ} \mu_j + \sum_{i \in \calI} \omega_i.
\end{align*}
Thus, we have $\sum_{i \in \calI} \omega_i = 0$. In addition, as $\|\bomega\|_1 < 2\delta_{\calT_{\calE}^{(\bnu, \bmu)}}$, we have $\sum_{i \in \calC} \omega_i < \delta_{\calT_{\calE}^{(\bnu, \bmu)}}$ for any $\calC \subseteq \calI$. 

Now, by \eqref{eq: robust_strict_ineq}, we conclude that $\calK_{\calE}^{(\hat{\bnu}, \bmu)} \subseteq \calK_{\calE \backslash \calE_r}^{(\bnu, \bmu)} = \calK_{\calE}^{(\bnu, \bmu)}$, where $\calK_{\calE}^{(\hat{\bnu}, \bmu)}$ is the SSC cone corresponding to $(\hat{\bnu}, \bmu, \calE)$ as defined in \eqref{eq: ssc_cone}. Furthermore, we have $\calH_{\calE}^{(\hat{\bnu}, \bmu)} \subseteq \calH_{\calE}^{(\bnu, \bmu)}$ which implies the ERP number of $\calT_{\calE}^{(\hat{\bnu}, \bmu)}$ is at most the ERP number of $\calT_{\calE}^{(\bnu, \bmu)}$ by Theorem \ref{theo: sub_space_SSC}. This completes the proof.
\hfill $\square$
\endproof
\subsubsection{Redundant Edges do not Help}
In this section, we prove Proposition \ref{proposition: redundant_edges_not_help} by using Lemma \ref{lemma: DAG}. So, we first present the proof of Lemma \ref{lemma: DAG} followed by the proof of Proposition \ref{proposition: redundant_edges_not_help}.
\proof{Proof of Lemma \ref{lemma: DAG}}
We prove this lemma by contradiction. Let $\{G_l=G(\calI_l \cup \calJ_l, \calE_l)\}_{l=1}^d$ be the CRP decomposition and $D(V_{crp}, \calE_{crp})$ be the corresponding CRP-graph. Let $\calA_{crp}$ be a cycle in the CRP-graph. Now, we show that the edges in $G$ corresponding to the cycle in the CRP-graph are not redundant. Let $\calT_v$ be the transportation polytope corresponding to demand $\{\nu_i\}_{i \in \calI_v}$, supply $\{\mu_j\}_{j \in \calJ_v}$, and flexibility $G_v$. As $
\calT_v$ satisfy the CRP condition for all $v \in V_{crp}$, we have $\bx^v \in \calT_v$ with $\calB(\bx^v) = \calE_l$. Now, define
\begin{align*}
x_{ij}^{\prime} = \begin{cases}
x_{ij}^v  &\textit{if } (i, j) \in \calE_v \quad \forall v \in V_{crp} \\
0 &\textit{otherwise}.
\end{cases} \quad \forall (i, j) \in \calE
\end{align*}
Note that, $\bx^{\prime} \in \calT_{\calE}^{(\bnu, \bmu)}$. Now, let $\calA_{crp} = \{u_1, \hdots, u_l, u_{l+1}\}$, where $u_{l+1} = u_1$. Then, there exists $i_{u_k} \in \calI_{u_k}$, $j_{u_k} \in \calJ_{u_k}$ such that $(i_{u_k}, j_{u_{k+1}}) \in \calE$ for all $k \in [l]$. In addition, as $G_v$ satisfy CRP condition, there exists a path $p_k$ in $G_{u_k}$ connecting $i_{u_k}$ to $j_{u_k}$. Now, consider the following cycle in $G$:
\begin{align*}
\calA = ((i_{u_{1}},j_{u_{2}}),p_2,(i_{u_{2}},j_{u_{3}}),p_3,\hdots,(i_{u_{l}},j_{u_{1}}),p_1). 
\end{align*}
For $\epsilon = \min_{(i,j): x_{ij}^\prime>0}\{x_{ij}^\prime\}/2$, define $\bx^{\prime\prime}$ as follows:
\begin{align*}
    x_{ij}^{\prime\prime}=\begin{cases}
    x_{ij}^\prime &\textit{if } (i,j) \notin \calA \\
    x_{ij}^\prime+\epsilon &\textit{if } (i,j) \text{ is at odd number position in} \ \calA \\
    x_{ij}^\prime-\epsilon &\textit{if } (i,j) \text{ is at even number position in} \ \calA.
    \end{cases}
\end{align*}
It is easy to verify that $\bx^{\prime \prime} \in \calT$ as we added and subtracted $\epsilon$ from $\bx^{\prime}$ corresponding to a cycle in $G(\calI \cup \calJ, \calE)$ and $x_{ij}^{\prime} > 0$ for all $(i, j) \in \bigcup_{k=1}^l p_k$. Note that, $x_{i_{u_k}j_{u_{k+1}}}^{\prime\prime} > 0$ for all $k \in [l]$ as $p_k$ have odd number of edges. This is a contradiction.
\hfill $\square$
\endproof

\proof{Proof of Proposition \ref{proposition: redundant_edges_not_help}}
First, note that for all $\calC$ such that $\sum_{j : \exists i \in \calC, (i, j)\in \calE \backslash \calE_r} \mu_j - \sum_{i \in \calC} \nu_i > 0$, we have
\begin{align*}
    \sum_{j : \exists i \in \calC, (i, j)\in \calE} \mu_j - \sum_{i \in \calC} \nu_i \geq \sum_{j : \exists i \in \calC, (i, j)\in \calE \backslash \calE_r} \mu_j - \sum_{i \in \calC} \nu_i > 0,
\end{align*}
which immediately implies that $\delta_{\calT_\calE^{(\bnu, \bmu)}} \geq \delta_{\calT_{\calE \backslash \calE_r}^{(\bnu, \bmu)}}$.

Now, we show that $\delta_{\calT_\calE^{(\bnu, \bmu)}} \leq \delta_{\calT_{\calE \backslash \calE_r}^{(\bnu, \bmu)}}$. Let $\calC^\star$ be such that $\delta_{\calT_{\calE \backslash \calE_r}^{(\bnu, \bmu)}} = \sum_{j : \exists i \in \calC^\star, (i, j)\in \calE \backslash \calE_r} \mu_j - \sum_{i \in \calC^\star} \nu_i$. Now, let $\{G_l \dfn G(\calI_l \cup \calJ_l, \calE_l)\}_{l=1}^d$ be the CRP decomposition of $\calT_\calE^{(\bnu, \bmu)}$ which is same as the connected components of $G(\calI \cup \calJ, \calE\backslash \calE_r)$. Then, as $\{G_l\}_{l=1}^d$ are disjoint and satisfy CRP condition, there exists $l^\star \in [d]$ such that $\calC^\star \subseteq \calI_{l^\star}$, as otherwise, the gap of $\calC^\star \cap \calI_{l^\star}$ will be smaller. Now, we construct a $\calC^\prime$ such that $\sum_{j : \exists i \in \calC^\prime, (i, j)\in \calE} \mu_j - \sum_{i \in \calC^\prime} \nu_i = \delta_{\calT_{\calE \backslash \calE_r}^{(\bnu, \bmu)}}$.

Let $D(V_{crp}, \calE_{crp})$ be the CRP-graph as defined in \eqref{eq: crp_graph}. Note that, the CRP-graph is a DAG by Lemma \ref{lemma: DAG}. Let $\{l_d, \hdots, l_1\}$ be a topological sorting of the CRP-graph. Now, let $k^{\prime} \in [d]$ be such that $l^\star = l_{k^\prime}$ and define $\calC^\prime = \calC^\star \cup \bigcup_{k=1}^{k^\prime-1} \calI_{l_k}$. By the definition of topological sorting and CRP-graph, we have 
\begin{align}
    \left\{j : \exists i \in \calC^\prime, (i, j) \in \calE\right\} &=\bigcup_{k=1}^{k^\prime-1}\left\{j : \exists i \in \calI_{l_k}, (i, j) \in \calE\right\} \cup \left\{j : \exists i \in \calC^\star, (i, j) \in \calE\right\} \nonumber\\
    &\overset{(a)}{=}\bigcup_{k=1}^{k^\prime-1} \calJ_{l_k} \cup \left\{j : \exists i \in \calC^\star, (i, j) \in \calE\right\} \nonumber\\
    &\overset{(b)}{=} \bigcup_{k=1}^{k^\prime-1} \calJ_{l_k} \cup \left\{j : \exists i \in \calC^\star, (i, j) \in \calE \backslash \calE_r\right\}, \label{eq: neighbor_constructed_set}
\end{align}
where $(a)$ follows as the neighbor of $\calI_l$ by considering non-redundant edges is $\calJ_l$ and the redundant edges are only of the form $(l_{k_1}, l_{k_2})$ for $k_1 > k_2$. Next, $(b)$ follows as the neighbor of $\calC^\star$ by considering redundant edges is a subset of $\bigcup_{k=1}^{k^\prime-1} \calJ_{l_k}$. Now, the above implies that
\begin{align*}
    \sum_{j : \exists i \in \calC^\prime, (i, j) \in \calE} \mu_j &= \sum_{k=1}^{k^\prime-1} \sum_{j \in \calJ_{l_k}} \mu_j + \sum_{j : \exists i \in \calC^\star, (i, j) \in \calE \backslash \calE_r} \mu_j \\
    &\overset{(a)}{=} \sum_{k=1}^{k^\prime-1} \sum_{i \in \calI_{l_k}} \nu_i + \sum_{i \in \calC^\star} \nu_i + \delta_{\calT_{\calE \backslash \calE_r}^{(\bnu, \bmu)}} \\
    &\overset{(b)}{=} \sum_{i \in \calC^\prime} \nu_i + \delta_{\calT_{\calE \backslash \calE_r}^{(\bnu, \bmu)}} \\
    &\overset{(c)}{\leq} \sum_{j : \exists i \in \calC^\prime, (i, j) \in \calE} \mu_j - \delta_{\calT_{\calE}^{(\bnu, \bmu)}} + \delta_{\calT_{\calE \backslash \calE_r}^{(\bnu, \bmu)}} \\
    \Rightarrow \delta_{\calT_{\calE}^{(\bnu, \bmu)}} &\leq \delta_{\calT_{\calE \backslash \calE_r}^{(\bnu, \bmu)}},
\end{align*}
where $(a)$ follows by the definition of $\calC^\star$ and using the fact that $G_{l_k}$ satisfy the CRP condition for all $k \in [d]$. Next, $(b)$ follows by \eqref{eq: neighbor_constructed_set}. Note that, $(b)$ also implies that $\sum_{i \in \calC^\prime} \nu_i < \sum_{j : \exists i \in \calC^\prime, (i, j) \in \calE} \mu_j$ as $\delta_{\calT_{\calE \backslash \calE_r}^{(\bnu, \bmu)}} > 0$ by definition. Using this and the definition of $\delta$, $(c)$ follows. This completes the proof.

\hfill $\square$
\endproof
\section{Improve an Existing Production System}
\subsection{Adding one Edge} \label{app: adding_one_edge}
\proof{Proof of Theorem \ref{theo: add_one_edge}}
Adding one edge to $G(\calI \cup \calJ, \calE)$ will introduce at max one additional edge in $D(V_{crp},\calE_{crp})$. Let $(i, j)$ be the edge added to $\calE$ and $(u_1, u_2)$ be the corresponding edge added to $\calE_{crp}$. Now, let $\{\calA_l\}_{l=1}^k$ be the set of cycles in $D(V_{crp},\calE_{crp} \cup (u_1, u_2))$. Note that, $(u_1, u_2) \in \calA_l$ for all $l \in [k]$ as $D(V_{crp},\calE_{crp})$ is a DAG by Lemma \ref{lemma: DAG}. Define $\calV_c = \bigcup_{l=1}^l \calA_l$ to be the set of all vertices that belong to a cycle. We claim that the CRP components of $\calT_\calE^{(\bnu, \bmu)}$ corresponding to $\calV_c$ results in a single CRP component by adding the edge $(i, j)$. In particular, we define $\calV_{crp}^{\prime} \dfn \calV_{crp} \backslash \calV_c \cup \{0\}$ and claim the following:
\begin{claim} \label{claim: CRP_add_one_edge}
The CRP decomposition of $\calT_{\calE \cup (i, j)}^{(\bnu, \bmu)}$ is given by $\{G_l^\prime\}_{l \in \calV_{crp}^{\prime}}$, where 
\begin{align*}
    G_l^\prime=\begin{cases}
    G(\calI_{l} \cup \calJ_l, \calE_l) &\textit{if } l \in V_{crp}^\prime \backslash \{0\} \\
    \bigcup_{v \in \calV_c} G(\calI_v \cup \calJ_v, \calE_v) \cup \{(i, j)\} &\textit{if } l = 0.
    \end{cases}
\end{align*}
\end{claim}
The proof of the claim follows by Proposition \ref{prop: characterization_CRP_decomp} and is deferred to Section \ref{sec: improve_claims}. Now, by Claim \ref{claim: CRP_add_one_edge}, the ERP number of $\calT_{\calE \cup (i, j)}^{(\bnu, \bmu)}$ is $|V|=|V_{crp}|-|\calV_c|+1$. This completes the proof. \hfill $\square$
\endproof
\proof{Proof of Corollary \ref{corollary: min_erp_number}}
Let $(i, j) \in \calI_{l_1} \times \calJ_{l_2}$ be an edge such that $l_1 \notin \calA_\star$. Now, we construct $(i^{\prime}, j^{\prime}) \in \calI_{l^\prime} \cup \calJ_{l_2}$ such that $n_{\calE \cup (i^\prime, j^\prime)}^{\operatorname{ERP}} \leq n_{\calE \cup (i, j)}^{\operatorname{ERP}}$, where $l^\prime \in \calA_\star$ is such that, there exists a path $P$ from $l_1$ to $l^\prime$. The proof follows similarly for the case of $l_2 \notin \calA^\star$ and we omit the details here. 

Let $\{\calA_l\}_{l=1}^k$ be the cycles in $D(V_{crp}, \calE_{crp} \cup (l_1, l_2))$ and note that $(l_1, l_2) \in \calA_l$ for all $l \in [k]$. Also, $l^\prime \notin \calA_l$ for all $l \in [k]$ as, otherwise, there exists a path from $l^\prime$ to $l_1$ in the CRP-graph which is a contradiction. Now, observe that, $\calA_l^\prime \dfn \calA_l[l_2, l_1] \cup P \cup (l^\prime, l_2)$ is a cycle in $D(V_{crp}, \calE_{crp} \cup (l^\prime, l_2))$ for all $l \in [k]$. This implies that
\begin{align*}
\bigg|\bigcup_{l=1}^k \calA_l^{\prime} \bigg| \geq \bigg|\bigcup_{l=1}^k \calA_l \bigg| + 1,
\end{align*}
as $l^\prime \notin \calA_l$ for all $l \in [k]$. This completes the proof by using Theorem \ref{theo: add_one_edge}. \hfill $\square$
\endproof
\subsection{Adding Multiple Edges}
\proof{Proof of Theorem \ref{theo: opt_sol_multiple_long_chains}}
Let $\{\tilde{G}_k\}_{k=1}^K$ be a feasible solution of \eqref{eq: min_erp_schedule}. In particular, let $\left\{(u_k, v_k)\right\}_{k=1}^K$ be such that $\tilde{G}_k = \tilde{G}_{k-1} \cup \{(u_{k}, v_{k})\}$. We first lower bound the ERP number of the graphs $\{\tilde{G}_k\}_{k=1}^K$ using induction. Let $\tilde{D}_k \dfn D(V_{crp}^k, \calE_{crp}^k)$ be the CRP-graph corresponding to $\tilde{G}_k$ and let $u_k \in \calI_{l_1^k}$, $v_k \in \calJ_{l_2^k}$, where $\{G(\calI_l^k \cup \calJ_l^k, \calE_l^k)\}_{l=1}^{n^{\operatorname{ERP}}_{\tilde{G}_k}}$ is the CRP decomposition of $\tilde{G}_k$. To state the induction hypothesis, let $\tilde{k}_0 = 0$ and define
\begin{align}
    \tilde{k}_l = \min\left\{k > \tilde{k}_{l-1} :  D(V_{crp}^{k-1}, \calE_{crp}^{k-1} \cup (l_{1}^{k}, l_2^{k})) \text{ contains a cycle}\right\} \quad \forall l \in [\tilde{p}]. \label{eq: def_k_l}
\end{align}
Now, the induction hypothesis is given as follows.
\begin{align}
    n^{\text{ERP}}_{\tilde{G}_k} \geq \begin{cases}
      n^{\text{ERP}}_{\tilde{G}_{k-1}} &\textit{if } k \notin \{\tilde{k}_l: l \in [\tilde{p}]\} \\
      \max\left\{n_{G_0}^{\text{ERP}} - \left(k - l\right), 1\right\} &\textit{if } k = \tilde{k}_l \ \forall l \in [\tilde{p}].
    \end{cases} \tag{IH} \label{eq: ind_hypothesis_add_edges}
\end{align}

\textbf{Base Case:} By the statement of the theorem, there does not exist redundant edges in $G_0$. Thus, by Theorem \ref{theo: add_one_edge}, $n^{\text{ERP}}_{\tilde{G}_1} = n^{\text{ERP}}_{G_0}$ as $\calE_{crp}^0 = \emptyset$. This completes the base case.

\textbf{Induction Step:} First consider the case when $k \neq \tilde{k}_l$ for all $l \in [\tilde{p}]$. Adding the edge $(u_k, v_k)$ to $\tilde{G}_{k-1}$ corresponds to adding the edge $(l_1^k, l_2^k)$ to $\tilde{D}_{k-1}$. By \eqref{eq: def_k_l}, $\tilde{D}_{k-1} \cup (l_1^k, l_2^k)$ contains no cycle. Thus, by Theorem \ref{theo: add_one_edge}, we have $n^{\text{ERP}}_{\tilde{G}_{k}} = n^{\text{ERP}}_{\tilde{G}_{k-1}}$. 

Now, let $k = \tilde{k}_l$ for some $l \in [\tilde{p}]$. Let $\calA_{\tilde{k}_i}$ be the set of vertices that belongs to a cycle in $\tilde{D}_{\tilde{k}_i-1} \cup (l_1^{\tilde{k}_i}, l_2^{\tilde{k}_i})$. Then, by Theorem \ref{theo: add_one_edge}, we have
\begin{align*}
n_{\tilde{G}_{\tilde{k}_l}}^{\operatorname{ERP}} &\geq n_{G_0}^{\operatorname{ERP}} - \sum_{i=1}^l |\calA_{\tilde{k}_i}| + l \\
&\geq n_{G_0}^{\operatorname{ERP}} - \tilde{k}_l + l,
\end{align*}
where, the last inequality follows as $\sum_{i=1}^l |\calA_{\tilde{k}_i}|$ is at most the number of edges added to $D_0$, which is equal to $\tilde{k}_l$.
Also, note that $n^{\text{ERP}}_{\tilde{G}_k} \geq 1$ by definition. This completes the induction step. Now, using \eqref{eq: ind_hypothesis_add_edges}, we get
\begin{align}
    n^{\text{ERP}}_{\tilde{G}_k} \geq \max\left\{n^{\text{ERP}}_{G_0} - \max\{k_l-l : k_l < k\}, 1 \right\} \quad \forall k \in [K]. \label{eq: ERP_given_solution}
\end{align}
Now, we define $\{G_l\}_{l=1}^p$ corresponding to \eqref{eq: def_k_l} by considering
\begin{align*}
    k_l = \min\left\{\eta-1+l, \tilde{k}_l\right\} \quad \forall l \in [p]
\end{align*}
with $p = \tilde{p}$. By construction, we have 
\begin{align}
     n^{\text{ERP}}_{G_k} &= \begin{cases}
    n^{\text{ERP}}_{G_{k-1}} &\textit{if } k \neq k_l \quad \forall l \in [p] \\
    \max\left\{n^{\text{ERP}}_{G_{0}} - k + l, 1 \right\}&\textit{if } \exists l \in [p], \ k = k_l.
     \end{cases} \nonumber \\
     &= \max\left\{n^{\text{ERP}}_{G_0} - \max\{k_l-l : k_l < k\}, 1 \right\} \quad \forall k \in [K]. \label{eq: erp_constructed_solution}
\end{align}
By comparing \eqref{eq: ERP_given_solution} and \eqref{eq: erp_constructed_solution}, we get
\begin{align*}
     n^{\text{ERP}}_{\tilde{G}_k} &\geq  n^{\text{ERP}}_{G_k} \quad \forall k \in [K] \\
     \Rightarrow f_k\left(n^{\text{ERP}}_{\tilde{G}_k}\right) &\geq  f_k\left(n^{\text{ERP}}_{G_k}\right) \quad \forall k \in [K] \\
     \Rightarrow \sum_{k=1}^K f_k\left(n^{\text{ERP}}_{\tilde{G}_k}\right) &\geq \sum_{k=1}^K f_k\left(n^{\text{ERP}}_{G_k}\right).
\end{align*}
This completes the proof. \hfill $\square$
\endproof
\proof{Proof of Corollary \ref{corollary: last_erp_number_min}}
By \eqref{eq: opt_soln}, it is immediate that $n^{\text{ERP}}_K = \max\{1, n-K+1\}$ for $p=1$ and $k_1 = \min\left\{n, K\right\}$. Now, as $G_0$ corresponds to an extreme point of $\calT$, all the edges in $G_0$ are not redundant. Thus, maximum number of redundant edges in $G_{K-1}$ is equal to $K-1$. So, $n^{\text{ERP}}_K \geq n-K+1$ as the maximum number of redundant edges in a cycle formed by adding an edge to $G_{K-1}$ is equal to $K-1$. In addition, we have $n^{\text{ERP}}_K \geq 1$ by definition. This completes the proof.   \hfill $\square$
\endproof
\proof{Proof of Corollary \ref{corollary: total_erp_number_min}}
Using Theorem \ref{theo: opt_sol_multiple_long_chains}, the optimization problem \eqref{eq: min_erp_schedule} can be reformulated as follows:
\begin{align}
    \gamma_1^\star = \min_{\mathbf{\tk}, p} \sum_{i=1}^{p+1} \tk_if\left(n - \sum_{j=1}^{i-1} \tk_j + (i-1)\right)-f(n) \nonumber \span \\
    \text{subject to} \ \sum_{i=1}^{p+1} \tk_i = K, \ \sum_{i=1}^p \tk_i \leq n-1+p, \ \tk_i \in \bbZ_+ \ \forall i \in [p]. \label{eq: constraints_long_chains} 
\end{align}
where $\sum_{l=1}^i \tk_l$ corresponds to $k_i$ in \eqref{eq: long_chain_set} and the constraints \eqref{eq: constraints_long_chains} are implied by \eqref{eq: long_chain_set}. Now, for the special case of $f(x) = x$, the objective function can be simplified as follows:
\begin{align*}
   \sum_{i=1}^{p+1} \tk_if\left(n - \sum_{j=1}^{i-1} \tk_j + (i-1)\right)-f(n) &= \sum_{i=1}^{p+1} \tk_i\left(n - \sum_{j=1}^{i-1} \tk_j + (i-1)\right)-n \\
    &= n\sum_{i=1}^{p+1} \tk_i - \sum_{i=1}^{p+1} \sum_{j=1}^{i-1} \tk_j\tk_i +\sum_{i=1}^{p+1} (i-1)\tk_i - n \\
    &= (K-1)n - \sum_{i=1}^{p+1} \sum_{j=1}^{i-1} \tk_j\tk_i +\sum_{i=1}^{p+1} (i-1)\tk_i \\
    &= (K-1)n - \frac{K^2}{2} + \frac{1}{2} \sum_{i=1}^{p+1} \tk_i^2 + \sum_{i=0}^p i\tk_{i+1}.
\end{align*}
So, we can reformulate \eqref{eq: min_erp_schedule} further to get
\begin{subequations} \label{eq: reformulated_obj}
\begin{align}
    \gamma_2^\star = \min_{\mathbf{\tk}, p} \frac{1}{2} \sum_{i=1}^{p+1} \tk_i^2 + \sum_{i=0}^p i\tk_{i+1} \label{eq: reformulated_obj_fn} \span \\
    \text{subject to} \ \sum_{i=1}^{p+1} \tk_i = K, \ \sum_{i=1}^p \tk_i \leq n-1+p, \ \tk_i \in \bbZ_+ \ \forall i \in [p].
\end{align}
\end{subequations}
Note that, $\gamma_1^\star = (K-1)n - K^2/2 + \gamma_2^\star$. By fixing $p$ and using the KKT conditions, the optimal solution of the linear relaxation of \eqref{eq: reformulated_obj} in terms of $p$ is given by
\begin{align} \label{eq: optimal_tk_function_of_p}
    \tk_i^\star = \min\left\{\frac{K}{p+1}, \frac{n-1}{p}+ \frac{1}{2}\right\} + \frac{p}{2} - i + 1 \quad \forall i \in [p],
\end{align}
Using the solution of the linear relaxation of \eqref{eq: reformulated_obj}, we construct the optimal solution of the integer program and present it in the following claim.
\begin{claim} \label{claim: opt_solution_nlip}
Let $\{i_j\}_{j=1}^m$ be an arbitrary subset of $\left[p+\mathbbm{1}\left\{\frac{pK}{p+1} < n-1+\frac{p}{2}\right\}\right]$. Define
\begin{align*}
    \tk_i^{\star\star} = \begin{cases}
    \left\lceil \tk_i^\star \right\rceil  &\textit{if } \exists j \in [m]: \ i = i_j \\
    \left\lfloor \tk_i^\star \right\rfloor  &\textit{otherwise}.
    \end{cases}
\end{align*}
where 
\begin{align*}
    m = \left(p+\mathbbm{1}\left\{\frac{pK}{p+1} < n-1+\frac{p}{2}\right\}\right) \operatorname{frac}\left(f(p)\right).
\end{align*}
Then
\begin{align*}
     \mathbf{\tk}^{\star\star}(p) = \arg\min_{\mathbf{\tk}} \frac{1}{2} \sum_{i=1}^{p+1} \tk_i^2 + \sum_{i=0}^p i\tk_{i+1} \span \\
    \text{subject to} \ \sum_{i=1}^{p+1} \tk_i = K, \ \sum_{i=1}^p \tk_i \leq n-1+p, \ \tk_i \in \bbZ_+ \ \forall i \in [p].
\end{align*}
\end{claim}
Let $i_j \overset{\Delta}{=} \min\{i : i \operatorname{frac}(f(p)) \geq j\}$ for all $j \in [m]$, define $k_i = \sum_{j=1}^i \tk_j^{\star \star}$, and observe that $k_i$ corresponds to \eqref{eq: k_i_optimal}. This completes one part of the corollary. To complete the proof, we now characterize the optimal value of $p$.

If, we have $\frac{pK}{p+1} < n - 1 + p/2$, then, by substituting the optimal $\mathbf{\tk}^{\star \star}(p)$ as a function of $p$ in \eqref{eq: reformulated_obj_fn} and defining $\delta_i = k^{\star \star}_i - k_i^\star$, we get
\begin{align*}
    \lefteqn{ \sum_{i=1}^{p+1} \left(\tk_i^{\star\star}\right)^2 + 2\sum_{i=0}^p i\tk_{i+1}^{\star \star}} \\
    &= \sum_{i=1}^{p+1} \left((k_i^\star +\delta_i)^2 + 2(i-1)\left(k_i^\star + \delta_i\right)\right) \\
    &= \sum_{i=1}^{p+1} \left(\left(k_i^\star\right)^2 + 2(i-1)k_i^\star\right) +  2\sum_{i=1}^{p+1} \delta_i\left(k_i^\star +i - 1\right) + \sum_{i=1}^{p+1} \delta_i^2 \\
    &= \sum_{i=1}^{p+1} \left(\left(k_i^\star\right)^2 + 2(i-1)k_i^\star\right) + \sum_{i=1}^{p+1} \delta_i^2 \\
    &=  \sum_{i=1}^{p+1} \left(f(p)^2 - (i-1)^2\right) +  \left(1 - \operatorname{frac}\left(f(p)\right)\right)^2 m + \operatorname{frac}\left(f(p)\right)^2 \left(p+1-m\right) \\
    &= (p+1)f(p)^2 - \frac{p(p+1)(2p+1)}{6} + (p+1)\operatorname{frac}(f(p))\left(1 - \operatorname{frac}(f(p)) \right)
\end{align*}
Similarly, if $\frac{pK}{p+1} < n - 1 + p/2$, then the objective function \eqref{eq: reformulated_obj_fn} can be simplified to get
\begin{align*}
     \sum_{i=1}^{p+1} \left(\tk_i^{\star\star}\right)^2 + 2\sum_{i=0}^p i\tk_{i+1}^{\star \star} ={}& p f(p)^2 - \frac{p(p+1)(2p+1)}{6} + p^2  + p\operatorname{frac}(f(p))\left(1 - \operatorname{frac}(f(p)) \right) \\
     & + (K-n-p+1)^2 + 2p(K-n-p+1) \\
     ={}& p f(p)^2 - \frac{p(p+1)(2p+1)}{6} + p\operatorname{frac}(f(p))\left(1 - \operatorname{frac}(f(p)) \right) + (K-n+1)^2
\end{align*}
Combining the above two cases, we get
\begin{align*}
    p =  \arg\min_{\bar{p} \in [n]} \left(\bar{p}+\mathbbm{1}\left\{\frac{K}{\bar{p}+1} < \frac{n-1}{\bar{p}} + \frac{1}{2} \right\}\right) \left(f(\bar{p})^2 + \operatorname{frac}\left(f(\bar{p})\right) - \operatorname{frac}\left(f(\bar{p})\right)^2\right) - \frac{1}{6}\bar{p}(\bar{p}+1)(2\bar{p}+1)
\end{align*}
This completes the proof.
\hfill $\square$
\endproof
\subsection{Proof of Claims} \label{sec: improve_claims}
\proof{Proof of Claim \ref{claim: CRP_add_one_edge}}
We first show that $G_l^\prime$ for all $l \in \calV_{crp}^\prime$ satisfy the CRP condition. Note that, if $l \neq 0$, then $G_l^\prime = G_l$ which satisfy the CRP condition by definition. Now, we consider $l=0$. Let $\calT_v$ be the transportation polytope corresponding to $(\{\nu_i\}_{i \in \calI_v}, \{\mu_j\}_{j \in \calJ_v}, \calE_v)$. As $\calT_v$ for all $v \in \calV_c$ satisfy the CRP condition, there exists $\bx^v \in \calT_v$ such that $\calB(\bx) = \calE_v$ by Lemma \ref{lemma: CRP_connectedness}. Now, denote the transportation polytope corresponding to $(\{\nu_i\}_{i \in G_0^{\prime}}, \{\mu_j\}_{j \in G_0^{\prime}}, \calE_0^{\prime})$ by $\calT^{\prime}$ and define
\begin{align*}
x_{ij}^{\prime} = \begin{cases}
x_{ij}^v  &\textit{if } (i, j) \in \calE_v \quad \forall v \in \calV_c \\
0 &\textit{otherwise}.
\end{cases} \quad \forall (i, j) \in \calE_0^{\prime}
\end{align*}
Note that, $\bx^{\prime} \in \calT^{\prime}$. This shows that the edges $\bigcup_{v \in \calV_c} \calE_v$ are not redundant in $\calT^{\prime}$. Now, let $(i_{v_1}, j_{v_2}) \in \calE_0^{\prime} \backslash \bigcup_{v \in \calV_c} \calE_v$ be such that $i_{v_1} \in \calI_{v_1}$, $j_{v_2} \in \calJ_{v_2}$ for $v_1, v_2 \in \calV_c$. We construct $\bx^{\prime \prime} \in \calT^{\prime}$ such that $x^{\prime \prime}_{i_{v_1}j_{v_2}}>0$. By the definition of $\calV_c$, there exists a cycle in $D(\calV_{crp}, \calE_{crp})$ containing $(v_1, v_2)$. Let the cycle be $\{v_1, v_2, \hdots, v_l, v_{l+1}\}$, where $v_{l+1}=v_1$, and $v_{k} \in \calV_c$ for all $k \in [l]$. Thus, there exists $i_{v_{k}} \in \calI_{v_{k}}$, $j_{v_{k}} \in \calJ_{v_{k}}$ for all $k \in [l]$ such that $(i_{v_{k}}, j_{v_{k+1}}) \in \calE_0^{\prime}$ for all $k \in [l]$. Now, as $G_{v}$ satisfy the CRP condition for all $v \in \calV_c$, it is a connected graph by Lemma \ref{lemma: CRP_connectedness}. Let $p_k$ be a path in $G_{v_k}$ from $j_{v_k}$ to $i_{v_k}$ for all $k \in [l]$. Using these paths, we define the following cycle in $G_0^{\prime}$:
\begin{align*}
\calA = ((i_{v_{1}},j_{v_{2}}),p_2,(i_{v_{2}},j_{v_{3}}),p_3,\hdots,(i_{v_{l}},j_{v_{1}}),p_1). 
\end{align*}
For $\epsilon = \min_{(i,j): x_{ij}^\prime>0}\{x_{ij}^\prime\}/2$, define $\bx^{\prime\prime}$ as follows:
\begin{align*}
    x_{ij}^{\prime\prime}=\begin{cases}
    x_{ij}^\prime &\textit{if } (i,j) \notin \calA \\
    x_{ij}^\prime+\epsilon &\textit{if } (i,j) \text{ is at odd number position in} \ \calA \\
    x_{ij}^\prime-\epsilon &\textit{if } (i,j) \text{ is at even number position in} \ \calA.
    \end{cases}
\end{align*}
It is easy to verify that $\bx^{\prime \prime} \in \calT^{\prime}$ as we added and subtracted $\epsilon$ from $\bx^{\prime}$ corresponding to a cycle in $G_0^{\prime}$ and $x_{ij}^{\prime} > 0$ for all $(i, j) \in \bigcup_{v \in \calV_c} \calE_v$. Also, note that $\bx^{\prime\prime}_{i_{v_1}j_{v_2}} > 0$ as required. This shows that the set of edges $\calE_0^{\prime} \backslash \bigcup_{v \in \calV_c} \calE_v$ are non redundant. As $G_0^{\prime}$ is a connected graph with no redundant edges, $\calT^\prime$ satisfies the CRP condition by Lemma \ref{lemma: existence_redundant}.

Now, define the graph $D(\calV_{crp}^{\prime}, \calE_{crp}^{\prime})$ with
\begin{align*}
    \calV_{crp}^{\prime} &= \calV_{crp} \backslash \calV_c \cup \{0\}, \quad \calE_{crp}^{\prime} = \left\{(l_1,l_2) \in  \calV_{crp}^{\prime} \times  \calV_{crp}^{\prime} :\exists i \in \calI_{l_1}^{\prime}, j \in \calJ_{l_2}^{\prime}, \textit{s.t. } (i,j) \in \calE\right\}. 
\end{align*}
Note that, $D(\calV_{crp}^{\prime}, \calE_{crp}^{\prime})$ is a DAG. We show this by contradiction. Let $\calA^{\prime}$ be a cycle in $D(\calV_{crp}^{\prime}, \calE_{crp}^{\prime})$. If $0 \notin \calA^{\prime}$, then $\calA^{\prime}$ is a cycle in $D(\calV_{crp}, \calE_{crp})$ as well. This is a contradiction. Now, if $0 \in \calA^{\prime}$, then there exists $v_1, v_2 \in \calV_c$ such that $\calA^{\prime}$ is a path from $v_1$ to $v_2$ in $D(\calV_{crp}, \calE_{crp})$. By definition of $\calV_c$, there exists cycles $\calA_1$ and $\calA_2$ in $D(\calV_{crp}, \calE_{crp} \cup (u_1, u_2))$ such that $v_1 \in \calA_1$, $v_2 \in \calA_2$, and $(u_1, u_2) \in \calA_1, \calA_2$. Now, define
\begin{align*}
\calA^{\prime\prime} = \calA^{\prime} \cup \calA_2[v_2, u_2] \cup \calA_1[u_2, v_1],
\end{align*}
where $\calA_2[v_2, u_2]$ is a path from $v_2$ to $u_2$, and $\calA_1[u_2, v_1]$ is a path from $u_2$ to $v_1$. Note that, $\calA^{\prime\prime}$ is a cycle in $D(\calV_{crp}, \calE_{crp} \cup (u_1, u_2))$ which implies $\calA^{\prime} = 0$ by the definition of $\calV_c$. This is a contradiction as $\calA^{\prime}$ is assumed to be a cycle. Thus, $D(\calV_{crp}^{\prime}, \calE_{crp}^{\prime})$ is a DAG. Denote a topological sorting of $D(\calV_{crp}^{\prime}, \calE_{crp}^{\prime})$ by $\{u_{\alpha(l)}\}_{l \in \calV_{crp}^{\prime}}$ where $\alpha$ is a bijection from $\calV_{crp}^{\prime}$ onto itself. Note that
\begin{align*}
\calJ_{\alpha(l)}^{\prime} \overset{*}{\subseteq} \{j : \exists i \in \calI_{\alpha(l)}^{\prime}, (i, j) \in \calE\} \overset{**}{\subseteq} \bigcup_{k \leq l} \calJ_{\alpha(k)}^{\prime},
\end{align*}
where $(*)$ follows as $G_{\alpha(l)}^{\prime}$ is connected as it satisfies the CRP condition. Next, $(**)$ follows by the definition of topological sorting. Now, consider a disjoint cover of $\calI$ given by $\{\calI^{\prime}_l\}_{l \in \calV_{crp}}$. The corresponding graphs defined as in Proposition \ref{prop: characterization_CRP_decomp} is exactly equal to $\{G_{\alpha(l)^{\prime}}\}_{l \in \calV_{crp}^{\prime}}$. Thus, the claim follows by Proposition \ref{prop: characterization_CRP_decomp}.
\hfill $\square$
\endproof
\proof{Proof of Claim \ref{claim: opt_solution_nlip}}
Let $\lambda$ and $\mu$ be the Lagrange multiplier of the equality and inequality constraints respectively. Then, by the KKT conditions, we have
\begin{subequations} \label{eq: kkt}
\begin{align}
    \tk_i^\star + i - 1 - \lambda - \mu\mathbbm{1}\left\{i \neq p+1\right\} = 0 \quad \forall i \in [p+1] \span \label{eq: kkt-1} \\
    \sum_{i=1}^{p+1} \tk_i^\star = K \span \label{eq: kkt-2} \\
    \text{if} \ \sum_{i=1}^p \tk_i^\star < n - 1 + p \ \text{then} \ \mu = 0 \span \label{eq: kkt-3}
\end{align}
\end{subequations}
First, consider the case when $\sum_{i=1}^p \tk_i^\star < n - 1 + p$, then we have $\mu = 0$. By \eqref{eq: kkt-1} and \eqref{eq: kkt-2}, we get
\begin{align*}
    \tk_i^\star = \frac{K}{p+1} + \frac{p}{2} - i + 1 \quad \forall i \in [p+1].
\end{align*}
To ensure $\sum_{i=1}^p \tk_i^\star < n - 1 + p$, we obtain the condition $Kp/(p+1) < n-1+p/2$. This completes the analysis for the first case.

Now, if $Kp/(p+1) \geq n-1+p/2$, then the inequality constraint must hold with equality as otherwise \eqref{eq: kkt} will have no feasible solutions. Thus, we have $\sum_{i=1}^p \tk_i^\star = n - 1 + p$. This immediately implies that $\tk_{p+1}^\star = K + 1 - n - p$. In addition, using \eqref{eq: kkt-1} and \eqref{eq: kkt-3}, we get
\begin{align*}
    \tk_i^\star = \frac{n-1}{p} + \frac{p+1}{2} - i + 1.
\end{align*}
Combining the two cases, we get that \eqref{eq: optimal_tk_function_of_p} is an optimal solution of the linear relaxation of \eqref{eq: reformulated_obj}. 

Now, first consider the case $\sum_{i=1}^p \tk_i^\star < n - 1 + p$ and define a feasible solution of the original problem given by $\tk_i = \tk_i^\star + \epsilon_i$ where $\sum_{i=1}^{p+1} \epsilon_i = 0$, and $\epsilon_i = \delta + b_i$ where $b_i \in \bbZ$. Then, \eqref{eq: reformulated_obj_fn} for this feasible solution is equal to
\begin{align*}
    \frac{1}{2}\sum_{i=1}^{p+1} \left(\tk_i^\star + \epsilon_i\right)^2 + \sum_{i=0}^{p} i\left(\tk_i^\star+\epsilon_i\right) &= \frac{1}{2} \sum_{i=1}^{p+1} \left(\tk_i^\star\right)^2 + \sum_{i=0}^p \tk_i^\star +\sum_{i=1}^{p+1} \epsilon_i \left(\tk_i^\star +i - 1\right) + \frac{1}{2}\sum_{i=1}^{p+1}\epsilon_i^2 \\
    &\overset{(a)}{=} \gamma_3^\star + \left(\frac{K}{p+1} + \frac{p}{2}\right)\sum_{i=1}^{p+1} \epsilon_i + \frac{1}{2}\sum_{i=1}^{p+1}\epsilon_i^2 \\
    &\overset{(b)}{=} \gamma_3^\star + \frac{1}{2}\sum_{i=1}^{p+1}\epsilon_i^2 \\
    &= \gamma_3^\star + \frac{(p+1)}{2}\delta^2 + \delta \sum_{i=1}^{p+1} b_i + \frac{1}{2}\sum_{i=1}^{p+1} b_i^2 \\
    &\overset{(c)}{=} \gamma_3^\star - \frac{(p+1)}{2}\delta^2 + \frac{1}{2}\sum_{i=1}^{p+1} b_i^2
\end{align*}
where $(a)$ follows by noting that the first two terms combined is equal to the optimal value of the linear relaxation $\gamma_3^\star$. Next, $(b)$ follows by noting that $\sum_{i=1}^{p+1} \epsilon_i = 0$. Further, $(c)$ follows by noting that $0 = \sum_{i=1}^{p+1} \epsilon = (p+1)\delta + \sum_{i=1}^{p+1} b_i$. Now, using this, we obtain the following relation between $\gamma_2^\star$ and $\gamma_3^\star$:
\begin{align*}
    \gamma_2^\star - \gamma_3^\star + \frac{(p+1)}{2}\delta^2 = \min_{\mathbf{b}} \frac{1}{2}\sum_{i=1}^{p+1} b_i^2 \span \\
    \text{subject to} \ \sum_{i=1}^{p+1} b_i = -(p+1) \delta, \ b_i \in \bbZ
\end{align*}
The above optimization problem attains optimality by setting $(p+1)\delta$ number of variables to $-1$ and others to $0$. This completes one part of the proof.

Now, consider the case $\sum_{i=1}^p \tk_i^\star \geq n - 1 + p$ and define a feasible solution of the original problem given by $\tk_i = \tk_i^\star + \epsilon_i$ where $\sum_{i=1}^{p+1} \epsilon_i = 0$, and $\epsilon_i = \delta + b_i$ for $i \in [b]$ and $\epsilon_{p+1} = b_{p+1}$ where $b_i \in \bbZ$. Then, similar to the previous case, \eqref{eq: reformulated_obj_fn} for this feasible solution is equal to
\begin{align*}
    \frac{1}{2}\sum_{i=1}^{p+1} \left(\tk_i^\star + \epsilon_i\right)^2 + \sum_{i=0}^{p} i\left(\tk_i^\star+\epsilon_i\right) &= \gamma_3^\star + \frac{1}{2}\sum_{i=1}^{p+1}\epsilon_i^2 \\
    &= \gamma_3^\star +  \frac{p}{2}\delta^2 + \delta \sum_{i=1}^p b_i + \frac{1}{2}\sum_{i=1}^{p+1} b_i^2 \\
    &= \gamma_3^\star - \frac{p}{2}\delta^2 - \delta b_{p+1} + \frac{1}{2}\sum_{i=1}^{p+1} b_i^2.
\end{align*}
Thus, we have
\begin{subequations} \label{eq: case_2_opt}
\begin{align}
    \gamma_2^\star - \gamma_3^\star + \frac{p}{2}\delta^2 = \min_{\mathbf{b}} \frac{1}{2}\sum_{i=1}^{p+1} b_i^2 - \delta b_{p+1} \span  \\
    \text{subject to} \ \sum_{i=1}^{p+1} b_i = -p \delta, \ b_{p+1} \geq 0 \ b_i \in \bbZ \quad i \in [p+1].
\end{align}
\end{subequations}
Let $\mathbf{b}^\star$ be the optimal solution of the above optimization. First, we show that $b_{p+1}^\star = 0$. Otherwise, if $b_{p+1}^\star > 0$, then pick $l \in [p]$ such that $b_l^\star < 0$ and define 
\begin{align*}
    \tilde{b}_i =\begin{cases}
    b_l^\star + b_{p+1}^\star &\text{if} \ i = l \\
    0 &\text{if} \ i = p+1 \\
    b_i^\star &\text{otherwise}.
    \end{cases}
\end{align*}
The objective function value at $\mathbf{\tilde{b}}$ is equal to
\begin{align*}
    \frac{1}{2}\sum_{i \neq l} (b_i^\star)^2 + \frac{1}{2}\left(b_l^\star + b_{p+1}^\star\right)^2 &= \frac{1}{2}\sum_{i=1}^{p+1} (b_i^\star)^2 + b_l^\star b_{p+1}^\star \\
    &\leq \frac{1}{2}\sum_{i=1}^{p+1} (b_i^\star)^2 - b_{p+1}^\star \\
    &< \frac{1}{2}\sum_{i=1}^{p+1} (b_i^\star)^2 - \delta b_{p+1}^\star.
\end{align*}
Thus, we obtain a contradiction, implying that $b_{p+1}^\star = 0$. Using this, we conclude that optimality of \eqref{eq: case_2_opt} is attained by setting $p \delta$ number of $\{b_i\}_{i \in [p]}$ to $-1$ and others to $0$. This completes the proof. 
 \hfill $\square$
\endproof
\end{APPENDICES}
\end{document}